\documentclass[fleqn,usenatbib]{mnras} 

\usepackage{newtxtext,newtxmath}

\usepackage[T1]{fontenc}
\usepackage{ae,aecompl}
\usepackage{newtxtext,newtxmath}

\usepackage{graphicx}	
\usepackage{amsmath}	
\usepackage{amssymb}	
 
\usepackage{hyperref}
\usepackage{graphicx}
\usepackage{comment}
\usepackage{bm}
\usepackage{wasysym}
\usepackage{cite}
\usepackage{multirow}

\DeclareRobustCommand{\appropto}{\mathrel{\vcenter{
			\offinterlineskip\halign{\hfil$##$\cr 
				\propto\cr\noalign{\kern2pt}\sim\cr\noalign{\kern-2pt}}}}}

\title[The KBC void and $H_0$ tension in $\Lambda$CDM and MOND]{The KBC void and Hubble tension contradict $\Lambda$CDM on a Gpc scale $-$ Milgromian dynamics as a possible solution} 
\author[M. Haslbauer et al.]
{Moritz Haslbauer$^{1,2}$\thanks{Email:	\href{mailto:mhaslbauer@astro.uni-bonn.de}{mhaslbauer@astro.uni-bonn.de} \newline $~~~~~~~~~~~~~~~~$
\href{mailto:mhaslbauer@mpifr-bonn.mpg.de}{mhaslbauer@mpifr-bonn.mpg.de} (Moritz Haslbauer)\newline $~~~~~~~~~~~~~~~~$
\href{mailto:ibanik@astro.uni-bonn.de}{ibanik@astro.uni-bonn.de} (Indranil Banik)}, Indranil Banik$^{1}$\thanks{Alexander von Humboldt Fellow}, and Pavel Kroupa$^{1,3}$\\
$^{1}$Helmholtz-Institut f\"ur Strahlen- und Kernphysik, University of Bonn, Nussallee 14-16, D-53115 Bonn, Germany\\
$^{2}$Max-Planck-Institut f\"ur Radioastronomie, Auf dem H\"ugel 69, D-53121 Bonn, Germany\\
$^{3}$Faculty of Mathematics and Physics, Astronomical Institute, Charles University, V Hole\v{s}ovi\v{c}k\'ach 2, CZ-180 00 Praha 8, Czech Republic\\}

\date{Accepted 2020 July 31. Received 2020 July 14; in original form 2020 May 8}

\pubyear{2020}
\begin{document}
\label{firstpage}
\pagerange{\pageref{firstpage}--\pageref{lastpage}}
\maketitle

\begin{abstract}

The KBC void is a local underdensity with the observed relative density contrast $\delta \equiv 1 - \rho/\rho_{0} = 0.46 \pm 0.06$ between 40 and 300 Mpc around the Local Group. If mass is conserved in the Universe, such a void could explain the $5.3\sigma$ Hubble tension. However, the MXXL simulation shows that the KBC void causes $6.04\sigma$ tension with standard cosmology ($\Lambda$CDM). Combined with the Hubble tension, $\Lambda$CDM is ruled out at $7.09\sigma$ confidence. Consequently, the density and velocity distribution on Gpc scales suggest a long-range modification to gravity. In this context, we consider a cosmological MOND model supplemented with $11 \, \rm{eV}/c^{2}$ sterile neutrinos. We explain why this $\nu$HDM model has a nearly standard expansion history, primordial abundances of light elements, and cosmic microwave background (CMB) anisotropies. In MOND, structure growth is self-regulated by external fields from surrounding structures. We constrain our model parameters with the KBC void density profile, the local Hubble and deceleration parameters derived jointly from supernovae at redshifts $0.023 - 0.15$, time delays in strong lensing systems, and the Local Group velocity relative to the CMB. Our best-fitting model simultaneously explains these observables at the $1.14\%$ confidence level (${2.53 \sigma}$ tension) if the void is embedded in a time-independent external field of ${0.055 \, a_{_0}}$. Thus, we show for the first time that the KBC void can naturally resolve the Hubble tension in Milgromian dynamics. Given the many successful a priori MOND predictions on galaxy scales that are difficult to reconcile with $\Lambda$CDM, Milgromian dynamics supplemented by $11 \, \rm{eV}/c^{2}$ sterile neutrinos may provide a more holistic explanation for astronomical observations across all scales.


\end{abstract}


\section{Introduction}
\label{sec:Introdcution}

The Cosmological Principle (CP) states that the Universe is homogeneous and isotropic on very large scales. This concept is the foundation of the current Lambda-Cold Dark Matter ($\Lambda$CDM) standard model of cosmology \citep{Ostriker_1995}, which assumes that Einstein's General Relativity is valid on all astrophysical scales. Applying it to the non-relativistic outskirts of galaxies yields nearly the same result as Newtonian dynamics $-$ the rotation curve should undergo a Keplerian decline beyond the extent of the luminous matter \citep{Almeida_2016}. The observed flat rotation curves of galaxies \citep[e.g.][]{Babcock_1939, Rubin_1970, Rogstad_1972} demonstrate that Newtonian gravity of the baryons alone is insufficient to hold them together, leading to the concept that each galaxy is surrounded by a CDM halo \citep{Ostriker_1973}. 
However, no experiment has ever confirmed the existence of CDM, with stringent upper limits coming from e.g. null detection of $\gamma$-rays from DM annihilation in dwarf satellites of the Milky Way \citep[MW;][]{Hoof_2020}. In addition to the hypothetical ingredient of CDM, the $\Lambda$CDM model also requires a cosmological constant $\Lambda$ in Einstein's gravitational field equations to explain the anomalous faintness of distant Type Ia supernovae \citep[SNe Ia;][]{Riess_1998,Schmidt_1998,Perlmutter_1999}. $\Lambda$ may be associated to a vacuum energy (dark energy).

This `concordance' flat $\Lambda$CDM model explains the cosmic microwave background (CMB) as relic radiation from the Universe at redshift $z \approx 1100$ \citep[e.g.][]{Bennett_2003, Planck_2018}. The temperature fluctuations within the CMB are of the order $\delta T / T \approx \delta \rho/\rho \approx 10^{-5}$ \citep{Wright_2004}. These are interpreted as tracers of density contrasts in the baryons alone, with the CDM being significantly more clustered by that time due to it not feeling radiation pressure. After recombination, baryons fell into the potential wells of the DM, starting the process of cosmic structure formation via gravitational instability.

Observations have shown that this widely used $\Lambda$CDM model faces several challenges, especially on galactic up to Mpc scales \citep[e.g.][and references therein]{Kroupa_2012, Kroupa_2015}. One of the most serious problems is the distribution of dwarf galaxies in the Local Group (LG). The MW is surrounded by a thin co-rotating disc of satellite galaxies \citep{Kroupa_2005}, which is part of the vast polar structure \citep{Pawlowski_2012} that also includes ultra-faint galaxies, globular clusters, and gas and stellar streams. Recently, \citet{Pawlowski_2020} showed that its kinematic coherence has increased further with Gaia Data Release 2 \citep{Gaia_2018}. A thin plane of co-rotating satellites is also observed around M31 \citep{Ibata_2013}.

It is very difficult to understand these structures if their member satellites are primordial \citep{Pawlowski_2014b}. However, such phase space-correlated structures can arise during an interaction between two disc galaxies, as observed e.g. in the Antennae galaxies \citep{Mirabel_1992}. Due to the higher velocity dispersion of the DM, such tidal dwarf galaxies (TDGs) should be free of DM in $\Lambda$CDM, as shown with simulations of galaxy interactions \citep{Barnes_1992, Wetzstein_2007} and in cosmological simulations \citep{Ploeckinger_2018, Haslbauer_2019a}. This would lead to very low internal velocity dispersions, which are in conflict with observations for satellites of the MW \citep{McGaugh_2010} and M31 \citep{McGaugh_2013}.

A disc of satellites has also been observed around Centaurus~A \citep[Cen~A;][]{Mueller_2018}, suggesting that such structures are ubiquitous and in any case not unique to the LG. Although they may well consist of TDGs, these are quite rare in $\Lambda$CDM due to their weak Newtonian self-gravity \citep{Haslbauer_2019b, Haslbauer_2019a}. This makes the Cen A satellite plane hard to explain even though we lack internal velocity dispersion measurements for its members \citep{Mueller_2018}. A review on satellite planes in the local Universe can be found in \citet{Pawlowski_2018}, who suggested that the TDG hypothesis could work in an alternative gravitational framework where all galaxies are DM-free. We consider this possibility further in Section~\ref{subsec: Milgromian dynamics}. Some evidence in favour of this scenario is the strong correlation between the bulge fractions and the number of satellite galaxies for the MW,  M31, M81, Cen A, and M101 \citep{Javanmardi_2020}. This is unexpected in standard cosmology \citep{Kroupa_2012, Kroupa_2015, Javanmardi_2019}, but may indicate that bulges and satellite galaxies formed simultaneously in galactic interactions.

Although $\Lambda$CDM is widely considered a successful theory in explaining large-scale structure, the observed Universe appears to be much more structured and organized than it predicts. In particular, \citet{Peebles_2010} reported that standard $\Lambda$CDM theory  is in conflict with the distribution of galaxies within $\approx 8 \, \rm{Mpc}$ of the LG. The local void contains much fewer galaxies than expected \citep[e.g.][]{Tikhonov_2009}, while massive galaxies are located away from the matter sheets where they ought to reside. These facts suggest a more rapid growth rate of structure (\citealt{Peebles_2010}; though see \citealt{Xie_2014}).

\citet{Karachentsev_2012} studied the matter distribution of the Local Volume in more detail, finding that the average density of matter within $\approx 50 \, \rm{Mpc}$ is only $\Omega_{\mathrm{m,loc}} = 0.08 \pm 0.02$, much lower than the global cosmic density at the present time \citep[$\Omega_{\mathrm{m},0} = 0.315$,][]{Planck_2018}. This is consistent with a more recent work which obtained $\Omega_{\mathrm{m,loc}} = 0.09-0.14$ within a sphere of radius $40 \, \rm{Mpc}$ around the LG \citep{Karachentsev_2018}. This is striking because the Harrison-Zeldovich spectrum and the current value of $\sigma_{8} = 0.811 \pm 0.006$ \citep{Planck_2018} predict root mean square (rms) density fluctuations of $23\%$ on this scale. Indeed, recent studies have questioned the assumption of homogeneity and isotropy \citep[e.g.][]{Javanmardi_2015, Kroupa_2015, Javanmardi_2017, Bengaly_2018, Colin_2019, Meszaros_2019, Migkas_2020}. 

Therefore, observations of the galaxy distribution on large scales can constrain various cosmological models and their different underlying gravitational theories. In this study, we investigate the local matter density and velocity field within 1~Gpc in $\Lambda$CDM and in a previously developed Milgromian cosmological model \citep{Angus_2009}. This allows us to assess the implications for the CP and Hubble tension.

\subsection{KBC void} \label{subsec:KBC void}

Several observations at different wavelengths have found evidence for a large local underdensity around the LG. The first indication for a deficiency in the galaxy luminosity density was observed in optical samples \citep[e.g.][]{Maddox_1990}. Using the ESO Slice Project galaxy survey that covers $\approx 23 \, \rm{deg^2}$ on the sky, \citet{Zucca_1997} found a local underdensity out to a distance of $\approx 140 h^{-1}\, \rm{Mpc}$ in the $b_{\mathrm{J}}$ band, where $h \approx 0.7$ is the present Hubble constant $H_0$ in units of $100 \, \rm{km\,s^{-1}\,Mpc^{-1}}$. 

Galaxy counts in the near-infrared (NIR) revealed that the local Universe is significantly underdense on a scale of $200-300 h^{-1} \, \rm{Mpc}$ around the LG \citep[e.g.][]{Huang_1997, Frith_2003, Busswell_2004, Frith_2005, Frith_2006, Keenan_2013, Whitbourn_2014}. NIR photometry accurately traces the stellar mass and is therefore a good proxy for the underlying matter distribution.

A local underdensity is also evident in the X-ray galaxy cluster surveys REFLEX II \citep{Boehringer_2015} and CLASSIX \citep{Boehringer_2020}. The latter work found a $15-30\%$ ($10-20\%$) underdensity in the matter distribution within a radius of $\approx 100 \, \rm{Mpc}$ ($140 \, \rm{Mpc}$).

At the opposite end of the spectrum, \citet{Rubart_2013} found that the cosmic radio dipole from the NRAO VLA Sky Survey is ${\approx 4\times}$ stronger than can be explained purely kinematically given the magnitude of the CMB dipole. Interestingly, the radio dipole points towards Galactic coordinates ($245^{\circ}, +43^{\circ}$) which, given the uncertainty of ${\approx 30^\circ}$, is consistent with the direction in which the LG moves with respect to (wrt.) the CMB \citep[$276^{\circ}\pm3^{\circ}, +30^{\circ}\pm3^{\circ};$][]{Kogut_1993}. In a subsequent study, \citet{Rubart_2014} showed that the unusually strong radio dipole could be explained by a single void with a size of $11\%$ of the Hubble distance and a density contrast of $\delta \equiv 1 - \rho/\rho_{0} = 1/3$, where $\rho$ is the local density and $\rho_0$ is the cosmic mean.

Moreover, \citet{Bengaly_2018} studied the dipole anisotropy of galaxy number counts over the redshift range $0.10 < z < 0.35$, revealing a large anisotropy for $z < 0.15$ that could be the imprint of a large local density fluctuation. Thus, a significant local underdensity is evident across the entire electromagnetic spectrum.

Here, we focus on the study by \citet*{Keenan_2013}, who found clear evidence for a large local underdensity by measuring the $K$-band galaxy luminosity function at different distances over a large part of the sky (see their figures 9 and 10). They used the 2M++ catalogue \citep{Lavaux_2011}, which combines photometry from the Two Micron All Sky Survey Extended Source Catalog (2MASS-XSC) with redshifts from the Sloan Digital Sky Survey (SDSS), the Two Micron Redshift Survey (2MRS), and the Six-degree Field Galaxy Redshift Survey (6DFGRS). This sample covers $37\,080 \, \rm{deg^{2}}$ ($90\%$ of the whole sky) and is $\approx 98\%$ complete to a limiting magnitude of $K_{\mathrm{s}} = 13.36$. Using this sample, \citet{Keenan_2013} estimated the luminosity density and derived a relative density contrast of $\delta \approx 0.5$ in the redshift range $0.0025 < z < 0.067$ compared to larger redshifts (see the pink down-pointing triangle in their figure~11, and their table~1). In addition, they also probed the density field to a deeper magnitude limit of $K_{\mathrm{s}} = 14.36$, but only in the SDSS and 6DFGRS regions. This yielded a slightly smaller density contrast of $\delta = 0.46 \pm 0.06$ between $z = 0.01$ ($\approx 40 \, \rm{Mpc}$) and $z = 0.07$ ($\approx 300 \, \rm{Mpc}$; see the light blue dot in their figure~11). In the following, we will show that the Keenan-Barger-Cowie (KBC) void is highly unexpected within the $\Lambda$CDM framework by virtue of its sheer size and depth. In order to minimize the tension, we assume for our analysis that $\delta = 0.46 \pm 0.06$, and refer to this as the KBC void. Calculating the $K$-band luminosity density in different regions suggests that it reaches the cosmic mean at a distance of $\approx 500 \, \rm{Mpc}$.

In the $\Lambda$CDM framework, the existence of such a deep and extended void is a puzzle given the expected Harrison-Zeldovich scale-invariant power spectrum, which states that the power $P \left( L \right)$ on some length scale $L$ varies as $P(L) \propto L^{-n_s}$, with $n_s = 1$ \citep{Harrison_1970, Zeldovich_1972}. Since the CMB anisotropies require a power of $\sigma_8 = 0.811 \pm 0.006$ on a scale of $8h^{-1}$~Mpc \citep{Planck_2018}, we expect density fluctuations of only $\approx 3.2\%$ between spheres of radius $L = 300$~Mpc.

Combining measurement errors with cosmic variance, we can estimate that the KBC void would falsify the $\Lambda$CDM model by well over $5\sigma$ because
\begin{eqnarray}
	\frac{0.46}{\sqrt{0.06^2 + 0.032^2}} ~=~ 6.8 \, .
\end{eqnarray}
In Section~\ref{sec:LCDM framework}, we provide a much more sophisticated analysis of how likely the KBC void is in standard cosmology. Since the measurement uncertainty of $6\%$ is much larger than the cosmic variance of $3.2\%$, the latter is not the main source of uncertainty in how far off $\Lambda$CDM is from matching the observations $-$ as explicitly calculated in Section~\ref{KBC void in LCDM}. Consequently, if we assume that $\Lambda$CDM is the correct model, the most likely explanation for the detection of such a deep void would be a measurement error. However, the KBC void is evident over the entire electromagnetic spectrum.

The above prediction of 3.2\% rests on two fundamental assumptions $-$ that the CMB reflects baryonic density fluctuations at $z = 1100$, and that General Relativity is valid on all scales. The existence of the KBC void might indicate that either or both of these assumptions must be relaxed. In this contribution, we focus on modifying gravity because the standard approach leads to problems in galaxies \citep[e.g.][and references therein]{Kroupa_2012, Kroupa_2015}.

A large local void should also have implications for local measurements of cosmological parameters such as the Hubble constant and deceleration parameter. If mass is conserved in the Universe and it was nearly homogeneous initially, a large fractional underdensity would show up in the velocity field. This is because the co-moving radius enclosing a fixed amount of mass must exceed its initial value, and changes in co-moving coordinates imply a peculiar velocity.

Suppose that we are living near the centre of a void whose true density relative to the cosmic mean is
\begin{eqnarray}
    \frac{\rho}{\rho_0} ~\equiv~ \alpha ~\equiv~ 1 - \delta \, .
\end{eqnarray}
This implies that the co-moving radius enclosing a fixed mass must exceed its initial value by a factor $\alpha^{-1/3}$. Depending on details of how the void grows, the impact on the locally measured Hubble parameter would be approximately the same. In other words,
\begin{eqnarray}
    \frac{H_{0}^{\mathrm{local}}}{{H_{0}^{\mathrm{global}}}} ~\approx~ \alpha^{-\frac{1}{3}} \, ,
    \label{eq:H_0_impact_a}
\end{eqnarray}
where $H_{0}^{\mathrm{local}}$ is the locally measured $H_{0}$, whose background (true) value is $H_{0}^{\mathrm{global}} \equiv \dot{a}/a$ at the present time, with $a$ the cosmic scale factor and an overdot indicating a time derivative. The mismatch between these $H_{0}$ values would create a redshift space distortion (RSD) effect whereby the physical volume of a survey with known redshift range would be reduced by a factor $\alpha$ compared to the case of no void. In this way, RSD would further reduce the observed $\alpha_{\mathrm{obs}}$ by a factor $\alpha$ if it is not accounted for and a constant $H_0$ is used to convert redshifts to distances \citep[as done in the work of][see their section 4.7]{Keenan_2013}. Thus, we expect that 
\begin{eqnarray}
    \alpha_{\mathrm{obs}} ~=~ \alpha^2 \, .
    \label{eq:alpha_impact}
\end{eqnarray}
Combining Equations~\ref{eq:H_0_impact_a} and \ref{eq:alpha_impact}, we get that
\begin{eqnarray}
    \frac{H_{0}^{\mathrm{local}}}{{H_{0}^{\mathrm{global}}}} ~\approx~ \alpha_{\mathrm{obs}}^{-\frac{1}{6}} \, .
    \label{eq:H_0_impact}
\end{eqnarray}
Given that $\alpha_{\mathrm{obs}} = 0.54$, the measured $H_{0}^{\mathrm{local}}$ should exceed the background value $H_{0}^{\mathrm{global}}$ by ${0.54}^{-1/6}$, i.e. by 11\%. This would raise $H_0$ from the Planck-based prediction of $67.4 \, \rm{km\,s^{-1}\,Mpc^{-1}}$ \citep{Planck_2018} to $74.7 \, \rm{km\,s^{-1}\,Mpc^{-1}}$, which is very close to the observed value (Section~\ref{subsec:Hubble tension}). This is unlikely to be a coincidence $-$ it is more parsimoniously explained as a consequence of the observed void under the standard assumption of matter conservation.

\subsection{Hubble tension} \label{subsec:Hubble tension}

In this context, we consider the Hubble tension, a statistically significant discrepancy between the locally measured cosmic expansion rate and the $\Lambda$CDM prediction based on the early universe properties needed to match the CMB power spectrum \citep[e.g.][]{Riess_2020}. The local Hubble constant can be determined through the distance ladder technique. Recently, the Supernova $H_0$ for the Equation of State (SH0ES) team \citep{Riess_2019} calibrated the distance ladder with eclipsing binaries in the Large Magellanic Cloud, masers in NGC 4258, and parallaxes of Galactic Cepheid variables via the Leavitt law. They derived a local Hubble constant of $H_{0}^{\mathrm{local}} = 74.03\pm1.42\,\rm{km\,s^{-1}\,Mpc^{-1}}$, which results in $4.4\sigma$ tension with the Planck-based prediction \citep[$H_{0}^{\mathrm{Planck}} = 67.4\pm0.5\,\rm{km\,s^{-1}\,Mpc^{-1}}$;][]{Planck_2018}.

The systematic error of the Cepheid background subtraction is only $0.029 \pm 0.037 \,\rm{mag}$, which is not sufficient to explain the $\approx 0.2\,\rm{mag}$ Hubble tension \citep{Riess_2020b}. Moreover, calibrating the SN Ia luminosity using instead Mira variables in the galaxy NGC 1559 with periods of $240 - 400$ d and using NGC 4258 (the Large Magellanic Cloud) as an anchor, \citet{Huang_2020} obtained $H_{0}^{\mathrm{local}} = 72.7\pm4.6\,\rm{km\,s^{-1}\,Mpc^{-1}}$ ($H_{0}^{\mathrm{local}} = 73.9\pm4.3\,\rm{km\,s^{-1}\,Mpc^{-1}}$; see also their table~6 and figure~11). Both values are consistent with $H_{0}^{\mathrm{local}}$ derived from Cepheid variables within the $1 \sigma$ confidence range, though the Mira-calibrated $H_{0}$ is less precise.

It is also possible to go beyond the traditional Cepheid-SN Ia route using Type II SNe as standard candles. These yield a high $H_{0}^{\mathrm{local}}$ of $75.8_{-4.9}^{+5.2}\,\rm{km\,s^{-1}\,Mpc^{-1}}$, which is very consistent with $H_{0}$ derived from Type Ia SNe $-$ albeit with larger uncertainties \citep{deJaeger_2020}. Thus, systematic errors in Type Ia SNe data are likely not driving the Hubble tension.

\citet{Camarena_2020} analysed the Pantheon SNe Ia sample without fixing the deceleration parameter ($q \equiv -a\ddot{a}/\dot{a}^2$) to the present $\Lambda$CDM prediction of $q_{_0} = -0.55$. They jointly derived $H_{0}^{\mathrm{local}} = 75.35\pm1.68\,\rm{km\,s^{-1}\,Mpc^{-1}}$ and $q_{_0} = -1.08\pm0.29$ from SNe in the redshift range $0.023 \leq z \leq 0.15$. This is in $4.54\sigma$ tension with $\Lambda$CDM. The unexpectedly low $q_{_0}$ is robust to the choice of data set \citep[table~5 of][]{Camarena_2020b}.

Interestingly, it is highly implausible to get such low $q_{_0}$ values at the background level. Even in a pure dark energy-dominated (de Sitter) universe, it is not possible to get $q_{_0} < -1$. Thus, first- and second-order effects in the local Hubble diagram seem to provide additional evidence for the KBC void. To quantify this, we compare the standard expansion rate history ($H_{0} = 67.4\,\rm{km\,s^{-1}\,Mpc^{-1}}$, $q_{_0}=-0.55$) with an extrapolation of the \citet{Camarena_2020} results. Approximating both as quadratic functions of time $t$ with $a(t_{_0}) \equiv 1$ at the present time $t_{_0}$, we get that the reconstructed $a(t)$ parabolas coincide $4.2\,\rm{Gyr}$ ago. This provides strong evidence for a Gpc-scale void independently of the galaxy luminosity density (discussed earlier in Section~\ref{subsec:KBC void}). 

A method of measuring $H_{0}$ independently of the cosmic distance ladder relies on time delays between multiple images of the same source, as occurs in strong gravitational lensing. \citet{Jee_2019} calibrated the SNe data with angular diameter distances to two gravitational lenses, obtaining $H_{0} = 82.4_{-8.3}^{+8.4}\,\rm{km\,s^{-1}\,Mpc^{-1}}$ for a flat $\Lambda$CDM cosmology. Although the uncertainties are quite large, their $H_{0}$ also exceeds the Planck prediction.

\citet{Shajib_2020} measured $H_{0} = 74.2_{-3.0}^{+2.7}\,\rm{km\,s^{-1}\,Mpc^{-1}}$ from the strong lens system DES J0408$-$5354, whose deflector lies at an angular diameter distance of $D_{d} = 1711_{-280}^{+376}\,\rm{Mpc}$ ($z = 0.597$). This is broadly consistent with measurements of the $H_{0}$ Lenses in COSMOGRAIL's Wellspring \citep[H0LiCOW;][]{Wong_2020}. Using a blinded analysis protocol (see their section~3.6), they obtained $H_{0} = 73.3_{-1.8}^{+1.7}\,\rm{km\,s^{-1}\,Mpc^{-1}}$ from six lensed quasar systems in the redshift range $z = 0.295-0.745$. Combining their results with the measurement of \citet{Riess_2019} leads to a $5.3\sigma$ discrepancy with $\Lambda$CDM expectations based on the CMB \citep{Wong_2020}. The latter work showed for the first time that the Hubble tension exceeds the $5 \sigma$ threshold typically used to judge the validity of scientific theories.

Although \citet{Kochanek_2020} suggested there might be biases in the strong lensing analysis causing $\approx 10\%$ uncertainties on the inferred $H_{0}$, \citet{Pandey_2019} showed that the SNe and strong lensing measurements are consistent and likely have systematics much smaller than the Hubble tension, as also found by \citet{Millon_2020}. Indeed, the near-perfect agreement between the SNe and lensing determinations despite the blinded protocol of the latter does suggest rather small uncertainties. Moreover, \citet{Wong_2020} found that $H_{0}$ measured from strong lensing decreases as a function of lens redshift at a significance of $1.9\sigma$. Their measurements converge towards the Planck prediction for more distant lenses (see their figure~A1). This again strongly suggests that the Hubble tension is indeed driven by a local environmental effect.

Another technique to determine $H_{0}$ uses maser-derived distance and velocity measurements, as done by the Megamaser Cosmology Project \citep{Reid_2009}. This method is independent of distance ladders, standard candles, and the CMB. It also faces rather different systematics to techniques that rely on gravitational lensing \citep{Pesce_2020}. They used measurements for the six maser galaxies UGC 3789, NGC 6264, NGC 6323, NGC 5765b, CGCG 074-064, and NGC 4258. Except for the well-studied case of NGC 4258 \citep[e.g.][]{Reid_2019}, these galaxies are located at distances between $51.5_{-4.0}^{+4.5}\,\rm{Mpc}$ and $132.1_{-17}^{+21}\,\rm{Mpc}$. The resulting $H_{0}^{\mathrm{local}} = 73.9 \pm 3.0 \, \rm{km\,s^{-1}\,Mpc^{-1}}$, consistent with \citet{Wong_2020} and again larger than predicted by Planck.

So far, we have distinguished between the Planck prediction and $H_{0}$ measurements from the local Universe that avoid assumptions about early Universe physics. Baryon acoustic oscillation (BAO) measurements combine the two through a CMB-based prior on the sound horizon at the time of last scattering. The co-moving length of this standard ruler is assumed to remain fixed, allowing its angular size at different epochs to constrain the expansion history \citep{Eisenstein_2005}. Such BAO-based $H_0$ measurements are available from redshift surveys at effective redshifts of $z_{\mathrm{eff}} = 0.38, 0.51$, and $0.61$ \citep{Alam_2017}, with the range recently extended to $z_{\mathrm{eff}} \approx 1.5$ \citep{Zhang_2019}. These yield a Hubble parameter consistent with the Planck prediction.

The combination of clustering and weak lensing data, BAO, and light element abundances gives $67.4_{-1.2}^{+1.1}\,\rm{km\,s^{-1}\,Mpc^{-1}}$ \citep{Abbott_2018}. Estimating $H_{0}$ using cosmic chronometers yields a nearly direct measure of the background cosmology. This is also consistent with Planck \citep{Gomez_2018}. Assuming spatial flatness of the Universe, \citet{Ruan_2019} combined cosmic chronometers with information on HII galaxies to show that the true value of $\dot{a}$ is much closer to the Planck value than the local value of \citet{Riess_2016}, with the latter discrepant at $\approx 3\sigma$.

\citet{Migkas_2020} inferred $H_{0}$ from the X-ray luminosity-temperature relation of galaxy clusters, finding that it ranges from $65.20 \pm 1.48$ to $76.64 \pm 1.41 \, \rm{km\,s^{-1}\,Mpc^{-1}}$ for different sky regions (see their figure~23). This range is similar to that between $H_{0}^{\mathrm{global}}$ \citep{Planck_2018} and $H_{0}^{\mathrm{local}}$ as found using SNe \citep[e.g.][]{Riess_2016, Riess_2019, Camarena_2020} or strong lensing systems \citep{Wong_2020}. The apparent anisotropy of the local velocity field could potentially be caused by our off-centre location within the KBC void, a non spherical void shape, or a combination of both. However, these considerations are beyond the scope of this work.

Remarkably, all these studies reveal that only the low-redshift probes prefer a high value for the Hubble constant, with high-redshift probes yielding similar results to the Planck-based prediction (see e.g. figure~12 in \citealt{Wong_2020}, or figure~1 in \citealt{Verde_2019}). Some recent reviews on the Hubble tension can be found in \citet{Verde_2019} and \citet{Riess_2020}. All these results point to the overall picture that the Hubble tension is driven by a local environmental effect like a void. In particular, the KBC void shows up not only in galaxy counts but also in the velocity field as an unexpected first and second time derivative of the apparent scale factor (as evidenced by the reported anomalies in $H_{0}$ and $q_{_0}$, respectively). As discussed in Section~\ref{subsec:KBC void}, a large local underdensity can potentially resolve the Hubble tension if mass conservation is assumed. Therefore, this would be a natural resolution to the Hubble tension that would minimize adjustments to the $\Lambda$CDM model on cosmological scales. In particular, there would be no need to assume a novel expansion rate history driven by yet more undetected sources such as early dark energy \citep[e.g.][]{Karwal_2016, Alexander_2019, Poulin_2019, Sakstein_2019}.\footnote{The work of \citet{Hill_2020} argues that early dark energy cannot resolve the Hubble tension due to constraints from other data.} Instead, the standard $\Lambda$CDM expansion rate history could be preserved. In Section \ref{subsec:Claimed_problems}, we discuss some of the objections to this approach.

The works of \citet{Enea_2018} and \citet{Shanks_2019b} constitute attempts to relate the Hubble tension and KBC void on the basis of mass conservation. In a next step, one has to perform more sophisticated dynamical modelling with reasonable initial conditions provided by the CMB. As we will argue, this is not possible with the standard governing equations of $\Lambda$CDM (Section~\ref{subsec:Comparison with observations}). In particular, \citet{Macpherson_2018} explicitly showed that cosmic variance caused by inhomogeneities of the underlying density field cannot resolve the Hubble tension. This is because the expected cosmic variance is too low, implying the Hubble tension and KBC void must both be measurement errors. Given the very different ways in which they are measured, this is highly implausible.

Thus, a large void and high $H_{0}^{\mathrm{local}}$ could well point to a different theory where both are explained by enhancing the long-range strength of gravity, which would promote the growth of structure. In principle, any alternative cosmological model that enhances cosmic variance through faster structure formation could explain the KBC void and Hubble tension, insofar as the model faces the Hubble tension. However, it is important for the model to explain phenomena in addition to those for which the model was explicitly designed, and to address observations on galaxy scales. Therefore, we concentrate on detailed dynamical modelling in the framework of an approach known to satisfy galaxy-scale constraints, and to promote the growth of structure on larger scales.

\subsection{Milgromian dynamics} \label{subsec: Milgromian dynamics}

\citet{Milgrom_1983} originally developed Milgromian dynamics (MOND) to explain the flattening of galactic rotation curves without the need of massive CDM haloes. MOND is a classical potential theory of gravity with a Lagrangian formalism \citep{Bekenstein_1984}. It explains the dynamical effects usually attributed to CDM by an acceleration-dependent modification to Newtonian gravity. In particular, the gravity at radius $r$ from an isolated point mass $M$ becomes 
\begin{eqnarray}
    g = \frac{\sqrt{GMa_{_0}}}{r} \quad \text{for} \quad r \gg r_{M} \equiv \sqrt{\frac{GM}{a_{_0}}} \, ,
    \label{eq:MOND_deep}
\end{eqnarray} 
where $G$ is the Newtonian gravitational constant, and $a_{_0}$ is Milgrom's constant. Empirically, $a_{_0} = 1.2 \times 10^{-10} \, \rm{m\,s^{-2}}$ to match galaxy rotation curves \citep[e.g.][]{Begeman_1991, McGaugh_2011}.

For a more complicated mass distribution, $\bm{g}$ follows a non-relativistic field equation \citep{Bekenstein_1984}. We use a more computer-friendly version known as quasilinear MOND \citep[QUMOND;][]{Milgrom_2010}. In this approach,
\begin{eqnarray}
    \nabla^{2} \Phi ~=~ - \nabla \cdot \left[ \nu \left( g_{_\mathrm{N}} \right) \bm{g}_{_\mathrm{N}} \right] \, ,
    \label{eq:MOND_basic}
\end{eqnarray}
where $\Phi$ is the gravitational potential, $\bm{g}_{_\mathrm{N}}$ is the Newtonian gravitational field, and $r \equiv \left| \bm{r} \right|$ for any vector $\bm{r}$. The function $\nu \left( g_{_\mathrm{N}} \right)$ interpolates between the Newtonian ($\left| \nabla \Phi \right| \gg a_{_0} $) and deep-MOND ($\left| \nabla \Phi \right| \ll a_{_0}$) regimes. Throughout this project, we apply the widely used `simple' interpolating function \citep{Famaey_2005}:
\begin{eqnarray}
    \nu(g_{_\mathrm{N}}) ~=~ \frac{1}{2} + \sqrt{\frac{1}{4} + \frac{a_{_0}}{g_{_\mathrm{N}}}} \, .
    \label{eq:MOND_simple_interpolation_function}
\end{eqnarray}
This closely approximates the empirically determined radial acceleration relation (RAR) between $\bm{g}_{_\mathrm{N}}$ obtained from photometry and $\bm{g} \equiv -\nabla \Phi$ obtained from rotation curves \citep{McGaugh_2016, Lelli_2017}. Our void models are not much affected by the choice of $\nu$ function as they are deep in the MOND regime. This is because any local void solution to the Hubble tension must generate peculiar velocities of $\approx 7 \, \rm{km\,s^{-1}\,Mpc^{-1}}$ in a Hubble time. For a void with size of 300 Mpc, this implies an acceleration of only ${0.04 \, a_{_0}}$. Since this is $\ll a_{_0}$, we expect MOND to have a significant effect on the void dynamics.

Equation~\ref{eq:MOND_deep} implies the baryonic Tully-Fisher relation \citep[BTFR;][]{McGaugh_2000a}, namely that
\begin{eqnarray}
M_{\mathrm{b}} \propto {v_{\mathrm{f}}}^{\xi} \, ,
    \label{eq:MOND_BTFR}
\end{eqnarray}
where $M_{\mathrm{b}}$ is the baryonic mass, $v_{\mathrm{f}}$ is the asymptotic rotation velocity of a disc galaxy, and the exponent $\xi = 4$. Empirically, a tight relation of this form is evident with $\xi \approx 3-4$ \citep[e.g.][]{McGaugh_2000a, McGaugh_2005, Stark_2009, McGaugh_2011, TorresFlores_2011, Ponomareva_2018}. The more recent investigations put $\xi$ very close to the MOND-predicted value of 4, which is also what we expect empirically based on the RAR. Since $v_{\mathrm{f}}$ can be measured independently of distance but $M_{\mathrm{b}}$ depends on the adopted distance, the BTFR provides another independent method to obtain $H_{0}^{\mathrm{local}}$. Recently, \citet{Schombert_2020} calibrated the BTFR with redshift-independent distance measurements from Cepheids and/or the tip magnitude of the red giant branch for 30 galaxies in the Spitzer Photometry and Accurate Rotation Curves catalogue \citep[SPARC;][]{Lelli_2016} 
and 20 galaxies from \citet{Ponomareva_2018}. The so-calibrated BTFR was then applied to 95 independent SPARC galaxies for which only the redshift is known. Since the SPARC catalogue contains galaxies up to distances of $\approx 130\,\rm{Mpc}$, \citet{Schombert_2020} derived $H_{0}$ of the very local Universe. They got $H_{0}^{\mathrm{local}} = 75.1\pm2.3(\mathrm{stat})\pm1.5(\mathrm{sys})\,\rm{km\,s^{-1}\,Mpc^{-1}}$ (see also their table~5). This is quite consistent with other measurements from the late Universe and significantly exceeds the $\Lambda$CDM prediction based on the CMB (Section~\ref{subsec:Hubble tension}). Interestingly, the dominant source of systematic uncertainty is how to correct redshifts of SPARC galaxies for peculiar velocities induced by large-scale structure. This points towards mis-modelled peculiar velocities as a possible cause for the entire Hubble tension.

According to Equation~\ref{eq:MOND_basic}, MOND is non-linear in the acceleration, which yields the interesting concept of the external field effect \citep[EFE;][]{Milgrom_1986}. In contrast to Newtonian gravity, the non-linearity of Milgrom's law causes the internal gravitational forces within a MONDian subsystem to be affected by the external gravitational field from its environment \emph{even without any tides}. This breaks the strong equivalence principle. The EFE has likely been observed in the declining rotation curves of some disc galaxies \citep{Haghi_2016} and the internal dynamics of dwarfs. For example, Crater~II is a diffuse dwarf satellite galaxy of the MW at a distance of $\approx 120 \, \rm{kpc}$ \citep{Torrealba_2016}. Its observed velocity dispersion of $2.7 \pm 0.3 \, \rm{km\,s^{-1}}$ \citep{Caldwell_2017} is below the isolated MOND prediction of $4 \, \rm{km\,s^{-1}}$ \citep{McGaugh_2016}. Taking into account the Galactic EFE reduces the MOND prediction to $2.1_{-0.6}^{+0.9} \, \rm{km\,s^{-1}}$, matching the observed value within uncertainties. Similar examples are the ultra-diffuse dwarf galaxies Dragonfly 2 (DF2) and DF4, where the MOND predictions agree with observations only if the EFE is included \citep{Kroupa_2018, Haghi_2019a}. For the more isolated galaxy DF44, the MOND prediction without the EFE is consistent with observations \citep{Bilek_2019, Haghi_2019b}.

The EFE is also important within the MW, whose MONDian escape velocity curve is similar to observations \citep{Banik_2018c}. Since Equation~\ref{eq:MOND_deep} yields a logarithmically divergent potential, escape from an isolated object is not possible in MOND unless the EFE is taken into account. Recently, \citet{Pittordis_2019} showed that MOND without an EFE is completely ruled out by the observed relative velocity distribution of wide binary stars in the Solar neighbourhood at separations of $\approx 10$ kAU. Including the EFE leads to nearly Newtonian behaviour, though the predicted 20\% difference is likely detectable in a more thorough analysis \citep{Banik_2018d} that must include contamination by undetected close companions \citep{Clarke_2020}.

In addition to its successes with internal dynamics of galaxies \citep[reviewed in][]{Famaey_2012}, MOND may also explain the discs of satellites around the MW and M31 as TDGs born out of a past MW-M31 flyby. A previous close interaction is required in MOND \citep{Zhao_2013} due to the almost radial MW-M31 orbit \citep{Van_der_Marel_2012, Van_der_Marel_2019}. In such an interaction, structures resembling satellite planes can be formed \citep{Bilek_2018}. Using restricted $N$-body models to explore a wide range of flyby geometries, \citet{Banik_2018a} identified models where the tidal debris around the MW and M31 align with their observed satellite planes and have a similar radial extent. A past MW-M31 interaction would naturally explain the apparent correlation between their satellite planes, and with other structures in the LG \citep{Pawlowski_2014}. It may also account for the anomalous kinematics of the NGC 3109 association, which is difficult to understand in $\Lambda$CDM \citep{Peebles_2017, Banik_2018b}.

Interestingly, there is an order of magnitude coincidence between the value of $a_{_0}$ and the cosmic acceleration rate:
\begin{eqnarray}
    2 \mathrm{\pi} a_{_0} ~\approx~ c H_{0} ~\approx~ c^{2} \sqrt{\Lambda/3} \, , 
    \label{eq:MOND_coincidence}
\end{eqnarray}
where $c$ is the speed of light \citep{Milgrom_1983}. This may indicate that MOND is related to a fundamental theory of quantum gravity \citep[e.g.][]{Milgrom_1999, Pazy_2013, Smolin_2017, Verlinde_2017}. A bigger clue would come from tighter empirical constraints on the time evolution of $a_{_0}$, which at present are still weak \citep{Milgrom_2017}. Even so, his work showed that current data are sufficient to rule out the $a^{-3/2}$ scaling required by the model of \citet{Zhao_2008}, which additionally would have a very significant impact on the CMB (Sections~\ref{subsubsec:Radiation dominanted era and CMB} and \ref{subsubsec:Theoretical_assumptions}).

Another intriguing coincidence is that the total matter density is very nearly $2 \mathrm{\pi}$ times the baryonic density, i.e. $\Omega_{\mathrm{m}} \approx 2 \mathrm{\pi} \Omega_{\mathrm{b}}$ \citep{Milgrom_2020b}. This could imply that the effective gravitational constant in a MONDian Friedmann equation is a factor of $2 \mathrm{\pi}$ larger than for a system decoupled from the cosmic expansion. However, we will not follow this interpretation here.

The first relativistic version of MOND was developed by \citet{Bekenstein_2004}. This was modified slightly by \citet{Skordis_2019} so that gravitational waves propagate at the speed of light, as required for consistency with the near-simultaneous detection of gravitational waves and their electromagnetic counterpart \citep{Abbott_2017}. The theory of \citet{Skordis_2019} allows solutions where the background cosmology follows the standard Friedmann equations to high precision (see their section 4). We discuss this further in Section \ref{subsec:nuHDM cosmological model}, where we explain why the expansion rate history and the power spectrum of the CMB should be nearly the same as in $\Lambda$CDM. Thus, MOND would suffer from the Hubble tension in just the same way as $\Lambda$CDM if $H_{0}^{\mathrm{local}} = \dot{a}$ at the sub-per cent level.

Fortunately, this might not be the case $-$ \citet{Sanders_1998} showed that due to the long-range modification to gravity, MOND produces much larger and deeper voids than predicted by $\Lambda$CDM cosmology. Thus, MOND could be a promising framework to explain both the KBC void and the Hubble tension. We therefore extrapolate Milgrom's law of gravity from sub-kpc to Gpc scales. For the first time, we study the Hubble tension and KBC void in the context of MOND. We emphasize that MOND was originally designed to address discrepancies on galactic scales \citep{Milgrom_1983}, so no new assumptions are made specifically to address the latest data on the low-$z$ distance-redshift relation and galaxy counts $-$ apart from the usual assumption that the background follows a standard evolution to high precision (Section \ref{subsubsec:Background cosmology}), and that MOND applies only to density deviations from the cosmic mean \citep[e.g.][]{Llinares_2008, Angus_2011, Angus_2013, Katz_2013, Candlish_2016}. In this context, we aim to provide a unified explanation for both the dynamical discrepancies on galaxy scales and the $z \la 0.2$ matter density and velocity field given current constraints from the CMB.

The layout of this paper is as follows: In Section~\ref{sec:LCDM framework}, we quantify the likelihood of the observed KBC void and how it might relate to the Hubble tension in a $\Lambda$CDM context. After introducing a cosmological MOND model in Section~\ref{sec:MOND framework}, we compare it to observations of the local Universe (Section~\ref{sec:Results of MOND simulation}). The implications for $\Lambda$CDM and MOND cosmologies are discussed in Section~\ref{sec:Discussion}. We finally conclude in Section~\ref{sec:Conclusion}. Throughout this paper, co-moving distances are marked with the prefix `c' (e.g. cMpc, cGpc).

\section{\texorpdfstring{$\Lambda$CDM}{LCDM} framework} \label{sec:LCDM framework}

In this section, we describe how we use a cosmological $\Lambda$CDM simulation to quantify cosmic variance and thereby determine the likelihood of finding ourselves inside the observed KBC void in standard cosmology. We also consider the implications of our results when combined with the Hubble tension.

\subsection{Cosmic variance in the Millennium XXL simulation} \label{subsec:MXXL simulation}

Millennium XXL \citep[MXXL;][]{Angulo_2012} is a standard $\Lambda$CDM cosmological simulation that evolves $6720^{3}$ DM particles from $z = 63$ forwards to $z = 0$. Though it only considers DM, baryonic physics should have a negligible role on the 300 Mpc scale we consider. The simulation box has a length of $3h^{-1} \, \rm{cGpc}$, resulting in a volume that is ${216 \times}$ larger than that of the Millennium simulation \citep{Springel_2005}. The mass of a particle is $8.456 \times 10^{9} \, \rm{M_{\odot}}$ and its Plummer-equivalent softening length is $13.7 \, \rm{kpc}$. The MXXL simulation assumes a flat $\Lambda$CDM cosmology consistent with WMAP-7 results, i.e. the present matter density parameter is $\Omega_{\mathrm{m,0}}=0.25$, that of dark energy is $\Omega_{\mathrm{\Lambda,0}}=0.75$, $\sigma_{8}=0.9$, $H_{0}=73 \, \rm{km\,s^{-1}\,Mpc^{-1}}$, and the power spectrum is assumed to be of the Harrison-Zeldovich form ($n_{\mathrm{s}} = 1$). The baryonic mass of each subhalo is obtained by applying the semi-analytic galaxy formation code \textsc{l-galaxies} \citep{Springel_2005} to the MXXL data \citep[see also section~2.2 in][]{Angulo_2014}.

We use MXXL to calculate the relative density contrast given by the stellar mass distribution in subhaloes with stellar mass $M_{*} > 10^{10} h^{-1} \, \rm{M_{\odot}}$ at $z = 0$. For this purpose, we consider $10^6$ vantage points distributed on a Cartesian grid with a spacing of $30 h^{-1} \, \rm{Mpc}$ in each direction. To maximize the accuracy of our results, we use the nearest subhalo as our final choice for the vantage point. Our adopted minimum mass avoids an excessive computational cost, but still leaves enough subhaloes to accurately determine the expected cosmic variance. Using only stellar masses makes our results more comparable to observations in the NIR.

We need to allow for the incomplete sky coverage of \citet{Keenan_2013}. Following their section 2.5, we adopt a sky area of $37\,080 \, \rm{deg^{2}}$, which in dimensionless units is
\begin{eqnarray}
	A ~=~ 37080 \times \left( \frac{\mathrm{\pi}}{180} \right)^{2} \, .
	\label{eq:mimic_observations_A}
\end{eqnarray}
We assume the incompleteness is caused by observational difficulties at low Galactic latitudes. Thus, we define a mock Galactic spin axis by randomly generating a unit vector $\widehat{\bm{n}}_i$ drawn from an isotropic distribution. We can then define an angle $\theta_j$ based on the direction towards another subhalo at position $\bm{r}_j$ relative to our vantage point.
\begin{eqnarray}
	\cos \theta_{j} ~\equiv~ \frac{\bm{r}_j \cdot \widehat{\bm{n}}_i}{r_j} \, .
	\label{eq:mimic_observations_costheta_i}
\end{eqnarray}
The subscript $i$ refers to the vantage point, while $j$ refers to another subhalo observed from there. We mimic incomplete sky coverage by requiring that
\begin{eqnarray}
    \label{eq:mimic_observations_condidition_1}
    \left| \cos \theta_{j} \right| &>& \cos \theta_{\mathrm{obs}} \, , \quad \mathrm{where} \\
    \cos \theta_{\mathrm{obs}} &=& 1 - \frac{A}{4 \mathrm{\pi}} \, .
\end{eqnarray}
Since most of the sky is surveyed, $\cos \theta_{\mathrm{obs}} = 0.10$.

The observed density contrast is calculated for galaxies in the redshift range $0.01<z<0.07$ \citep[table~1 in][]{Keenan_2013}. Therefore, we further require selected subhaloes to satisfy
\begin{eqnarray}
	r_{\mathrm{min}} ~<~ r_j ~<~ r_{\mathrm{max}} \, ,
	\label{eq:mimic_observations_condidition_2}
\end{eqnarray}
where $r_{\mathrm{min}} = 40 \, \rm{Mpc}$ and $r_{\mathrm{max}} = 300 \, \rm{Mpc}$. The relative density contrast around vantage point $i$ is then
\begin{eqnarray}
    \delta_i &\equiv& 1 - \frac{ \sum_j M_j}{V \rho_{0}} \, , \quad \mathrm{with}\\
    V &=& \frac{4 \mathrm{\pi}}{3} \left(1 - \cos \theta_{\mathrm{obs}} \right) \left(r_{\mathrm{max}}^3 - r_{\mathrm{min}}^3 \right) \, .
    \label{eq:mimic_observations_density_contrast}
\end{eqnarray}
The sum is taken over all subhaloes with $M_{*} > 10^{10} h^{-1} \, \rm{M_{\odot}}$ that satisfy Equations \ref{eq:mimic_observations_condidition_1} and \ref{eq:mimic_observations_condidition_2}. These conditions restrict us to a volume $V$. The cosmic mean density $\rho_{0}$ is found by relaxing the position-related conditions and dividing the much larger sum by the whole simulation volume.

\subsection{Comparison with observations} \label{subsec:Comparison with observations}

We now compare our so-obtained list of $\delta_i$ with the observed local matter distribution. By combining our results with prior analytic work in $\Lambda$CDM, we also assess the implications for the Hubble tension and conduct a joint analysis.

\subsubsection{KBC void} \label{KBC void in LCDM}

As discussed in Section~\ref{subsec:KBC void}, \citet{Keenan_2013} discovered a large local underdensity with an apparent density contrast of $\delta_{\mathrm{obs}} = 0.46 \pm 0.06$ around the LG assuming a fixed distance-redshift relation with $H_{0} = 70 \, \rm{km\,s^{-1}\,Mpc^{-1}}$ (see their section~4.7). To compare their reported $\delta_{\mathrm{obs}}$ with $\Lambda$CDM expectations, we need to account for the fact that any underdensity $\delta$ would also affect the local Hubble parameter by
\begin{eqnarray}
    \frac{\Delta H}{H} ~\equiv~ f \delta \, ,
    \label{eq:Hubbe_RSDcorrection}
\end{eqnarray}
where e.g. \citet{Marra_2013} showed that for $\delta \ll 1$ in $\Lambda$CDM,
\begin{eqnarray}
    f ~=~ \frac{{\Omega_{\mathrm{m}}}^{0.6}}{3b} \, ,
\end{eqnarray}
with the bias factor $b = 1$ (see also Section \ref{Other_voids}). As a result, the volume within a fixed redshift would be reduced below that assumed in \citet{Keenan_2013} by a fraction
\begin{eqnarray}
	\frac{\Delta V}{V} ~=~ -3f\delta \, .
\end{eqnarray}
The apparent underdensity $\tilde{\delta}_i$ uncorrected for RSD would then be
\begin{eqnarray}
    1 - \tilde{\delta}_i ~=~ \frac{1 - \delta_i}{1 - 3f\delta_i} \, .
\end{eqnarray}
For the small underdensities expected in $\Lambda$CDM (see below), this approximately implies
\begin{eqnarray}
    \tilde{\delta}_{i} ~=~ \delta_{i} \left( 1 + 3f \right) \, .
    \label{eq:density_RSDcorrection}
\end{eqnarray}
In other words, the apparent (RSD-uncorrected) underdensity would be $1.5\times$ larger than the actual value.

\begin{figure}
    \includegraphics[width=\linewidth]{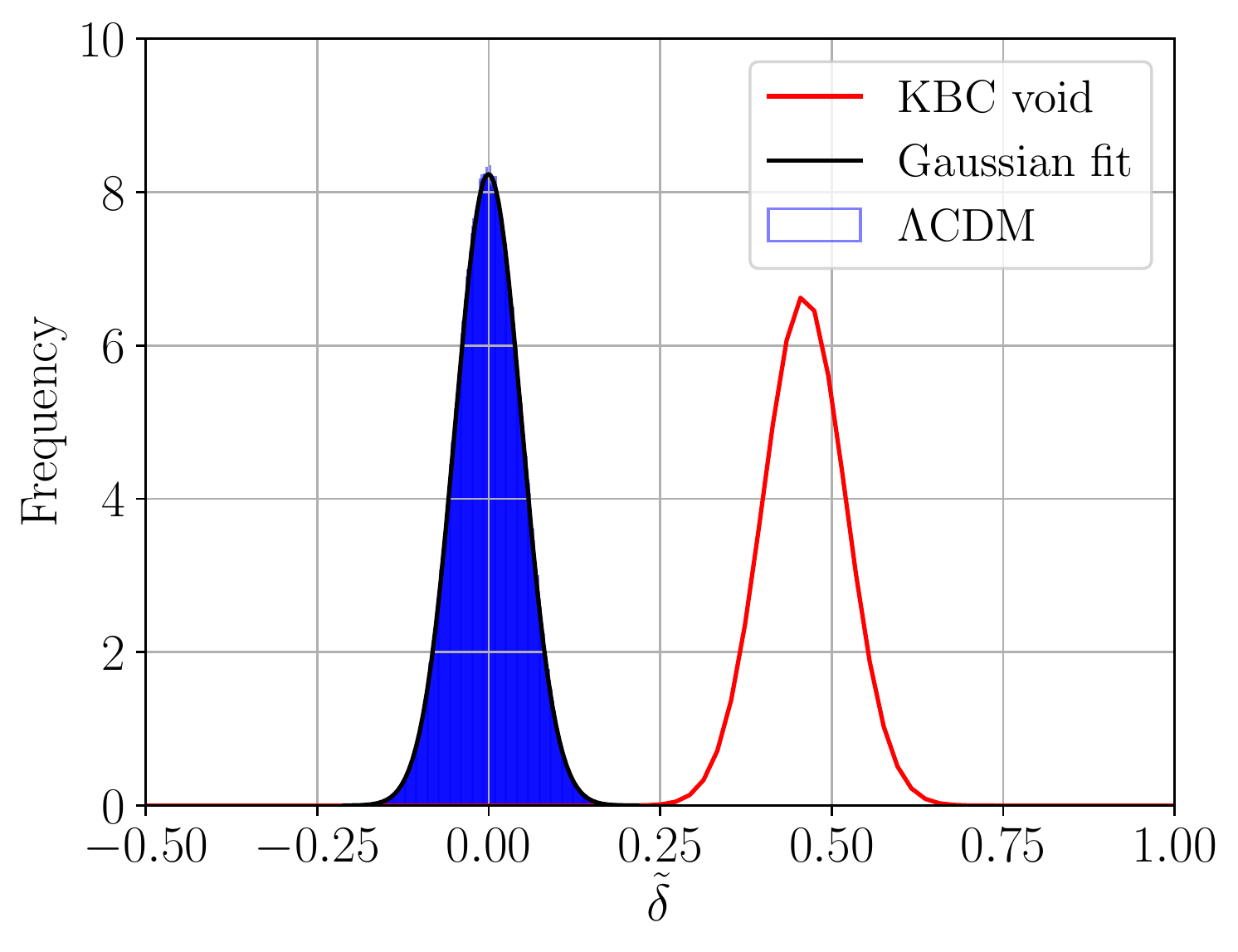}
    \caption{Distribution of the apparent relative density contrast $\tilde{\delta}$ (Equation~\ref{eq:density_RSDcorrection}) of spheres with a $300 \, \rm{Mpc}$ radius less an inner $40 \, \rm{Mpc}$ hole in the $\Lambda$CDM MXXL simulation, calculated at redshift $z = 0$ (Section~\ref{subsec:MXXL simulation}). The red solid curve shows the observed density contrast of $\delta_{\mathrm{obs}} = 0.46 \pm 0.06$ with Gaussian errors \citep[see also figure~$11$ and table~$1$ in][]{Keenan_2013}. The $\tilde{\delta}$ values closely follow a Gaussian distribution with a dispersion of $\sigma_{\mathrm{\Lambda CDM}} = 0.048$ (black curve). A more detailed Gaussianity test is performed in Appendix~\ref{Appendix_normality_tests}. Both curves are normalised to the same area.}
    \label{figure_LCDM_results}
\end{figure}

Figure~\ref{figure_LCDM_results} shows the distribution of $\tilde{\delta}_{i}$ in the standard $\Lambda$CDM MXXL simulation. This yields true rms density fluctuations of $3.2\%$, so observations uncorrected for RSD should exhibit fluctuations of $4.8\%$. To a very good approximation, these should be normally distributed, as demonstrated in Appendix \ref{Appendix_normality_tests}. Since $46/\sqrt{6^2 + 4.8^2} \approx 6.0$, we expect the discrepancy to be at the $\approx 6\sigma$ level.

Comparing the density contrast predicted by standard cosmology with the observed KBC void reveals a very significant discrepancy (Figure~\ref{figure_LCDM_results}). This is usually quantified by finding the likelihood $P$ of observing a more severe discrepancy, which we find for each vantage point and then average:
\begin{eqnarray}
	P &=& \frac{1}{N} \sum_{i = 1}^{N} f_{\chi \mapsto P} \left( \left| \frac{\widetilde{\delta_{i}} - \delta_{\mathrm{obs}}}{\sigma_{\mathrm{obs}}} \right| \right) \, , \quad \mathrm{with}\\
	f_{\chi \mapsto P} \left( \chi \right) &\equiv& 1 - \frac{1}{\sqrt{2 \mathrm{\pi}}} \int_{-\chi}^\chi \exp \left( -\frac{x^2}{2} \right) dx \, .
	\label{eq:f_chi_to_P}
\end{eqnarray}
Here, $N = 10^6$ is the number of vantage points, $\delta_{\mathrm{obs}} = 0.46$ is the observed underdensity, and $\sigma_{\mathrm{obs}} = 0.06$ is its uncertainty. The function $f_{\chi \mapsto P}$ gives the likelihood that a 1D Gaussian is more than $\chi$ standard deviations away from its mean. We use the inverse function $f_{P \mapsto \chi}$ to convert the so-obtained $P$-value into a more easily understood form, as will usually be done throughout this article. In this way, we find that the KBC void is in $6.04 \sigma$ tension with $\Lambda$CDM cosmology if it is accurately represented by the MXXL simulation on a 300 Mpc scale.

\subsubsection{Implications for the Hubble tension} \label{subsub:Implications for the Hubble tension}

\begin{figure}
    \includegraphics[width=\linewidth]{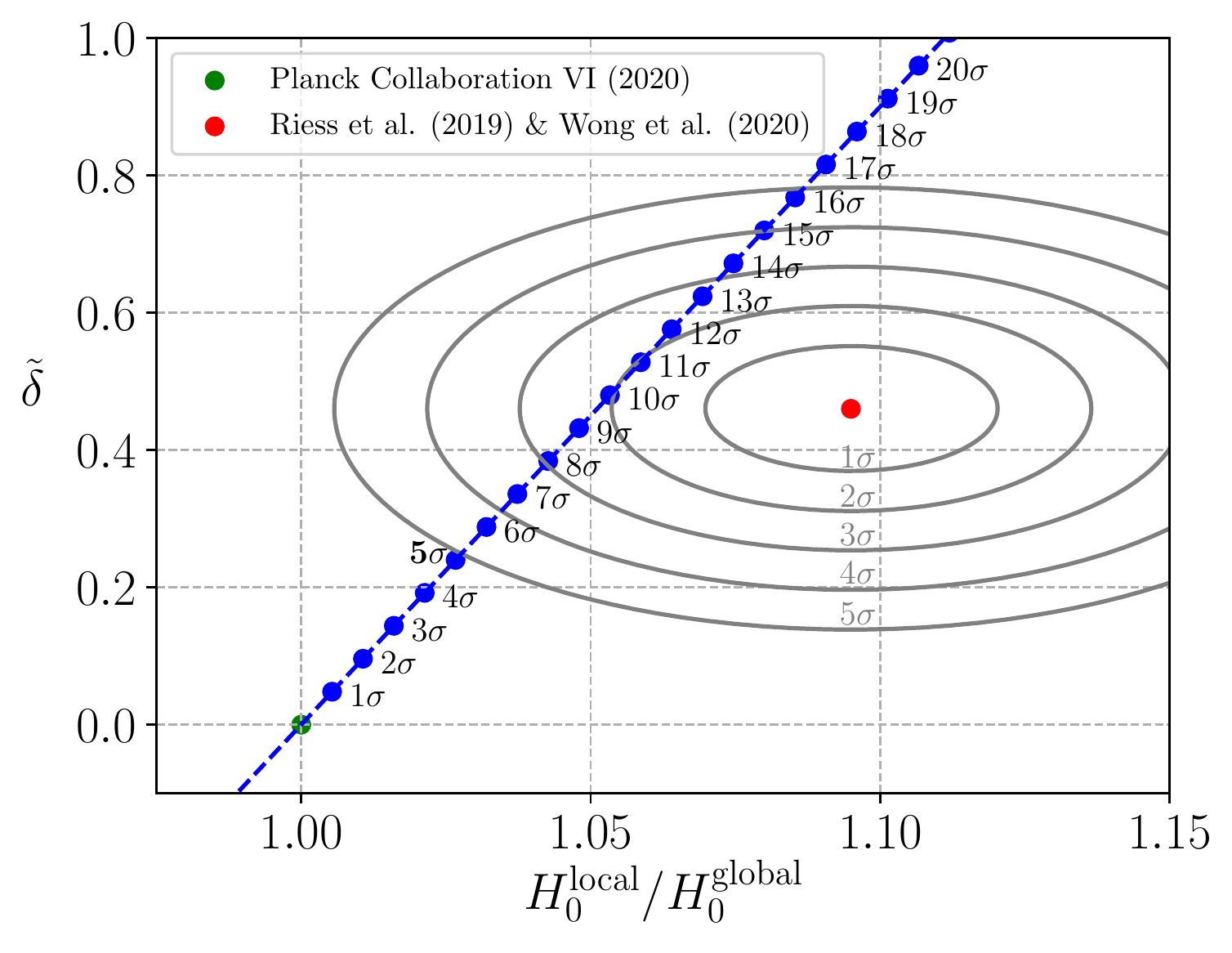}
    \caption{The local underdensity and Hubble constant in the $\Lambda$CDM framework and as found observationally. The green point shows $H_{0}^{\mathrm{global}} = 67.4 \pm 0.5 \, \rm{km\,s^{-1}\,Mpc^{-1}}$ \citep{Planck_2018} and a local density equal to the cosmic mean ($\tilde{\delta} = 0$). The red data point is the local Hubble constant combined from the SH0ES and H0LiCOW projects \citep[$H_{0}^{\mathrm{local}} = 73.8 \pm 1.1 \, \rm{km\,s^{-1}\,Mpc^{-1}}$;][]{Riess_2019, Wong_2020} and the locally observed $\delta_{\mathrm{obs}} = 0.46 \pm 0.06$ \citep{Keenan_2013}. The grey contour lines show the indicated confidence levels assuming the measurements are independent. The blue points show the expected cosmic variance in $\Lambda$CDM corrected for RSD (Equations~\ref{eq:Hubbe_RSDcorrection} and \ref{eq:density_RSDcorrection}) at the indicated confidence level. Notice that a $5 \sigma$ fluctuation is not enough to get within $5 \sigma$ of the local observations.}
    \label{figure_LCDM_results_cosmic_variance}
\end{figure}

In any matter-conserving cosmological model, we expect an underdensity to be associated with some change in the local expansion rate (Equation~\ref{eq:H_0_impact}). Figure~\ref{figure_LCDM_results_cosmic_variance} illustrates the manner in which this occurs for $\Lambda$CDM. In principle, the KBC void can boost the global Hubble constant to its local value observed by the SH0ES and H0LiCOW teams \citep[$H_{0}^{\mathrm{local}} = 73.8 \pm 1.1 \, \rm{km\,s^{-1}\,Mpc^{-1}}$,][]{Riess_2019,Wong_2020}. In fact, the straight line drawn on Figure~\ref{figure_LCDM_results_cosmic_variance} should curve to the right for large $\delta$ because as $\delta \to 1$, we expect that $H_{0}^{\mathrm{local}}/H_{0}^{\mathrm{global}} \to \infty$ due to mass conservation \citep[Equation~\ref{eq:H_0_impact}, see also figure 1 of][]{Marra_2013}. Thus, the expected relation between $H_{0}^{\mathrm{local}}$ and $\tilde{\delta}$ would pass rather close to the observations (red point). However, a $10 \sigma$ density fluctuation would be necessary to reduce the Hubble tension to the $2 \sigma$ level. Moreover, even a $5 \sigma$ underdensity in $\Lambda$CDM is still not enough to get within $5 \sigma$ of the local observations. This suggests that combining the KBC void and Hubble tension leads to a discrepancy with $\Lambda$CDM that slightly exceeds $5\sqrt{2} \sigma = 7.07 \sigma$. We next perform a more detailed joint analysis.

\subsubsection{Combined implications for $\Lambda$CDM} \label{Combined implications for LCDM}

As discussed in Section~\ref{subsec:KBC void}, the locally measured $H_0$ is discrepant at the $5.3 \sigma$ level with the Planck-based $\Lambda$CDM prediction \citep{Wong_2020} if we neglect the small expected impact of cosmic variance \citep{Wojtak_2014}. In the previous section, we have shown that the KBC void is in $6.04 \sigma$ tension with $\Lambda$CDM (Figure~\ref{figure_LCDM_results_cosmic_variance}). Therefore, both the KBC void and Hubble tension are difficult to explain within the $\Lambda$CDM framework $-$ we can explain both simultaneously, but this would require a $10\sigma$ density fluctuation (Figure~\ref{figure_LCDM_results_cosmic_variance}). In this context, the most plausible explanation is that both are caused by measurement errors. If so, we would have to assume two independent $>5 \sigma$ errors, an unlikely scenario. The combined tension would correspond to $\chi^{2} = 5.30^{2} + 6.04^{2}$ for $2$ degrees of freedom. This results in a probability of $P = \mathrm{exp} \left( -\chi^{2}/2 \right) = 9.4 \times 10^{-15}$, which is equivalent to $7.75 \sigma$ for one variable.

Measurements of the local density and velocity fields rely on rather different techniques, justifying our assumption of independence. For instance, a miscalibration of SNe magnitudes would affect $H_{0}^{\mathrm{local}}$ but not $\delta_{\mathrm{obs}}$ as the latter is a relative density contrast between different redshift bins. Thus, it is extremely unlikely that both phenomena are caused purely by measurement errors. Moreover, the KBC void is evident at different wavelengths as well as independently on smaller ($<50~\,\rm{Mpc}$) scales \citep{Karachentsev_2012}, while several independent teams have measured a higher local expansion rate than the Planck-based $\Lambda$CDM prediction (Sections~\ref{subsec:KBC void} and \ref{subsec:Hubble tension}, respectively).

A more rigorous way to estimate the combined tension is to average the $P$-values across different vantage points considering their individual $\delta_i$, how this would perturb the local expansion rate, and how the resulting RSD would lead to an enhanced apparent $\tilde{\delta}_i$. The average $P$-value is thus
\begin{eqnarray}
    P  &=& \frac{1}{N} \sum_{i = 1}^{N}  \exp \left( -\frac{\chi_{i}^{2}}{2} \right) \, ,\quad \mathrm{where} \\
    \chi_{i}^{2} &=& \left(\frac{ \tilde{\delta}_{i} - \delta_{\mathrm{obs}}}{\sigma_{\delta}} \right)^{2} + \left( \frac{\tilde{H}_{0,i} - H_{0}^{\mathrm{local}}}{\sigma_{H_{0}}} \right)^{2} \quad \mathrm{and} \\
    \tilde{H}_{0,i} &=&  H_{0}^{\mathrm{global}} \left( 1 + f \delta_i \right)
	\label{eq:generalization_LCDM_2}
\end{eqnarray}
is the apparent local Hubble constant. Here, $H_{0}^{\mathrm{local}} = 73.8 \,\rm{km\,s^{-1}\,Mpc^{-1}}$ and $\sigma_{H_{0}} = 1.2\,\rm{km\,s^{-1}\,Mpc^{-1}}$, with the latter including an allowance for the $0.5\,\rm{km\,s^{-1}\,Mpc^{-1}}$ uncertainty from \citet{Planck_2018}. This procedure reveals that the KBC void and Hubble tension falsify the $\Lambda$CDM framework at $7.09 \sigma$, in agreement with our earlier estimate.

Our calculation of the cosmic variance in $\Lambda$CDM is derived from the stellar masses of subhaloes with $M_{*} > 10^{10} h^{-1} \, \rm{M_{\odot}}$, which should be more than sufficient to accurately trace the matter distribution on a 300 Mpc scale. Moreover, our results are consistent with expectations from the Harrison-Zeldovich spectrum \citep{Harrison_1970, Zeldovich_1972} and its early Universe normalisation required to match the CMB \citep{Planck_2018}. In a $\Lambda$CDM context, this is parametrized using $\sigma_{8}$, which implies rms fluctuations of $3.2\%$ on a $300 \, \rm{Mpc}$ scale at the present epoch. This agrees with our much more rigorous estimate using MXXL (Section~\ref{KBC void in LCDM}).

Therefore, the KBC void is not a consequence of random measurement errors or density fluctuations expected in standard cosmology. Structure formation mainly depends on the underlying gravitational law, strongly suggesting that the observed KBC void cannot be explained by treating baryonic physics differently on galaxy scales.

Although cosmic variance in a standard context is insufficient to explain the KBC void and $H_{0}$ from low-redshift probes \citep[e.g.][]{Macpherson_2018}, Figure~\ref{figure_LCDM_results_cosmic_variance} indicates that a large local void appears to be a promising explanation for these local observations. Consequently, we next consider a long-range modification to gravity which should enhance cosmic variance while accurately explaining observations on galactic scales with a fixed acceleration threshold \citep{Famaey_2012}. Section~\ref{subsec:Claimed_problems} discusses some commonly used arguments for why the KBC void cannot solve the Hubble tension.

\section{MOND framework} \label{sec:MOND framework}

As shown in the previous section, the cosmic variance expected within the $\Lambda$CDM framework is insufficient to explain the KBC void and Hubble tension. Thus, we aim to investigate structure formation and the velocity field in MOND \citep{Milgrom_1983}. In this section, we first introduce a conservative MOND cosmology that has the same expansion rate history and overall matter content as $\Lambda$CDM, but with CDM replaced by hot dark matter (HDM) to account for light element abundances, galaxy clusters, and the CMB without much affecting galaxies \citep{Angus_2009}. We then explain how we parametrize the initial void density profile and evolve it forwards to the present time (Section~\ref{subsec:Governing equations}). Finally, we describe how predictions for local observables are extracted from our models (Section~\ref{subsec:Observational constraints}).

\subsection{The \texorpdfstring{$\nu$HDM}{vHDM} cosmological model} \label{subsec:nuHDM cosmological model}

Any viable cosmological model has to explain the angular power spectrum of the CMB and the primordial abundances of light elements. \citet{Angus_2009} provided a promising cosmological model that seeks to address the shortcomings of MOND on galaxy cluster and larger scales using an extra sterile neutrino species with a mass of $m_{\nu_{s}} = 11 \, \rm{eV}/c^2$. Thermally produced neutrinos of this mass would have the same relic abundance as CDM particles in standard cosmology, but would behave as HDM in the sense of not clustering on galaxy scales.\footnote{In $\Lambda$CDM, sterile neutrinos with $m_{\nu_{s}} \approx 7 \, \rm{ke
V}/c^2$ are often considered as DM candidates \citep[e.g.][]{Bulbul_2014, Boyarsky_2014}. Like $11 \, \rm{eV}/c^2$ sterile neutrinos, these would also be relativistic during the nucleosynthesis era (Section~\ref{section:BBN}), but would cluster in galaxies.} The composition of the universe as a whole would be similar to $\Lambda$CDM $-$ baryons would still comprise $\approx 5\%$ of the present critical density of the universe, sterile neutrinos would replace the $\approx 25 \%$ contribution of CDM, and dark energy would yield the remaining $\approx 70\%$ (i.e. $\Omega_{\mathrm{m},0} = \Omega_{\mathrm{b},0} + \Omega_{\mathrm{\nu_{s}},0} \approx 0.3$ and $\Omega_{\mathrm{\Lambda},0} \approx 0.7$). We refer to this model as the $\nu$HDM paradigm, where $\nu$ stands for both the interpolating function in QUMOND (Equation~\ref{eq:MOND_simple_interpolation_function}) and sterile neutrinos, maximizing the chance that it is physically meaningful. The observed expansion history of the Universe seems broadly consistent with $\Lambda$CDM cosmology \citep[e.g.][]{Joudaki_2018}. As shown by \citet{Angus_2009}, $\nu$HDM yields the same expansion history as $\Lambda$CDM due to the same overall matter content and the same Friedmann equations at the background level \citep{Skordis_2006}. This issue is discussed further in Section \ref{subsubsec:Background cosmology}.

Although the existence of sterile neutrinos is not experimentally confirmed yet, they are theoretically consistent with standard particle physics \citep{Merle_2017}. Observationally, the $\nu$HDM model is motivated mainly by galaxy clusters, where the dynamical discrepancy cannot be explained in MOND without DM \citep{Sanders_2003}. Furthermore, DM is necessary to address the offset between X-ray and lensing peaks in the Bullet Cluster \citep{Clowe_2006}, since MOND acting on the baryons alone is unable to fully replace the role played by CDM in standard cosmology \citep{Angus_2007}. We emphasize that these observations do not uniquely require CDM since they are on a much larger spatial scale than the hypothesized CDM haloes of individual galaxies \citep{Ostriker_1973}.

In this context, \citet{Angus_2010} analysed $30$ of the most virialized galaxy groups and clusters in the $\nu$HDM paradigm. They found that the required HDM density in all cases reaches the so-called Tremaine-Gunn limit \citep{Tremaine_Gunn_1979} at the centre for sterile neutrinos with $m_{\nu_{s}} = 11 \, \rm{eV}/c^2$. This is a strong indication that the DM density in galaxy cluster cores is limited by quantum degeneracy pressure (the Pauli Exclusion Principle). Note that MOND fits to galaxy rotation curves are hardly affected by sterile neutrinos with $m_{\nu_{s}} \la 100 \, \rm{eV}/c^2$, even if their number density reaches the Tremaine-Gunn limit \citep[section 4.4 of][]{Angus_2010}. As a result, $\nu$HDM is likely to explain the internal dynamics of both galaxies and galaxy clusters. Introducing sterile neutrinos is thus well consistent with astronomical observations and almost consistent with the standard model of particle physics (unlike CDM particles), but they nevertheless require experimental verification.

In the following, we address the background evolution of $a \left( t \right)$ in the $\nu$HDM framework, allowing us to address the primordial abundances of light elements and the CMB. We also consider the implications for large-scale structure, where substantial differences are expected from $\Lambda$CDM. The theoretical uncertainties of the here applied MOND approach are summarized 
in Section~\ref{subsubsec:Theoretical_assumptions}, which focuses on how density perturbations should be treated in MOND.

\subsubsection{Background cosmology}
\label{subsubsec:Background cosmology}

The background evolution $a \left( t \right)$ requires a relativistic theory that yields the appropriate MOND limit in galaxies. In this contribution, we make certain assumptions about the parent relativistic theory that gives rise to MOND. These assumptions are based on prior work, in particular with the tensor-vector-scalar (TeVeS) theory that was the first covariant framework with an appropriate MOND limit \citep{Bekenstein_2004}. His section 7 indicates that the background evolution should be very similar to General Relativity at all epochs for the same matter-energy content.

The background evolution and perturbations in TeVeS were addressed in detailed calculations done by \citet{Skordis_2006_theory}. To avoid detectable departures from the standard expansion history during the nucleosynthesis era, the free dimensionless parameter $\mu_0$ must be rather large \citep{Skordis_2006}.\footnote{$\mu_0$ is related to the TeVeS parameter $\kappa$ \citep[equation~16 of][]{Bekenstein_2004} via $\mu_{0} \equiv 8\mathrm{\pi}/\kappa$.} In particular, if we allow the extra energy density contributed by the scalar field to comprise a fraction $X$ of the critical density during the radiation-dominated era, then the contribution in the matter and $\Lambda$-dominated eras would be $X/9$. Primordial light element abundances then imply that the standard Friedmann equation would differ from the TeVeS cosmology at only the sub-per cent level (see their figure~1). The very small contribution of the scalar field density was also demonstrated in figure~2 of \citet{Dodelson_2006}. Therefore, we will assume that the background cosmology is identical to that of $\Lambda$CDM. Since the CMB is also expected to have similar properties in both frameworks (Section \ref{subsubsec:Radiation dominanted era and CMB}), they both lead to the Hubble tension in a similar manner provided that $\dot{a} = H_{0}^{\mathrm{local}}$, i.e. if cosmic variance in the local measurements is much smaller than the Hubble tension. Our main argument is that this assumption is valid in $\Lambda$CDM but need not be in MOND.

While the original version of TeVeS is inconsistent with gravitational waves travelling at $c$, a slightly modified version does have this property, even in the presence of perturbations \citep{Skordis_2019}. The above-mentioned results should carry over to the updated version of TeVeS, though this should be carefully demonstrated in future work. The preliminary results of \citet{Skordis_2020} are an important step in this direction.

Throughout this work, we assume dark energy not to be an artefact of an observer in an underdense region seeing an apparently accelerating expansion due to the developing inhomogeneities \citep{Buchert_2001}. However, we emphasize that proper time-averaging of global properties of the universe would be required to further study the present model \citep{Wiltshire_2007}.

\subsubsection{Big Bang nucleosynthesis} \label{section:BBN}

Big Bang nucleosynthesis (BBN) occurred at a temperature of $kT \approx 1 \, \rm{MeV}$, where $k$ is the Boltzmann constant. A review on BBN can be found e.g. in \citet{Cyburt_2016}. In the $\nu$HDM framework, \citet{Skordis_2006} showed that it is possible to have essentially no departure from the standard expansion history during the radiation-dominated era. However, the model would still have an effect on BBN because at $kT \approx 1 \, \rm{MeV}$, sterile neutrinos with $m_{\nu_{s}} \approx 11 \, \rm{eV}/c^2$ would be relativistic. Their weaker interactions would cause them to decouple earlier, so they would add an extra $7/8$ to $g_{*}$, the number of effective relativistic degrees of freedom. Since the Hubble parameter $H \equiv{\dot{a}}/a$ scales as $H \propto \sqrt{g_{*}}$ and standard physics predicts $g_{*} = 10.75$, this would increase $H$ by only $4\%$, causing a slight impact on the primordial abundances of light elements. As shown in equation~13 of \citet{Cyburt_2016}, any increase in $H$ raises the primordial $He$-4 mass fraction $Y_{\mathrm{p}}$ because free neutrons have less time to decay. Their detailed calculations have shown that this dependence can be fitted with a power law of the form
\begin{eqnarray}
    Y_{\mathrm{p}} ~\appropto~ N_{\nu}^{0.163} \, .
   	\label{eq:BBN_He}
\end{eqnarray}
In standard cosmology, the effective neutrino number is $N_{\nu} = 3.046$, which slightly exceeds $3$ because neutrinos decouple only slightly before electron-positron annihilation at $kT = 511 \, \rm{keV}$. Thus, an extra sterile neutrino species would increase $Y_{\mathrm{p}}$ by a factor of $(4.046/3.046)^{0.163} = 1.047$, implying the standard value of $Y_{\mathrm{p}} = 0.247$ would rise to $0.259$. This is only a small effect, so observations of the primordial $He$ abundance in ancient gas clouds currently do not set a strong constraint on the existence of an extra sterile neutrino. For instance, measurements of the $He$ abundance of a gas cloud at $z = 1.724$ backlit by a quasar yield $Y = 0.250_{-0.025}^{+0.033}$ \citep{Cooke_2018b}. Using a sample of H \textsc{ii} regions, \citet{Aver_2012} derived $Y_{\mathrm{p}} = 0.2534 \pm 0.0083$. Even if their reported uncertainty is taken at face value, $Y_{\mathrm{p}} = 0.259$ is quite possible.

Measurements of the primordial abundances of $D$ and $Li$-7 are less sensitive to $N_{\nu}$ \citep{Cyburt_2002}. However, primordial $D$ abundances are relatively well known. \citet{Cooke_2018a} obtained $N_{\nu} = 3.41 \pm 0.45$ based on $(D/H)_{\mathrm{p}}$ derived from a metal-poor damped Ly~$\alpha$ system. Therefore, both $D$ and $He$ measurements allow an extra sterile neutrino, which was actually favoured by the earlier analysis of \citet{Steigman_2012}. We do not consider the more problematic case of $Li$-7, though see \citet{Howk_2012} for a gas phase measurement in the Small Magellanic Cloud that seems to resolve the lithium problem.

These considerations only hold for sterile neutrinos in thermal equilibrium during the nucleosynthesis era. However, if sterile neutrinos decoupled much earlier, their number density could be lower depending on whether any other particle subsequently became non-relativistic. If so, $\Delta N_\nu$ would be lower, reducing the impact on $g_*$ and on BBN. This scenario would require a higher sterile neutrino mass to recover the standard value of $\Omega_{\mathrm{m}}$.

\subsubsection{Radiation-dominated era and the CMB} \label{subsubsec:Radiation dominanted era and CMB}

After BBN, the next major constraint on any cosmological model comes from the CMB. This occurred shortly after the epoch of matter-radiation equality at $z_{\mathrm{eq}} = 3411 \pm 48$ \citep{Planck_2018}.\footnote{$z_{\mathrm{eq}}$ is tightly constrained by the acoustic oscillations in the CMB because during the earlier radiation-dominated era, perturbations in the sub-dominant matter component are unable to grow through gravitational instability.} This corresponds to a photon temperature of $kT \approx 0.80 \, \rm{eV}$, which is much less than the mass of the here considered sterile neutrinos. Consequently, they would behave just like non-relativistic CDM, causing $z_{\mathrm{eq}}$ to be the same as in the $\Lambda$CDM model.

The CMB was emitted at $z_{_\mathrm{CMB}} \approx 1100$, corresponding to $kT \approx 0.26~\rm{eV}$. At this time, matter dominated the energy budget of the universe. Since the background cosmology of the $\nu$HDM model is the same as for $\Lambda$CDM and the plasma physics is unchanged, the sound horizon at recombination would still have the standard value of $147.09 \pm 0.26 \, \rm{cMpc}$ \citep{Planck_2018}. This is directly related to the angular scale of the first acoustic peak in the CMB, which should thus be unaffected in our model.

$11 \, \rm{eV}/c^2$ sterile neutrinos would be non-relativistic at the time of last scattering. Since both $T$ and the peculiar velocity $v_{\mathrm{pec}}$ should decline $\propto 1/a$, we expect the sterile neutrinos to typically have
\begin{eqnarray}
  v_{\mathrm{pec}} ~\approx~ \frac{0.26 \, \rm{eV}}{11 \, \rm{eV}} c ~=~ 0.024 \, c \, .
\end{eqnarray}
This implies a free-streaming length of $L_{\mathrm{fs}} \approx 3.5 \, \rm{cMpc}$, which is much shorter than the horizon scale. Since the first acoustic peak of the CMB occurs at a multipole moment of $\ell \approx 200$ \citep{Jaffe_2001}, free-streaming becomes important only for $\ell \ga 200/\left( 0.024\sqrt{3}\right) = 4900$, beyond the range accessible by \citet{Planck_2018}. This is consistent with section~6.4.3 of \citet{Planck_2015}, which explicitly states that any particles with $m > 10 \, \rm{eV}/c^2$ ``are so massive that their effect on the CMB spectra is identical to that of CDM.''

The $\nu$HDM paradigm does more than simply replace CDM with HDM. Because of the Milgromian force law, the paradigms differ with regards to the evolution of sub-horizon perturbations. In the following, we estimate the gravitational field from inhomogeneities around $t_{_\mathrm{CMB}}$, the time of recombination.

The peculiar velocities are of order $v_{\mathrm{pec}} \approx c \delta$ and were built up over a duration of $t_{_\mathrm{CMB}} = 380 \, \rm{kyr}$. Assuming rms density fluctuations of $\delta_{_\mathrm{CMB}} = 10^{-5}$ as observed in the baryons, we can obtain a lower bound on the peculiar acceleration $g_{_\mathrm{CMB}}$ sourced by inhomogeneities.
\begin{eqnarray}
    g_{_\mathrm{CMB}} ~\geq~ \frac{c \delta_{_\mathrm{CMB}}}{t_{_\mathrm{CMB}}} ~\approx~ 2.1 \, a_{_0} \, .
   	\label{eq:EFE_CMB_1}
\end{eqnarray}
This already exceeds Milgrom's constant $a_{_0}$. However, the gravity must have been significantly stronger to compensate for resistance from radiation pressure. In order to estimate the density fluctuations in the HDM component at $t_{_\mathrm{CMB}}$, we consider the value of $\sigma_{8} = 0.811 \pm 0.006$ on a scale of $8 h^{-1} \approx 12 \,\rm{cMpc}$ that is required to fit the CMB anisotropies \citep{Planck_2018}. For a scale-invariant power spectrum, the density fluctuations on the $147 \, \rm{cMpc}$ scale of the first acoustic peak in the CMB are $12\sigma_{8}/147 \approx 0.065$ at the present epoch, as can also be seen by scaling our results of Section~\ref{subsec:Comparison with observations} for fluctuations on a $300 \, \rm{cMpc}$ scale.\footnote{The here used MXXL simulation is calibrated to the CMB data gathered by WMAP-1 \citep{Angulo_2012}.} Since $\Lambda$CDM predicts that $\delta \propto a$ in the matter-dominated era and neglecting the effect of dark energy, we would expect density fluctuations of $\delta_{_\mathrm{CMB}} \approx 5.9 \times 10^{-5}$ at $t_{_\mathrm{CMB}}$. Taking into account that structure formation slowed down when the Universe became dark energy-dominated at $z \leq 0.7$ and was slower around the time of recombination due to the still significant amount of radiation, we estimate that
\begin{eqnarray}
	\delta_{_\mathrm{CMB}} ~\approx~ 10^{-4} \, .
\end{eqnarray}
Thus, the typical gravitational field at recombination was
\begin{eqnarray}
	g_{_\mathrm{CMB}} ~\approx~ 21 a_{_0} \, ,
	\label{eq:EFE_CMB_2}
\end{eqnarray}
implying that MOND had only a minor impact at that time.

In the matter-dominated era ($a \gg a_{\mathrm{eq}}$), the density perturbations grow $\propto a$ after their mode enters the horizon. Therefore, the Harrison-Zeldovich power spectrum predicts that the power of the density perturbations scales inversely with their length $L$ \citep{Harrison_1970, Zeldovich_1972}, i.e.
\begin{eqnarray}
    P \left( L \right) ~\propto~ L^{-1} \, .
   	\label{eq:MOND_scaling_relations_4}
\end{eqnarray}
Since the mass enclosed by the mode is $M \propto L^{3}$, the mass perturbation must scale as
\begin{eqnarray}
    \Delta M ~\propto~ L^{2} \, .
   	\label{eq:MOND_scaling_relations_6}
\end{eqnarray}
Therefore, the perturbation's Newtonian gravity is independent of $L$, i.e.
\begin{eqnarray}
    g_{_\mathrm{N}} ~=~ \text{const.}
   	\label{eq:MOND_scaling_relations_7}
\end{eqnarray}
The Harrison-Zeldovich power spectrum breaks down for length-scales that enter the horizon before $a_{\mathrm{eq}}$. Since no modes would be able to grow during the radiation-dominated era, these short-wavelength modes would have much less power than predicted by a $1/L$ scaling relation. Thus, $g_{_\mathrm{N}}$ would be smaller. However, in MOND, these short-wavelength modes would be embedded in the EFE generated especially by long-range modes (Section~\ref{subsec: Milgromian dynamics}). This would severely limit the MOND boost to the internal gravity of shorter modes, since their total $g_{_\mathrm{N}}$ depends on both their internal gravity and any external field. For this reason, we expect that modes of any $L$ were unaffected by MOND around the epoch of recombination.

We next consider how this picture changes with time. Since Newtonian density perturbations are expected to grow as $\delta \propto a$ in the matter-dominated era, the mass perturbation should also scale as
\begin{eqnarray}
    \Delta M ~\propto~ a \, .
   	\label{eq:MOND_scaling_relations_8}
\end{eqnarray}
For linear ($\delta \ll 1$) perturbations whose co-moving size hardly changes, the Newtonian gravity should scale as
\begin{eqnarray}
    g_{_\mathrm{N}} ~\propto~ a^{-1} \, .
   	\label{eq:MOND_scaling_relations_9}
\end{eqnarray}
Our previous estimation showed that the gravitational field sourced by inhomogeneities is $g \gg a_{_0}$ at $t_{_\mathrm{CMB}}$ (Equation~\ref{eq:EFE_CMB_2}). We now see that even larger gravitational fields are expected at earlier times, further justifying our assumption that MOND would have little effect then.\footnote{MOND effects can be further reduced at early times if $a_{_0}$ was smaller, or if density perturbations couple to the background in a non-trivial way (Section \ref{subsubsec:Theoretical_assumptions}).}

We can combine Equations~\ref{eq:EFE_CMB_2} and \ref{eq:MOND_scaling_relations_9} to deduce that MOND does not play a significant role in structure formation until $z \la z_{\mathrm{MOND}} = 50$. This underpins the commonly used assumption that MOND does not play a role in the very early universe, but would promote the formation of the first galaxies \citep{Sanders_1998}.

The high accelerations around the time of recombination strongly suggest that the MOND gravity law would not by itself affect the acoustic oscillations in the CMB. This issue was investigated further by \citet{Skordis_2006}, who considered a covariant formulation of MOND. Their figure~2 confirms our conclusion that the modification to gravity has by itself only a very small effect for plausible choices of the model parameters consistent with BBN. However, their use of three ordinary neutrino species with a much lower mass of $2 \, \rm{eV}/c^2$ led to significant free streaming effects that are totally inconsistent with the latest observations \citep{Planck_2013}. If instead a single $11 \, \rm{eV}/c^2$ sterile neutrino is used, a very good fit can be obtained to the CMB power spectrum for the reasons just discussed \citep[figure~1 of][]{Angus_2009}. Note also that with a standard $a \left( t \right)$, the angular diameter distance to the CMB would be the same as in $\Lambda$CDM, placing the acoustic peaks at the correct angular scales. Indeed, figure~1 of \citet{Angus_2011} shows that the CMB power spectra in the $\nu$HDM and $\Lambda$CDM models agree quite closely, so both paradigms are consistent with observations taken by WMAP-7, the Atacama Cosmology Telescope (ACT), and the Arcminute Cosmology Bolometer Array Receiver up to $\ell = 2500$.

\subsubsection{Evolution of perturbations and large-scale structure} \label{subsubsec:Large-scale structure}

Even if the CMB power spectrum is correct in our framework, the observed CMB is also influenced by foreground structures. Section~\ref{subsubsec:CMB_monopole} discusses the gravitational redshift of the entire last scattering surface due to the rather high MOND potential of the KBC void. Foreground lensing of the CMB by large scale structures and the integrated Sachs-Wolfe (ISW) effect would also be stronger in MOND. There are some observational hints that these effects are stronger than expected in $\Lambda$CDM (Section~\ref{Other_voids}). These tensions could be eased in a theory where structure formation is more efficient. However, it is possible that $\nu$HDM overcorrects the problem and produces too much foreground lensing and/or a Sachs-Wolfe effect in disagreement with observations. These issues are beyond the scope of our work, but should be addressed before the $\nu$HDM framework can be considered to fully account for all observed aspects of the CMB. This would almost certainly require numerical simulations of structure formation. In addition, photon propagation through such a simulation would need to be handled with care, taking account of inhomogeneities and their time evolution \citep[e.g.][]{Wiltshire_2007}.
 
\citet{Nusser_2002} considered the growth of density perturbations in a Milgromian framework. Their section~2 introduced the basic principle used in all subsequent MOND cosmological simulations \citep{Llinares_2008,Angus_2011, Angus_2013, Katz_2013, Candlish_2016}. These simulations make the ansatz that a MONDified Poisson equation (usually Equation~\ref{eq:MOND_basic}) is applied only to the density perturbations about the mean background value, as evident e.g. in equation~2 of \citet{Candlish_2016}.\footnote{Equation~4 of \citet{Nusser_2002} assumes the deep-MOND limit, but we generalize it to an arbitrary acceleration using an interpolating function (Equation~\ref{eq:MOND_simple_interpolation_function}). Note that the deep-MOND limit is a reasonable assumption for the KBC void (Section~\ref{subsubsec:Theoretical_assumptions}).} This `Jeans swindle' \citep{Binney_1987_book} approach to MOND was justified using the earlier work of \citet{Sanders_2001}, who showed its validity in a non-relativistic Lagrangian formulation of MOND (see his section~2). The approach is certainly valid for systems such as galaxies that are much denser than the cosmic mean. The use of non-relativistic gravitational equations should be sufficient when dealing with structures such as the KBC void that are much smaller than the cosmic horizon, since gravity travel time effects would not be too significant.

\citet{Falco_2013} showed that the Jeans swindle is formally correct in Newtonian gravity $-$ including the background would simply add on the force required to maintain the time-dependent Hubble flow velocity. However, it still needs to be rigorously demonstrated that the swindle remains mathematically valid in a MONDian model with a non-linear gravity law. Therefore, although this ansatz is commonly used by the MOND community, it is one of the strongest assumptions in the here presented cosmological model.

One of the few works that does not make this assumption is \citet{Sanders_2001}, whose model is a non-relativistic two-field Lagrangian-based theory of MOND. The coupling between these two fields is described by an adjustable parameter $\beta$ in his modified Poisson equation~8. Setting $\beta = 0$ is equivalent to applying the Jeans swindle approach. However, if $\beta \neq 0$, there exists a coupling between the peculiar acceleration sourced by inhomogeneities and the zeroth-order Hubble flow acceleration $\bm{g}_{\mathrm{Hubble}}$ (Equation~\ref{eq:governing_equation_Force_1}). \citet{Sanders_2001} adopted $\beta = 3.5$ for his main analysis. As discussed in the cosmology section of \citet{Sanders_2002}, $\bm{g}_{\mathrm{Hubble}}$ essentially contributes an extra source of gravity to the total entering the $\nu$ calculation in Equation~\ref{eq:MOND_basic}, limiting the MOND boost to gravity. We call this the `Hubble field effect' (HFE), since it is similar to but distinct from the usual EFE in MOND $-$ both make the behaviour more Newtonian. In Section~\ref{subsubsec:Theoretical_assumptions}, we address theoretical uncertainties arising from the HFE, which is neglected in our main analysis. A non-zero HFE would substantially affect large-scale structures especially at scales $\ga 100 \, \rm{cMpc}$, which could be used to constrain it in future studies (Section~\ref{subsubsec:Theoretical_assumptions}). However, we argue there that even with a strong HFE, cosmic variance would still be enhanced $3\times$ compared to $\Lambda$CDM expectations on a $300\,\rm{Mpc}$ scale under conservative assumptions, enough to reproduce the KBC void. 

\citet{Nusser_2002} built on the model of \citet{Sanders_2001} but assumed instead that $\beta = 0$ because he could not find any physical justification for coupling both fields, i.e. for the HFE. This uncoupled (Jeans swindle) approach is generally the one adopted in MOND cosmological simulations \citep[e.g.][]{Llinares_2008, Angus_2013, Katz_2013, Candlish_2016}. In particular, \citet{Angus_2013} used it in a cosmological $N$-body simulation designed to address the formation of large-scale structure in MOND supplemented by sterile neutrinos. Although their work was novel and very advanced for its time, it faces some conceptual and numerical problems. In particular, they concluded that their model with $11 \, \rm{eV}/c^2$ sterile neutrinos significantly underestimates the number of low-mass galaxy clusters and slightly overestimates the number of very massive clusters (see e.g. their figure~4). This inconsistency between the model and observational data could arise for several reasons. Their conclusion is based on a simulation with a box size of $256 h^{-1}\,\rm{cMpc}$ and a particle resolution of only $\approx 3.78 \times 10^{10} \, \rm{M_{\odot}}$. The underproduction of low-mass galaxy clusters could be explained by the low particle resolution and therewith by an absence of low-mass particles needed to form such systems. In addition, they do not use a grid with adaptive mesh refinement (AMR), which causes that the potential wells especially of the smaller clusters may not be resolved properly, making them difficult to form. Therefore, it would be highly valuable to revisit their cosmological simulations with an AMR grid code such as \textsc{phantom} of \textsc{ramses} \citep{Lueghausen_2015}, which adapts the potential solver of the widely used \textsc{ramses} algorithm \citep{Teyssier_2002}.

In general, small simulation boxes lack large-scale modes. Since the EFE is mainly sourced by very massive objects, a too small simulation box would potentially underestimate the EFE on MONDian subsystems. Thus, the internal gravitational field would be too strong, which could also explain the efficient formation of massive galaxy clusters in \citet{Angus_2013}.

As already discussed at the beginning of this section, \citet{Angus_2010} demonstrated that the required neutrino density in $30$ virialized galaxy groups and clusters reaches the Tremaine-Gunn limit at the centre, which supports the $\nu$HDM model. However, the neutrino degeneracy pressure in the cores of galaxy clusters has not been included in the simulations of \citet{Angus_2013}. If one would account for this effect, it would be more difficult to form massive galaxy clusters because gravity is resisted by neutrino degeneracy pressure.

Finally, \citet{Angus_2013} compared their simulated halo mass functions with cluster mass functions derived from observations at $z \leq 0.3$ \citep{Reiprich_2002} and $z \leq 0.1$ \citep{Rines_2008}. As we have seen in Section~\ref{subsec:KBC void}, the KBC void has a similar extent. It is evident in X-ray galaxy cluster surveys \citep[e.g.][]{Boehringer_2015, Boehringer_2020}. Therefore, local observations are biased against high-mass clusters, e.g. the massive merging galaxy cluster El Gordo \citep[ACT-CL J0102-4915,][]{Marriage_2011} with a mass of $3 \times 10^{15}\,\rm{M_{\odot}}$ \citep{Jee_2014} at $z = 0.87$ \citep{Menanteau_2012} would almost certainly not be evident in local observations from within a deep void. Thus, local observations do not provide a representative cluster mass function of the whole Universe, so cannot be compared with the entire simulated halo population.

Consequently, the \citet{Angus_2013} cosmological model has never been tested in full detail on large scales. An object similar to El Gordo was identified in the $\nu$HDM simulation of \citet{Katz_2013}, so initial results seem promising. It would be highly valuable to revisit their analysis in more physically and numerically advanced large-scale simulations. This is because the $\nu$HDM framework provides a viable explanation for BBN and the CMB, but also works on galaxy cluster scales while recovering the successes of MOND in galaxies. At present, there is no $N$-body or hydrodynamical simulation with a large enough box size to study the KBC void in a MONDian framework. Therefore, we develop a semi-analytic simulation for this purpose. In the following, we introduce the governing equations and parameters of the here discussed $\nu$HDM cosmological model.

\subsection{Governing equations} \label{subsec:Governing equations}

We develop a simplified simulation in which the trajectories of particles are integrated up to the present time from $z = 9$, which corresponds to $\approx 0.5 \, \rm{Gyr}$ after the Big Bang (Equation~\ref{eq:governing_equation_a}). As derived from General Relativity in section~2.2 of \citet{Banik_2016}, the particle's trajectory is described by the background cosmological acceleration term and any additional gravity sourced by inhomogeneities:
\begin{eqnarray}
	\label{eq:governing_equation_Force_1}
    \ddot{\bm{r}} &=&  \bm{g}_{\mathrm{void}} +  \overbrace{\frac{\ddot{a}}{a} \bm{r}}^{\bm{g}_{\mathrm{Hubble}}} \, , \\
    \dot{\bm{r}_{i}} &=& H_{i} \bm{r}_{i} \, ,
\end{eqnarray}
where $\bm{r}$ is the particle's position relative to the void centre, $\bm{g}_{\mathrm{void}}$ is the local gravitational acceleration sourced only by density deviations from the cosmic mean, $\bm{g}_{\mathrm{Hubble}}$ is the acceleration in a homogeneously expanding spacetime, and $i$ subscripts denote initial values when $a = 0.1$. At that time, particles are assumed to be on the Hubble flow. However, the initial matter distribution is assumed to be inhomogeneous. A spherically symmetric underdensity causes a Newtonian gravitational force of
\begin{eqnarray}
	 g_{_\mathrm{N}} &\equiv& \frac{G \Delta M}{r^2} \, , \quad \mathrm{with} \\
	    \Delta M &\equiv& \frac{4 \mathrm{\pi}}{3} \rho_{0} \left( \frac{r}{a} \right)^3 - M_{\mathrm{enc}} \, ,
		\label{eq:governing_equation_gN}
\end{eqnarray}
where $\Delta M$ is the mass deficit within radius $r$, $\rho_{0}$ is the present cosmic mean density of matter, and $M_{\mathrm{enc}}$ is the enclosed mass. Since we assume mass conservation and no shell crossing, $M_{\mathrm{enc}}$ remains constant for an individual particle. In the case of no void, $g_{_\mathrm{N}} = 0$ since $\Delta M = 0$. The exact set-up of the initial void profile is described in Section~\ref{subsubsec:Void initial profiles} and Appendix \ref{Appendix: KBC void mass profile}.

Applying the Jeans swindle approach to MOND (Section \ref{subsubsec:Large-scale structure}), the gravitational force $g$ is calculated with the `simple' interpolation function (Equation~\ref{eq:MOND_simple_interpolation_function}) between the Newtonian and deep-MOND regimes \citep{Famaey_2005}. The EFE is included by quadrature summing $g_{_\mathrm{N}}$ and the Newtonian-equivalent external field $g_{_\mathrm{N,ext}}$ \citep{Famaey_2007}:
\begin{eqnarray}
    g ~=~ g_{_\mathrm{N}} \left( \frac{1}{2} + \sqrt{\frac{1}{4} + a_{_0} \left(g_{_\mathrm{N}}^{2}+g_{_\mathrm{N,ext}}^{2}\right)^{-\frac{1}{2}}}  \right) \, .  
    \label{eq:governing_equation_g}
\end{eqnarray}
The EFE and its impact on the void will be described in more detail in Sections~\ref{subsubsec:External field history} and \ref{subsubsec:Motion of the void}, respectively. Milgrom's constant $a_{_0} = 1.2 \times 10^{-10} \, \rm{m\,s^{-2}}$ is taken to be constant over cosmic time. Substantially higher values in the past may conflict with the CMB (Section~\ref{subsubsec:Radiation dominanted era and CMB}) and high-redshift rotation curves \citep{Milgrom_2017}.

Solving Equation~\ref{eq:governing_equation_Force_1} requires knowledge of the background cosmology. As argued in Section \ref{subsubsec:Background cosmology}, assuming this follows a standard Friedmann equation should be accurate at the sub-per cent level. We therefore apply the second Friedmann equation and assume a standard flat background cosmology ($\Omega_{\mathrm{m,0}} + \Omega_{\mathrm{\Lambda,0}} = 1$), yielding
\begin{eqnarray}
	\frac{\ddot{a}}{a} ~&=&~ - \frac{4 \mathrm{\pi} G}{3} (\rho_{\mathrm{m}} - 2 \rho_{\mathrm{\Lambda}}) \\
	~&=&~ H_{0}^{2} \left(- \frac{1}{2} \Omega_{\mathrm{m,0}} a^{-3} + \Omega_{\mathrm{\Lambda,0}} \right) \, , 
	\label{eq:ddot_a_to_a}
\end{eqnarray}
where $\rho_{\mathrm{m}}$ and $\rho_{\mathrm{\Lambda}}$ are the cosmic mean densities of matter and dark energy, respectively. We assume that $\rho_{\mathrm{m}} \propto a^{-3}$ while $\rho_{\mathrm{\Lambda}} = \text{const}$. The parameters $\Omega_{\mathrm{m},0}$ and $\Omega_{\mathrm{\Lambda},0}$ are the present-day matter and the dark energy densities in units of the critical density $\rho_{\mathrm{c}} = 3 H_{0}^{2}/\left( 8 \mathrm{\pi} G \right)$. We set $\Omega_{\mathrm{m,0}} = 0.315$, $\Omega_{\mathrm{\Lambda},0} = 0.685$, and choose a global Hubble constant of $H_{0} = 67.4 \, \rm{km\,s^{-1}\,Mpc^{-1}}$, consistently with the latest Planck data \citep{Planck_2018}. Imposing the boundary conditions $a = 0$ when $t = 0$ and $\dot{a} = H_0$ at $a = 1$, we get that
\begin{eqnarray}
	a \left( t \right) ~=~ \left( \frac{\Omega_{\mathrm{m},0}}{\Omega_{\mathrm{\Lambda},0}} \right)^{\frac{1}{3}} \sinh^{\frac{2}{3}} \left( {\frac{3}{2} \sqrt{\Omega_{\mathrm{\Lambda},0}} H_{0} t} \right) \, .
	\label{eq:governing_equation_a}
\end{eqnarray}

\subsubsection{Initial void profile} \label{subsubsec:Void initial profiles}

The implemented void in the fiducial simulation run is initialized with a Maxwell-Boltzmann radial density profile. This is motivated by the observed Local Volume, where the density increases inwards for distances $\la 40 \, \rm{Mpc}$ \citep[see e.g. figure~3 in][]{Karachentsev_2018}. The enclosed mass within co-moving radius $r_{\mathrm{com}}$ from the void centre is thus given by
\begin{eqnarray}
	\label{eq:governing_equation_Menc_MB}
	M_{\mathrm{enc}} &=& 4 \mathrm{\pi} \rho_{0} r_{\mathrm{void}}^{3} \left( \frac{x^{3}}{3} - \alpha_{\mathrm{void}} \epsilon \right) \, , ~ \mathrm{where}\\
	\epsilon &=& \int_0^x x'^{4} \exp \left( -\frac{x'^2}{2} \right) dx' \\
	&=& 3 \sqrt{\frac{\mathrm{\pi}}{2}} \mathrm{erf}\left( \frac{x}{\sqrt{2}} \right) - x \left( x^2 + 3 \right) \exp \left(-\frac{x^{2}}{2}\right) \, .
\end{eqnarray}
The dimensionless radius $x \equiv r_{\mathrm{com}}/r_{\mathrm{void}}$, while $\alpha_{\mathrm{void}}$ is the initial void strength and $r_{\mathrm{void}}$ is the parameter determining its co-moving size at $z = 9$. The first term in Equation~\ref{eq:governing_equation_Menc_MB} is the mass within a sphere of co-moving radius $r_{\mathrm{com}}$ if the density were equal to the cosmic mean, with the void arising from the mass deficit imposed by the second term.

We run different simulations with $\alpha_{\mathrm{void}}$ ranging from $10^{-5}$ to $10^{-2}$ and $r_{\mathrm{void}}$ ranging from $\left(50 - 1030\right) \, \rm{cMpc}$. The parameter range of the initial void strength is motivated by the expected density fluctuations at $z = 9$ based on CMB data. In addition, we also run simulations in which the void is modelled with a Gaussian or an exponential initial density profile (Appendices \ref{Appendix: KBC void mass profile} and \ref{Appendix:Joint probabilities for different void profiles}).

\subsubsection{External field history} \label{subsubsec:External field history}

As stated in Section~\ref{subsec: Milgromian dynamics}, the EFE is a consequence of the non-linearity of Milgrom's law of gravity \citep{Milgrom_1986}. Thus, we allow for the possibility that the void as a whole is embedded in an EF from even larger scales. We follow the usual approach of assuming the EF is sourced by a distant point-like object. This allows us to obtain the present-day Newtonian-equivalent external field using the simple interpolation function \citep{Famaey_2005}:
\begin{eqnarray}
    \frac{g_{_\mathrm{N,ext}}}{a_{_0}} ~=~ \frac{{\tilde{g}}^{2}_{\mathrm{ext}}}{1 + \tilde{g}_{\mathrm{ext}}} \, , 
	\label{eq:presentday_Newtonian_external_field}
\end{eqnarray}
where $\tilde{g}_{\mathrm{ext}}$ is the external field in units of $a_{_0}$.

\begin{table*}
    \caption{Constants and parameters of the here used cosmological MOND models. Our fiducial model assumes a Maxwell-Boltzmann initial density profile for the void (Section~\ref{subsubsec:Void initial profiles}) and a time-independent external field ($n_{\mathrm{EFE}} = 0$ in Equation~\ref{eq:time_dependent_external_field_history}).}
    \label{tab:parameters_MOND_model}
    \begin{tabular}{lll} \hline 
    Constants & Description & Value \\ \hline 
    $H_{0}^{\mathrm{global}}$ & Present-day global Hubble constant & $67.4 \, \rm{km\,s^{-1}\,Mpc^{-1}}$ \\
    $\Omega_{\mathrm{m},0}$ & Present-day matter density in units of $\rho_{\mathrm{c}}$ & $0.315$ \\
    $\Omega_{\mathrm{\Lambda},0}$ & Present-day dark energy density in units of $\rho_{\mathrm{c}}$ & $0.685$ \\
    $a_{i}$ & Cosmic scale factor at the start of the simulation & $0.1$ \\
    $a_{_0}$ & Milgrom's constant& $1.2 \times 10^{-10} \, \rm{m\,s^{-2}}$ \\ \hline
    External field parameters & & Parameter range \\ \hline
    $\tilde{g}_{\mathrm{ext}}$ & Present-day external field in units of $a_{_0}$ & $\left( 0, 0.5 \right)$\\
    $n_{\mathrm{EFE}}$ & Time dependence of the external field (Equation~\ref{eq:time_dependent_external_field_history})& $\left( -2, +2 \right)$\\ \hline
    Void parameters & & \\ \hline
    $\alpha_{\mathrm{void}}$ & Initial void strength at  $z = 9$ & $(10^{-5},10^{-2})$ \\
    $r_{\mathrm{void}}$ & Initial void size at $z = 9$ & $(50 \, \rm{cMpc}, 1030 \, \rm{cMpc})$ \\ \hline 
    \end{tabular}
\end{table*}

The evolution of the EFE over cosmic time is unknown due to the lack of a fully self-consistent MONDian framework. Since the EFE depends on the environment in which the MONDian system is embedded and thus on the formation of structure, we assume that the external field has a power-law dependence on the cosmic scale factor:
\begin{eqnarray}
    g_{_\mathrm{N,ext}}(t) ~=~ g_{_\mathrm{N,ext}}(t_{_0})a^{n_{\mathrm{EFE}}}(t) \, ,
    \label{eq:time_dependent_external_field_history}
\end{eqnarray}
where $t_{_0} = 13.8 \, \rm{Gyr}$ is the present time, and $n_{\mathrm{EFE}}$ is a free parameter ranging from $-2$ to $+2$ in steps of $0.5$ for different models. For our fiducial simulation run, we adopt a time-independent external field ($n_{\mathrm{EFE}} = 0$). The results for different external field histories are discussed in Section~\ref{subsubsec:Structure formation in MOND}. Table~\ref{tab:parameters_MOND_model} summarizes the fixed and free parameters of our models.

\subsection{Extracting mock observables} \label{subsec:Observational constraints}

Our cosmological MOND models are constrained by the observed density contrast of the KBC void \citep{Keenan_2013}, the local Hubble constant and deceleration parameter derived jointly from SNe data \citep{Camarena_2020}, the Hubble constant from strong lensing \citep{Wong_2020, Shajib_2020}, and the motion of the LG wrt. the CMB \citep{Kogut_1993}. In the following, we explain how we obtain the corresponding simulated quantities.

Our approach involves comparing the void models described in Section~\ref{subsec:Governing equations} with a control simulation of a void-free standard cosmology. The control trajectories have a fixed co-moving radius:
\begin{eqnarray}
	r(t) ~=~ r \left( t_{_0} \right) a \left( t \right).
	\label{eq:void_free_dynamics}
\end{eqnarray}
Since the lookback time can be derived from SNe luminosities or angular diameter distances in a standard background cosmology, we fix this variable between the void and control models, allowing us to analyse the difference in other variables. The main advantage of this approach is that in the absence of a local void, our calculated late-time cosmological parameters (e.g. $H_{0}$ and $q_{_0}$) would revert to their values in standard cosmology.

Local observations imply that we are located close to the void centre \citep{Keenan_2013, Karachentsev_2018}. Therefore, as a simplification we assume in our analysis that we are at the void centre (Sections~\ref{subsubsec:Density contrast}$-$\ref{subsubsec:Hubble constant from strong lensing}), except when calculating the likelihood of the observed LG peculiar velocity (Section~\ref{subsubsec:Motion of the void}). It is beyond the scope of our work to analyse the Hubble diagram and density field that might be seen by a substantially off-centre observer.

\subsubsection{Apparent scale factor}

The main quantity we extract is the redshift experienced by a photon as it travels from a particle to the void centre. This is given by
\begin{eqnarray}
	\frac{\lambda_{\mathrm{obs}}}{\lambda_{\mathrm{emit}}} ~=~ \frac{1}{a(t)} \overbrace{\sqrt{\frac{c + v_{\mathrm{int}}}{c - v_{\mathrm{int}}}}}^{\text{Doppler}}  \overbrace{\exp \left( \frac{1}{c^2} \int g_{\mathrm{void}} \, dr \right)}^{\text{GR}} \, ,
	\label{eq:governing_equation_Lambda_ratio_GR}
\end{eqnarray}
where $\lambda_{\mathrm{{obs}}}$ and $\lambda_{\mathrm{{emit}}}$ are the wavelengths of the light as measured by the observer and at the source of emission, respectively, $v_{\mathrm{int}}$ is the peculiar velocity of the particle relative to the void centre, and $g_{\mathrm{void}}$ is the gravity in the radially outwards direction. The factor of $a^{-1}$ arises from expansion of the universe while light from the particle is travelling towards us. This is the only factor that needs to be considered even without the void. The term marked `Doppler' is the special relativistic Doppler effect, while the exponential factor (marked `GR') is the gravitational redshift that arises because photons must climb up the void potential to reach its centre. As discussed in Section~\ref{subsubsec:Hubble constant from strong lensing}, relativistic lensing in MOND should yield similar results to General Relativity for the same $g$.

To limit the complexity of our algorithm and because we are dealing with a void at low $z$, we approximate the GR contribution by assuming the final density profile of the void is also applicable at earlier times. This leads to a time-independent gravitational field $g_{\mathrm{void}} \left( r \right)$. We use this to calculate the integral in Equation~\ref{eq:governing_equation_Lambda_ratio_GR} out to the co-moving distance where our past lightcone intersects the particle's trajectory (Section~\ref{subsubsec:Lightcone analysis}).

Since the observed SNe and lensing Hubble diagrams reported by observers are not corrected for the large peculiar velocities we expect in our model, the apparent scale factor is simply
\begin{eqnarray}
	a_{\mathrm{app}} ~\equiv~ \frac{\lambda_{\mathrm{emit}}}{\lambda_{\mathrm{obs}}} \, .
	\label{eq:governing_equation_Lambda_ratio_GR_reciproc}
\end{eqnarray}
We compare the behaviour of this $a_{\mathrm{app}}$ with the corresponding values in our control simulations, which are governed by Equation~\ref{eq:void_free_dynamics}. Since we run a finite number of trajectories for each model, we interpolate between them to ensure the comparison is done at fixed lookback time.

\subsubsection{Density contrast and redshift space distortion} \label{subsubsec:Density contrast}

In our models, the fractional underdensity inside a shell between radii $r_{\mathrm{min, now}}$ and $r_{\mathrm{max, now}}$ at the present time is
\begin{eqnarray}
	\label{eq:governing_equation_density_model}
	&&1 - \delta_{\mathrm{model}} ~= \\
    &&\left[1 - 3 \alpha_{\mathrm{void}} \left( \frac{I_{\mathrm{max}} - I_{\mathrm{min}}}{x_{\mathrm{max}}^{3} - x_{\mathrm{min}}^{3}} \right) \right] \left( \frac{r_{\mathrm{max, initial}}^{3} - r_{\mathrm{min, initial}}^{3}}{r_{\mathrm{max, now}}^{3} - r_{\mathrm{min, now}}^{3}} \right) \, ,\mathrm{with} \nonumber \\
    &&I_{\mathrm{min}} = \int_{_0}^{x_{\mathrm{min}}} x^{4} \exp \left( -\frac{x^{2}}{{2}} \right) dx \, .
\end{eqnarray}
Here, $r_{\mathrm{min, initial}}$ and $r_{\mathrm{max, initial}}$ are the initial co-moving distances of particles which are currently at $r_{\mathrm{min, now}}$ and $r_{\mathrm{max, now}}$, respectively, and $x_{\mathrm{min}} \equiv r_{\mathrm{min, initial}}/r_{\mathrm{void}}$. Similar procedures are used to calculate $x_{\mathrm{max}}$ and $I_{\mathrm{max}}$. The first term represents the initial density contrast, while the second accounts for expansion of the co-moving volume enclosed by the two shells.

As discussed in Section~\ref{subsec:KBC void}, the analysis of \citet{Keenan_2013} used a distance-redshift relation based on the assumption of no void (see their section 4.7). Therefore, we apply an RSD correction to the observed relative density contrast in order to estimate the true value:
\begin{eqnarray}
    \left(1 - \delta_{\mathrm{obs, corr}} \right) &=& \left(1 - \delta_{\mathrm{obs}} \right) f_{\mathrm{model}} \, , \quad \mathrm{with} \\
    f_{\mathrm{model}} &=& \left( \frac{r_{\mathrm{control,out}}^{3} - r_{\mathrm{control,in}}^{3}}{r_{\mathrm{void,out}}^{3} - r_{\mathrm{void,in}}^{3}} \right) \, . 
	\label{eq:redshift_distortion}
\end{eqnarray}
Here, $\delta_{\mathrm{obs}}$ is the observed relative density contrast between the distances $r_{\mathrm{void,in}}$ and $r_{\mathrm{void,out}}$ at the present time. However, observations uncorrected for RSD are reported as if the known redshift range of the survey covers the distance range $r_{\mathrm{control,in}} - r_{\mathrm{control,out}}$, which are the corresponding distances to the same $z$ in a void-free universe. The number of galaxies counted by the observers thus corresponds to a different $\delta$ than what they report, which is the RSD effect. Note that its magnitude will depend on the void model, so it is not possible to know the true density contrast in a model-independent way. This is because it is not possible to convert redshifts to distances without a dynamical model of the void. As a result, the uncertainty $\sigma_{\mathrm{obs,corr}}$ is also model-dependent.

We can compare the so-corrected observed $\delta$ to the model prediction (Equation~\ref{eq:governing_equation_density_model}). This leads to a $\chi^2$ contribution of
\begin{eqnarray}
    \chi_{\delta}^{2} ~=~ \left( \frac{\delta_{\mathrm{model}} - \delta_{\mathrm{obs,corr}}}{\sigma_{\mathrm{obs,corr}}} \right)^{2} \, ,
    \label{eq:redshift_distortion_chi_sq}
\end{eqnarray}
which is calculated for the relative density contrasts in the redshift range $0.01 < z < 0.07$ and between distances of $600 \, \rm{Mpc}$ and $800 \, \rm{Mpc}$ at the present time. According to \citet{Keenan_2013}, we estimate that $\delta_{\mathrm{obs,in}} = 0.46 \pm 0.06$ in the inner part of the void, while $\delta_{\mathrm{obs,out}} = 0.0 \pm 0.1$ in its outer part (see their table~1 and figure~11).

\subsubsection{Lightcone analysis} \label{subsubsec:Lightcone analysis}

To determine exactly when we would observe a test particle, we need to determine the intersection between its trajectory and our past lightcone. This occurs when the co-moving distance travelled by a light ray emitted from a particle equals the time-dependent co-moving distance to the particle. In other words,
\begin{eqnarray}
	c \int_{t_{_\mathrm{LC}}}^{t_{_0}} \frac{dt}{a \left( t \right)} ~=~ \frac{r(t_{_\mathrm{LC}})}{a(t_{_\mathrm{LC}})} \, ,
	\label{eq:governing_equation_past_lightcone}
\end{eqnarray}
where $t_{_\mathrm{LC}}$ is the cosmic time when our past lightcone intersects a particle's trajectory. This is obtained by solving Equation~\ref{eq:governing_equation_past_lightcone} using the Newton-Raphson algorithm. We can then calculate relevant quantities at that time, which is used in our analyses related to the Hubble diagram (Section~\ref{subsubsec:Hubble constant from SNe}).

However, when comparing the simulated $v_{\mathrm{pec}}$ with the observed LG peculiar velocity (Section~\ref{subsubsec:Motion of the void}), we need to extract $v_{\mathrm{pec}}$ at the present epoch since the measurement relates to the LG motion today. To limit the complexity of our analysis, we also use the present positions of particles when determining the density field of the void (Section~\ref{subsubsec:Density contrast}). This should be valid if the void has not appreciably changed in the time needed for light to cross it, which is reasonable for a void much smaller than the Hubble distance $c/H_0 = 4.4$~Gpc.

\subsubsection{Hubble constant and deceleration parameter from SNe} \label{subsubsec:Hubble constant from SNe}

We constrain our models with the results of \citet{Camarena_2020}, who derived the local Hubble constant and deceleration parameter jointly from Pantheon SNe in the redshift range $0.023 \leq z \leq 0.15$. As discussed earlier, we first find the difference in the apparent scale factor between our void model and a control void-free model.
\begin{eqnarray}
    \Delta a \left( t \right) ~\equiv~ a_{\mathrm{app}} - a_{\mathrm{control}} \, ,
   	\label{eq:difference_scalefactor}
\end{eqnarray}
where $a_{\mathrm{app}}$ is the apparent scale factor (Equation~\ref{eq:governing_equation_Lambda_ratio_GR_reciproc}), and $a_{\mathrm{control}}$ is the scale factor at the same cosmic time in the control model of a void-free standard cosmology. Expanding Equation~\ref{eq:difference_scalefactor} as a Taylor series in the vicinity of the present time $t_{_0}$, we get that
\begin{eqnarray}
    \Delta a \left(t \right) = \Delta \dot a\left(t_{_0} \right) \left(t - t_{_0} \right) + \frac{1}{2} \Delta \ddot{a} \left(t_{_0} \right) \left(t - t_{_0} \right)^2 + \mathcal{O} \left( t - t_{_0} \right)^3.
\end{eqnarray}
Dividing the above equation by $a \left(t_{_0} \right) \equiv 1$ and using the definitions of the Hubble parameter ($H \equiv \dot{a}/a$) and deceleration parameter ($q \equiv - a\ddot{a}/\dot{a}^{2}$), we obtain that
\begin{eqnarray}
    \frac{\Delta{a} \left( t \right)}{a \left( t_{_0} \right)} = \frac{\Delta \dot{a} \left(t_{_0}\right)}{a(t_{_0})} \left(t- t_{_0} \right) + \frac{\Delta \ddot a \left( t_{_0} \right) }{2 a \left( t_{_0} \right)} \left(t - t_{_0}\right)^{2} + \mathcal{O} \left(t-t_{_0} \right)^{3} \nonumber \\ 
    = \Delta H_{0} \left(t - t_{_0} \right)  - \frac{\Delta \left(q_{_0} H_{0}^{2} \right)}{2} \left(t - t_{_0} \right)^{2} + \mathcal{O} \left( t-t_{_0}\right)^{3} \, ,
	\label{eq:Taylor_expansion_scalefactor_divided}
\end{eqnarray}
where $\Delta H_{0}$ and $\Delta \left(q_{_0} H_{0}^{2} \right)$ are the boosts to these parameters due to the void, and the $_0$ subscripts denote present-day quantities. We find these by fitting $\Delta a \left( t \right)$ using a parabola forced to pass through $\Delta a = 0$ at $t = t_{_0}$. The local Hubble and deceleration parameters are thus
\begin{eqnarray}
	H_{0}^{\mathrm{local}} &=& H_{0}^{\mathrm{global}} + \Delta H_{0}  \, , \\
    q_{_0}^{\mathrm{local}} &=& \frac{ \left( H_{0}^{\mathrm{global}} \right)^2 \left( \frac{\Omega_{\mathrm{m,0}}}{2} - \Omega_{\mathrm{\Lambda,0}} \right) +\Delta \left( q_{_0} H_{0}^{2} \right)}{ \left( H_{0}^{\mathrm{local}} \right)^{2}} \, .
	\label{eq:q0}
\end{eqnarray}
Because of a historical accident where it was assumed that the expansion of the Universe should decelerate, $q_{_0}$ was defined as the present deceleration parameter $-a \ddot{a}/\dot{a}^2$. It was subsequently shown that the Universe accelerates, implying $q_{_0} < 0$ \citep{Riess_1998, Schmidt_1998, Perlmutter_1999}. In order to minimize confusion from unnecessary use of $-$ signs, we introduce from now on the acceleration parameter:
\begin{eqnarray}
    \overline{q}_{_0} ~\equiv~ - q_{_0} ~\equiv~ \frac{a \ddot{a}}{\dot{a}^2} \left( t = t_{_0} \right) \, .
	\label{eq:q0_new}
\end{eqnarray}
The $\Lambda$CDM theory with the parameters obtained from \citet{Planck_2018} predicts $\overline{q}_{_0} = \Omega_{\mathrm{\Lambda},0} - \frac{1}{2} \Omega_{\mathrm{m},0} = 0.53$. In the absence of a void, $H_{0}^{\mathrm{local}}$ and $\overline{q}_{_0}^{\mathrm{local}}$ become identical to the Planck values since we use those for the background cosmology (Equation~\ref{eq:governing_equation_a}).

The combined $\chi^{2}$ contribution from $H_{0}$ and $\overline{q}_{_0}$ is
\begin{eqnarray}
	\chi_{H_{0}, \overline{q}_{_0}}^{2} &=& \frac{1}{2} \left[ \frac{\left( A + B \right)^{2}}{1 + C} + \frac{\left( A - B \right)^{2}}{1 - C} \right] \, , \quad \mathrm{where} \\
	A &\equiv& \frac{ H_{0} - H_{0}^{\mathrm{local}} }{\sigma_{H_{0}}} \quad \mathrm{and}  \\
	B &\equiv& \frac{ \overline{q}_{_0} - \overline{q}_{_0}^{\mathrm{local}}}{\sigma_{\overline{q}_{_0}}} \, .
	\label{eq:chi_sq_H0_q0}
\end{eqnarray}
Observationally, $\overline{q}_0^{\mathrm{local}} = 1.08 \pm 0.29$ and $H_{0}^{\mathrm{local}} = 75.35 \pm 1.68 \, \rm{km\,s^{-1}\,Mpc^{-1}}$, with a mutual correlation coefficient of $C = 0.515$ \citep{Camarena_2020}. Their section~4 mentions that their posterior inference is very close to Gaussian, justifying our $\chi^2$ approach.

\subsubsection{Hubble constant from strong lensing} \label{subsubsec:Hubble constant from strong lensing}

Empirically, it has been shown that light deflection in strong lenses works similar to General Relativity for the same non-relativistic $\bm{g}$ \citep{Collett_2018}. In the H0LiCOW lenses, $\bm{g}$ is constrained using the positions and time delays between images and also with velocity dispersion data. Their analysis should remain valid even in a MOND context, since the results of \citet{Collett_2018} can be reproduced in relativistic versions of MOND \citep{Milgrom_2013}. The latter work showed that this approach works well empirically even in the deep-MOND regime, which can only be probed using weak lensing. This is because strong lensing always occurs in the Newtonian regime due to MOND's cosmological coincidence (Equation~\ref{eq:MOND_coincidence}), as explained in \citet{Sanders_1999}. Hence, strong lensing is little affected by MOND. None the less, we discuss in Section~\ref{subsubsec:Excluding strong lensing time delays} how $H_0$ measurements from strong lensing impact our analysis, and consider the effect of excluding these measurements.

\begin{table}
    \caption{Measurements of $H_{0}$ from lensed quasars with the deflector at redshift $z_{d}$, as reported by \citet{Wong_2020} and \citet{Shajib_2020} for a flat $\Lambda$CDM cosmology. Their data are used to constrain the MOND models in Section~\ref{sec:Results of MOND simulation}, where low and high error bars are averaged to get a single Gaussian uncertainty for each lens. }
    \label{tab:observed_H0_strong_lensing}
    \begin{tabular}{llll} \hline 
    Lens system & $z_{d}$ & $H_{0}$ & Reference \\ 
                &       & $[\rm{km\,s^{-1}\,Mpc^{-1}}]$ & \\ \hline
    B1608+656       & $0.6304$  & $71.0_{-3.3}^{+2.9}$ & \citet{Wong_2020} \\ [5pt]
    RXJ1131$-$1231   & $0.295$   & $78.2_{-3.4}^{+3.4}$ & '' \\ [5pt]
    HE 0435$-$1223     & $0.4546$  & $71.7_{-4.5}^{+4.8}$ & '' \\ [5pt]
    SDSS 1206+4332  & $0.745$   & $68.9_{-5.1}^{+5.4}$ & '' \\ [5pt]
    WFI2033$-$4723    & $0.6575$  & $71.6_{-4.9}^{+3.8}$ & '' \\ [5pt]
    PG 1115+080     & $0.311$   & $81.1_{-7.1}^{+8.0}$ & '' \\ [5pt]
    DES J0408$-$5354  & $0.597$   & $74.2_{-3.0}^{+2.7}$ & \citet{Shajib_2020} \\ \hline
    \end{tabular}
\end{table}

In our main analysis, we constrain our MOND models with $H_{0}$ measured from seven strong-lens systems with the deflector at redshift $z_{d}$. Our data set is derived from \citet{Shajib_2020} and \citet{Wong_2020}. In our models, the Hubble constant at redshift $z_{d}$ is estimated as
\begin{eqnarray}
    H_{0,\mathrm{lensing}}^{\mathrm{model}} ~=~ H_{0}^{\mathrm{global}} + \frac{\Delta a_{d}}{t_{d} - t_{0}} \, ,
    \label{eq:H0_lensing}
\end{eqnarray}
where $\Delta a_{d}$ is the difference in $a_{\mathrm{app}}$ between the void and control models, and $t_{d}$ is the cosmic age at redshift $z_{d}$. The $\chi^{2}$ contribution from all seven lenses is
\begin{eqnarray}
    \chi_{H_{0}, \mathrm{lensing}}^{2} ~=~ \sum_{i=1}^{7} \left( \frac{H_{0,\mathrm{lensing},i}^{\mathrm{model}} - H_{0,\mathrm{lensing},i}^{\mathrm{obs}}}{\sigma_{\mathrm{obs},i}^{\mathrm{lensing}}} \right)^{2},
	\label{eq:lensing_Hubble_chi_sq}
\end{eqnarray}
where $H_{0,\mathrm{lensing},i}^{\mathrm{obs}}$ and $\sigma_{\mathrm{obs},i}^{\mathrm{lensing}}$ are the derived Hubble constant and corresponding uncertainty for lens system $i$ as reported by \citet{Shajib_2020} or \citet{Wong_2020}, which we summarize in Table~\ref{tab:observed_H0_strong_lensing}.

\subsubsection{Local Group peculiar velocity} \label{subsubsec:Motion of the void}

An important constraint on our model is the observed motion of the LG relative to the surface of last scattering. The observed CMB dipole indicates that the LG moves with a peculiar velocity of $v_{\mathrm{LG}} = 627 \pm 22 \, \rm{km\,s^{-1}}$ towards Galactic coordinates $\left( l, b \right) = \left( 276^{\circ} \pm 3^{\circ}, 30^{\circ} \pm 3^{\circ} \right)$ \citep{Kogut_1993}.

\begin{figure}
    \includegraphics[width=\linewidth]{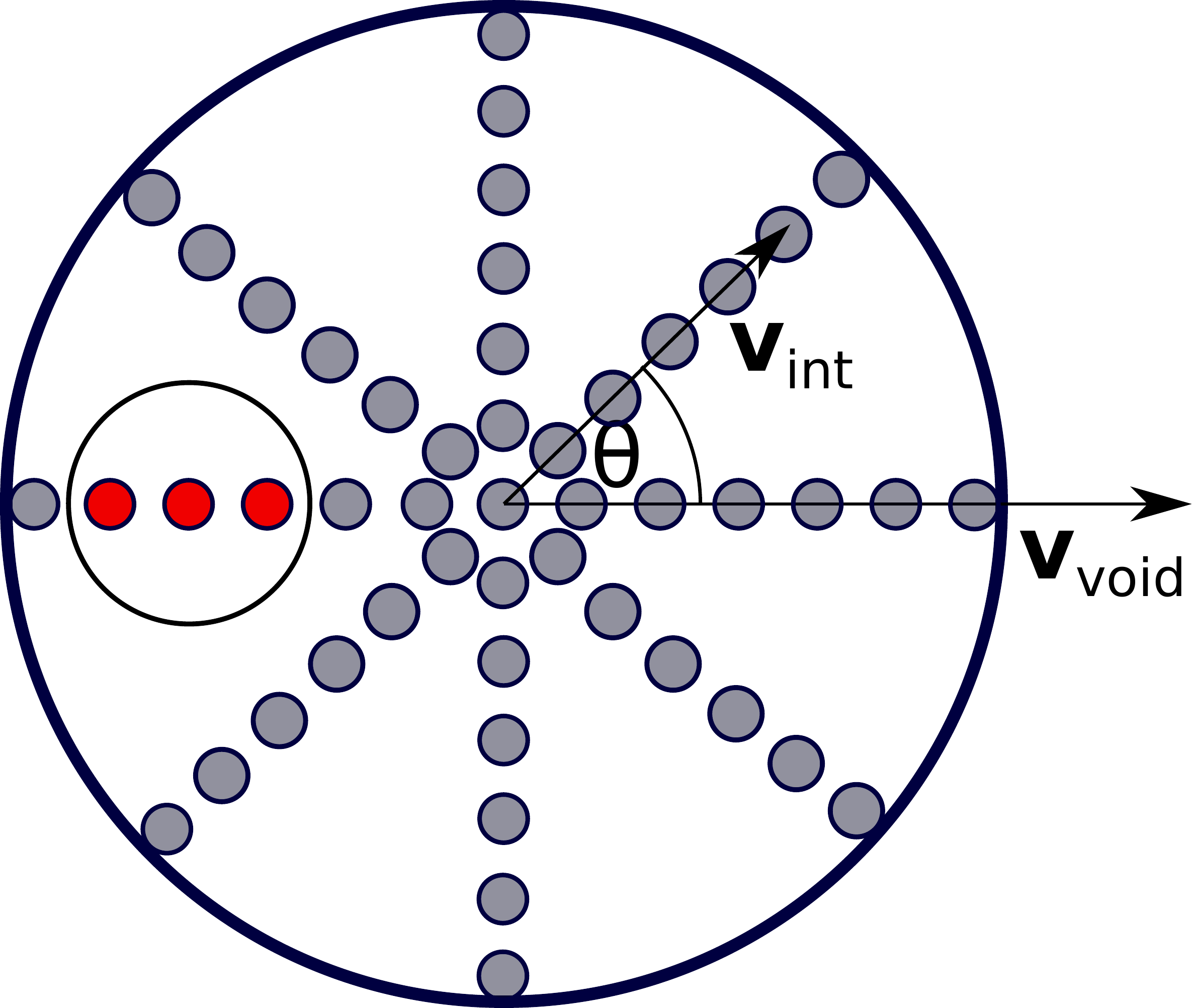}
    \caption{Schematic of the KBC void, which as a whole moves with velocity $v_{\mathrm{void}}$ due to the time-integrated EFE accounting for Hubble drag \citep[Equation~\ref{eq:motion_void}, see also section~2.2 in][]{Banik_2018a}. The total peculiar velocity of a particle wrt. the CMB, $v_{\mathrm{tot}}$, is calculated by combining $v_{\mathrm{void}}$ with the internal velocity $v_{\mathrm{int}}$ of a particle relative to the void centre (Equation~\ref{eq:total_peculiar_velocity}). The inner circle illustrates the region in which $v_{\mathrm{tot}} \leq v_{\mathrm{LG}} = 627 \, \rm{km\,s^{-1}}$. We estimate its volume by adding the volumes of the red cells.}
    \label{figure_void_motion_schematic}
\end{figure}

To calculate the expected peculiar velocity in different parts of the void, we first need to consider the motion of the void as a whole. The void peculiar velocity $\bm{v}_{\mathrm{void}}$ arises from the time-integrated EFE. Using the approach stated in section~2.2 of \citet{Banik_2018a}, we get that		
\begin{eqnarray}
    a \left( t_{_0} \right) \bm{v}_{\mathrm{void}} ~=~ \int_{t_{i}}^{t_{_0}} \bm{g}_{\mathrm{ext}} \left( t \right) \, a \left( t \right) dt \, , 
	\label{eq:motion_void}
\end{eqnarray}
where $\bm{g}_{\mathrm{ext}}$ is the external field, and the integrating factor $a$ accounts for Hubble drag. The total velocity of a particle wrt. the CMB is
\begin{eqnarray}
    v_{\mathrm{tot}}^{2} ~=~ v_{\mathrm{int}}^{2} + v_{\mathrm{void}}^{2} + 2 v_{\mathrm{int}} v_{\mathrm{void}} \cos{\theta} \, ,
	\label{eq:total_peculiar_velocity}
\end{eqnarray}
where $\bm{v}_{\mathrm{int}}$ is the `internal' velocity of the particle relative to the void centre, and $\theta$ is the angle between $\bm{v}_{\mathrm{void}}$ and the void-centric position of the particle. A schematic representation of this situation is depicted in Figure~\ref{figure_void_motion_schematic}.

For numerical purposes, the simulated void is divided into cells. The volume of cell $i$ is
\begin{eqnarray}
	V_{i} ~=~ \frac{\Delta \left(r^3 \right)}{3} \times 2 \mathrm{\pi} \Delta \left( \cos \theta \right) \, ,
	\label{eq:Volume_element}
\end{eqnarray}	
where $\Delta r$ and $\Delta \theta$ are the radial and angular bin size, respectively. $V_i$ is determined by the change in $r^3$ and $\cos\theta$ across the cell.

In order to quantify how the observed $v_{\mathrm{LG}} = 627 \, \rm{km\,s^{-1}}$ affects the relative probability of a model, we define $f_{\mathrm{motion}}$ as the proportion of cells which satisfy
\begin{eqnarray}
	v_{\mathrm{LG}} - \epsilon ~\leq~ v_{\mathrm{tot}} ~<~ v_{\mathrm{LG}} + \epsilon \, ,
	\label{eq:f_motion_condition}
\end{eqnarray}
where $\epsilon$ is a numerical parameter whose choice should have no bearing on our final results. To get a good balance between reducing numerical noise and increasing the accuracy, we choose $\epsilon = 50 \, \rm{km\,s^{-1}}$. The resulting error should be of order $\left(50/630 \right)^2$, which is acceptable given other uncertainties. On the other hand, $50  \, \rm{km\,s^{-1}}$ is much larger than the change in $v_{\mathrm{tot}}$ between adjacent cells. We obtain similar results if $\epsilon = 30  \, \rm{km\,s^{-1}}$ is used instead. Using this discretized scheme, we get that
\begin{eqnarray}
	f_{\mathrm{motion}} ~\equiv~ \sum_{i} V_{i} \div \frac{4 \mathrm{\pi} \left( n r_{\mathrm{void}}^{\mathrm{rms}} \right)^{3}}{3} \, , 
	\label{eq:P_value_motion_void}
\end{eqnarray}
where $r_{\mathrm{void}}^{\mathrm{rms}}$ is the rms size of the void, and $n$ is a dimensionless factor of order unity that sets our prior expectation for how close we are to the void centre (we must be within a distance of $n r_{\mathrm{void}}^{\mathrm{rms}}$). Since observations suggest that we are located quite close to the centre \citep[e.g.][]{Keenan_2013, Karachentsev_2018}, we adopt $n = 0.5$ for our probability calculations. We estimate the void size as
\begin{eqnarray}
	r_{\mathrm{void}}^{\mathrm{rms}} ~\equiv~ \sqrt{\frac{\int_0^\infty r^2 \delta \left( r \right) \, dr}{\int_0^\infty \delta \left( r \right) \, dr}} \, ,
	\label{r_void_rms}
\end{eqnarray}
with the void profile $\delta \left( r \right)$ found using Equation~\ref{eq:governing_equation_density_model}. In practice, we cut off the integrals at a very large distance much beyond the possible extent of the void. Since $\delta \to 0$ at large $r$, this is sufficient to accurately estimate the limiting values of both integrals.

In Section~\ref{subsec:Results The best-fitting model}, we apply a less sophisticated probability calculation where we assume that $v_{\mathrm{tot}}$ follows a Gaussian distribution. The extent to which the observed $v_{\mathrm{LG}}$ is an outlier to the simulated $v_{\mathrm{tot}}$ distribution is given by the proportion of the void volume with $v_{\mathrm{tot}} \leq v_{\mathrm{LG}}$. This allows an easier comparison with the other constraints.

Table~\ref{tab:comparison_bestfittingmodel_with_observations} summarizes the here presented observational constraints, which are used to test our cosmological MOND model in the following section.

\begin{table*}
    \caption{Parameters used to quantify the tension of different MOND models with observations, along with a brief description. More information can be found in the indicated section.}
    \label{tab:comparison_bestfittingmodel_with_observations}
    \begin{tabular}{lll} \hline 
    Parameter & Description & Section  \\ \hline 
    $\chi_{\delta_{\mathrm{in}}}^{2}$ & Density contrast of the void in the redshift range $0.01 < z < 0.07$& Sec. \ref{subsubsec:Density contrast} \\ 
    $\chi_{\delta_{\mathrm{out}}}^{2}$ & Density contrast of the void between $600 \, \rm{Mpc}$ and $800 \, \rm{Mpc}$& Sec. \ref{subsubsec:Density contrast} \\ 
    $\chi_{H_{0},\overline{q}_{_0}}^{2}$ & Hubble constant and acceleration parameter derived jointly from SNe with $0.023 \leq z \leq 0.15$ & Sec.~\ref{subsubsec:Hubble constant from SNe} \\
    $\chi_{H_{0},\mathrm{lensing}}^{2}$ & Hubble constant using time delays from seven strong lenses & Sec.~\ref{subsubsec:Hubble constant from strong lensing} \\    
    $f_{\mathrm{motion}}$ & Fraction of void volume whose velocity wrt. the CMB is similar to that of the LG & Sec. \ref{subsubsec:Motion of the void} \\ 
    \hline
    \end{tabular}
\end{table*}

\section{Results of MOND simulations} \label{sec:Results of MOND simulation}

In this section, we perform a detailed parameter study of our cosmological MOND models. This includes an estimation of the tension between our best-fitting model and observations of the local Universe. We focus on a void initialized with a Maxwell-Boltzmann profile (Section~\ref{subsubsec:Void initial profiles}). Results for Gaussian and exponential starting profiles are presented in Appendix~\ref{Appendix:Joint probabilities for different void profiles}. We first quantify the relative probabilities of different models (Section~\ref{subsec:Joint probabilities}), and then check how well our best-fitting model agrees with observations (Section~\ref{subsec:Results The best-fitting model}).

\subsection{Relative probabilities of different models} \label{subsec:Joint probabilities}

The observational constraints can mostly be assumed to have a Gaussian distribution, allowing a standard $\chi^2$-based analysis. This is due to the central limit theorem and the fact that e.g. many SNe are used in the study of \citet{Camarena_2020}. However, the expected distribution of $v_{\mathrm{tot}}$ (Equation~\ref{eq:total_peculiar_velocity}) is based on just one void, so we cannot assume Gaussianity. This constraint requires a more careful treatment, as explained in Section~\ref{subsubsec:Motion of the void}.

Combining the different constraints, we get that the joint probability of each model is
\begin{eqnarray}
    && P(\mathrm{Model} \left| \right. \mathrm{Observations}) \nonumber \\
    && \propto \left( \prod_{i} \frac{1}{\sigma_{\mathrm{obs},i}} \right) \exp \left(-\frac{\chi^{2}}{2} \right) \times f_{\mathrm{motion}} \, , \quad \mathrm{with} \\
    && \chi^{2} ~=~ \chi_{\delta_{\mathrm{in}}}^{2} + \chi_{\delta_{\mathrm{out}}}^{2} + \chi_{H_{0},\overline{q}_{_0}}^{2} + \chi_{H_{0}, \mathrm{lensing}}^{2} \, .
    \label{eq:P_value_total}
\end{eqnarray}
We use $i$ to label different observational constraints, each of which has uncertainty $\sigma_{\mathrm{obs},i}$. The only model-dependent uncertainties are the density contrasts of the inner and outer parts of the KBC void, a consequence of the applied RSD correction (Equation~\ref{eq:redshift_distortion}). 

\begin{figure*}
	\includegraphics[width=\linewidth]{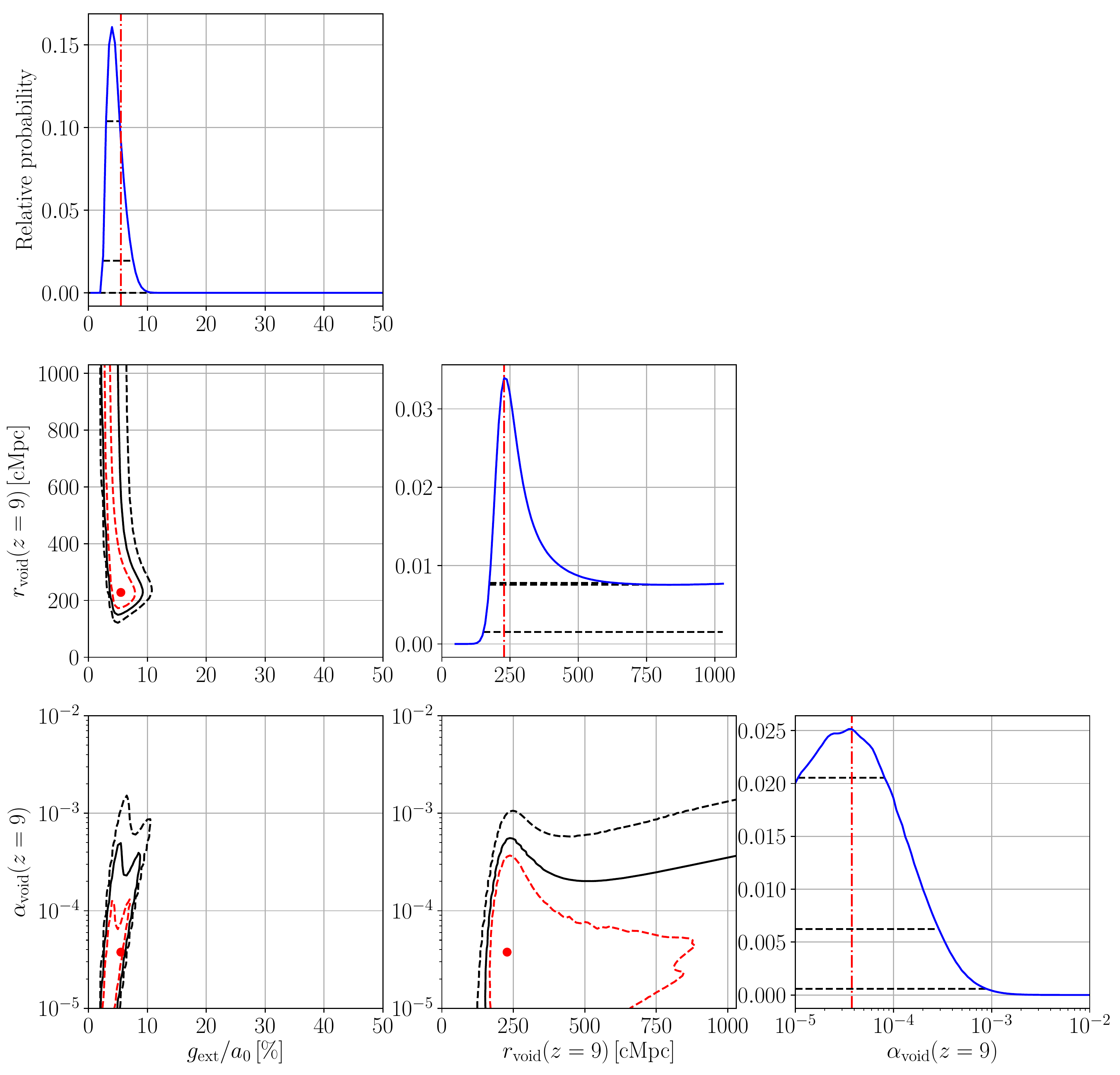}
	\caption{Marginalized posterior distribution of the indicated model parameters based on $10^{6}$ Maxwell-Boltzmann MOND void models. The red dashed, black solid, and black dashed contours mark the $1 \sigma$, $2 \sigma$, and $3 \sigma$ confidence levels, respectively. For 1D posteriors, these are shown using the horizontal black lines. The $1 \sigma$ and $2 \sigma$ lines are at almost the same level for $r_{\mathrm{void}}$. The red dot or vertical line marks the best-fitting model with an external field strength of $g_{\mathrm{ext}} = 0.055 \, a_{_0}$, and an initial void size and strength of $r_{\mathrm{void}} = 228.2 \, \rm{cMpc}$ and $\alpha_{\mathrm{void}} = 3.76 \times 10^{-5}$, respectively, at $z = 9$. This model is analysed in more detail in Section~\ref{subsec:Results The best-fitting model}.}
	\label{figure:marginalization_MBprofile}
\end{figure*}

Figure~\ref{figure:marginalization_MBprofile} shows the marginalized posterior distributions of the model parameters and parameter pairs based on $10^{6}$ MOND models. The assumed external field strength has a significant impact on individual models. On the one hand, increasing the EFE typically makes the MONDian subsystem more Newtonian, suppressing the growth of structures. This results in less pronounced voids at the present time, and consequently a local Hubble constant and acceleration parameter closer to the Planck predictions. On the other hand, some EFE is required because otherwise structure formation would be too efficient, causing a very high local Hubble constant and acceleration parameter. These considerations restrict $g_{\mathrm{ext}}$ to the range $ \left( 0.024 - 0.076 \right) a_{_0}$ at $2\sigma$ confidence, with the most likely value being $0.04 \, a_{_0}$.

In contrast, the initial void size and strength are not strongly constrained $-$ our analysis merely yields $2 \sigma$ limits of $\alpha_{\mathrm{void}} =  10^{-5} - 2.91 \times 10^{-4}$ and $r_{\mathrm{void}} = \left( 173.4 - 818.6 \right) \, \rm{cMpc}$. This is because the local Hubble constant and acceleration parameter are estimated only for SNe in the redshift range $0.023 \leq z \leq 0.15$, which does not constrain the outer part of the void. Although some constraints are available at higher $z$ from lensed quasars, the uncertainties of $H_{0}$ measured in this way are relatively large, allowing for a wide range of possible model parameters (Table~\ref{tab:observed_H0_strong_lensing}).

The best-fitting model is that for which the joint probability (Equation~\ref{eq:P_value_total}) becomes maximal. We mark this as a red dot in Figure~\ref{figure:marginalization_MBprofile} and consider it our fiducial model. It has an initial void strength of $\alpha_{\mathrm{void}} = 3.76 \times 10^{-5}$ at $z = 9$, an initial void size of $r_{\mathrm{void}} = 228.2 \, \rm{cMpc}$, and an external field strength of $g_{\mathrm{ext}} = 0.055 \, a_{_0}$, causing the void as a whole to move with $v_{\mathrm{void}} = 1586 \, \rm{km\,s^{-1}}$. We analyse this particular model in more detail in the subsequent section. 
    
The marginalized posterior distributions for MOND models with Gaussian and exponential initial profiles are shown in Appendix~\ref{Appendix:Joint probabilities for different void profiles}. Those models still assume a time-independent EFE. In Section~\ref{subsubsec:Structure formation in MOND}, we present and discuss an analysis demonstrating that allowing time-dependence of the EFE reveals no strong preference for a time-varying EFE, though some variation is expected on theoretical grounds.

\subsection{The fiducial model} \label{subsec:Results The best-fitting model}

In the following, we discuss the results of our best-fitting (fiducial) model.

\subsubsection{Density profile} \label{subsubsec:Density profile}

\begin{figure*}
    \includegraphics[width=88mm]{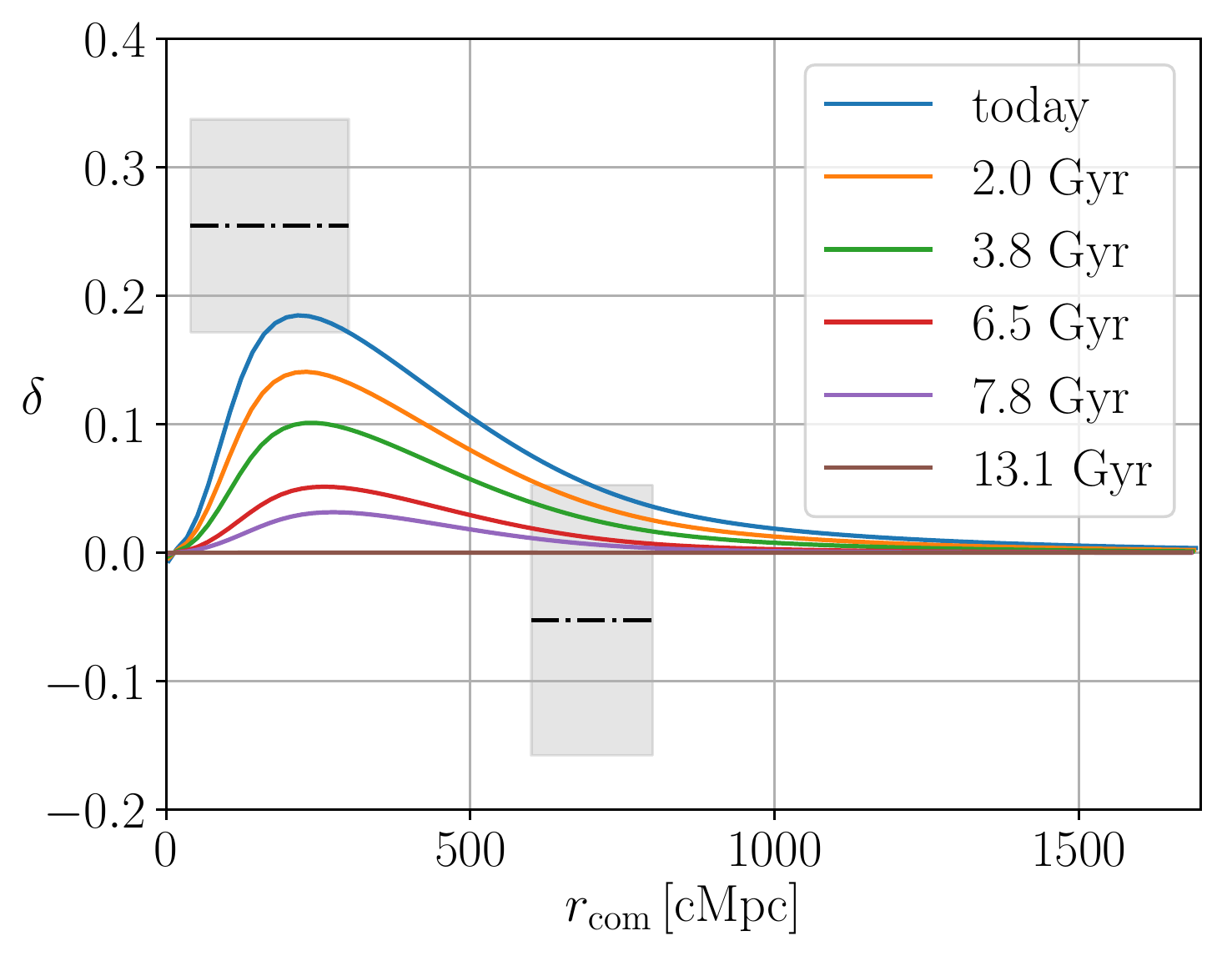}
    \includegraphics[width=88mm]{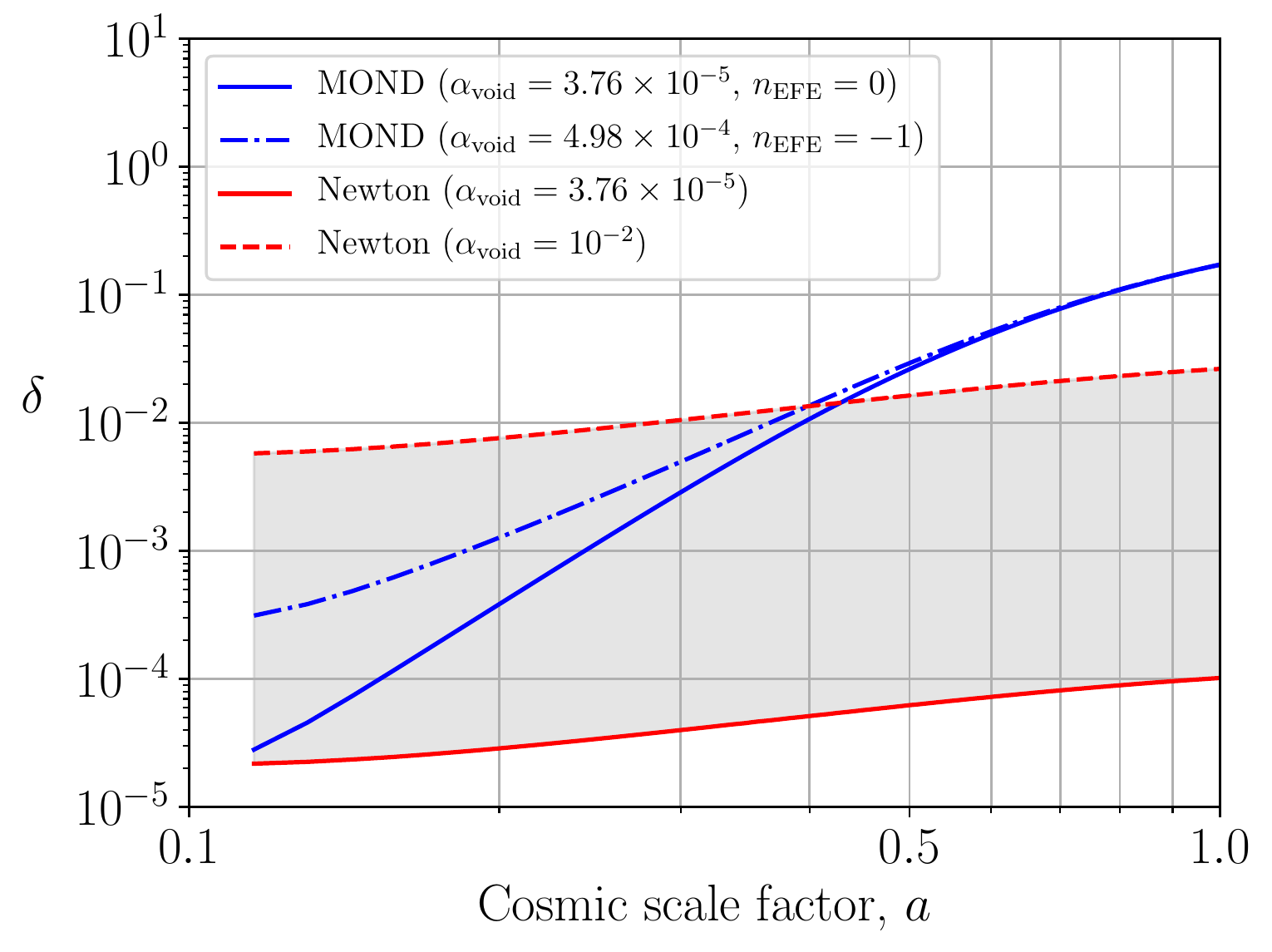}
    \caption{\emph{Left}: Time evolution of the radial density profile of the fiducial MOND model ($g_{\mathrm{ext}} = 0.055 \, a_{_0}$, $r_{\mathrm{void}} = 228.2 \, \rm{cMpc}$, $\alpha_{\mathrm{void}} = 3.76 \times 10^{-5}$). Different line colours refer to different lookback times, as indicated in the legend. The two black dot-dashed lines and the grey-shaded areas mark the RSD corrected observed density contrast of the KBC void, i.e. $\delta _{\mathrm{obs, corr}}^{\mathrm{in}} = 0.254 \pm 0.083$ between $40$ and $300 \, \rm{Mpc}$, and $\delta _{\mathrm{obs, corr}}^{\mathrm{out}} = -0.052 \pm 0.105$ between $600$ and $800 \, \rm{Mpc}$. \emph{Right}: Evolution of the density contrast within a sphere of radius $300 \, \rm{cMpc}$ for the fiducial MOND model (the blue solid line, $\delta ~\appropto~  a^{3.8}$ for $0.3 \leq a \leq 0.7$), a MOND model with approximately the same $\delta$ and EFE today but with higher EFE in the past according to $n_{\mathrm{EFE}} = -1$ in Equation~\ref{eq:time_dependent_external_field_history} (the blue dot-dashed line, $\delta \appropto a^{3.3}$ for $0.3 \leq a \leq 0.7$), and Newtonian models (the red lines, $\delta \appropto a^{0.8}$ for $0.3 \leq a \leq 0.7$). The red solid line shows the Newtonian model with the same initial void parameters as the fiducial MOND model, while the red dashed line refers to another Newtonian model with the same $r_{\mathrm{void}}$ but where $\alpha_{\mathrm{void}} = 10^{-2}$.}
    \label{figure_void_density_profile_results}
\end{figure*}

We begin by studying the density contrast of the fiducial model at different times. This is plotted in the left-hand panel of Figure~\ref{figure_void_density_profile_results}. The void starts with an initial size of $r_{\mathrm{void}} = 228.2 \, \rm{cMpc}$ and a very small initial strength of $\alpha_{\mathrm{void}} = 3.76 \times 10^{-5}$ at $z = 9$. Equation \ref{r_void_rms} implies that $r_{\mathrm{void}}^{\mathrm{rms}} = r_{\mathrm{void}} \sqrt{3} = 395.2 \, \rm{cMpc}$ at that time. At present, the void has grown to a size of $r_{\mathrm{void}}^{\mathrm{rms}} = 528.7 \, \rm{Mpc}$ and has a density contrast of $\delta_{\mathrm{in}} = 0.172$ in the redshift range $0.01 < z < 0.07$. Correcting the corresponding observed density contrast from \citet{Keenan_2013} by the model-dependent RSD correction factor $f_{\mathrm{model}}^{\mathrm{in}} = 1.38$ yields $\delta _{\mathrm{obs, corr}}^{\mathrm{in}} = 0.254 \pm 0.083$. This agrees with the simulated value at the $0.99\sigma$ level. The calculated density contrast between $600$ and $800 \, \rm{Mpc}$ is $\delta_{\mathrm{out}} = 0.050$, which also compares favourably with the RSD-corrected observed density contrast $\delta _{\mathrm{obs, corr}}^{\mathrm{out}} = -0.052 \pm 0.105$ ($f_{\mathrm{model}}^{\mathrm{out}} = 1.05$). The tension in this case is only $0.97 \sigma$.

The long-range modification to gravity in MOND causes structure formation to be much more efficient than in $\Lambda$CDM cosmology \citep[e.g.][and references therein]{Sanders_1999, Famaey_2012}. This can be seen in the right-hand panel of Figure~\ref{figure_void_density_profile_results}, which shows the density contrast of a $300 \, \rm{cMpc}$ sphere for the best-fitting MOND model (the blue solid line) and two Newtonian models (the red lines, with shaded grey region between them) over cosmic time. The solid red and blue lines correspond to the same initial conditions, but end up with very different $\delta$ at the present time. As expected, Newtonian models with different initial $\delta$ show a similar pattern of structure growth since they are all in the linear regime. The density contrast scales as $\delta ~\appropto~ a^{3.8}$ and $\delta ~\appropto~ a^{0.8}$ over the interval $0.3 \leq a \leq 0.7$ in our best-fitting MONDian model and equivalent Newtonian model, respectively. The $\Lambda$CDM scaling is slightly $<1$ because dark energy slows down the growth of structure at late times. The very rapid structure growth in our MOND model can be reduced by applying a higher EFE in the past. In the case of a time-dependent EFE with $n_{\mathrm{EFE}} = -1$ in Equation~\ref{eq:time_dependent_external_field_history}, the growth rate reduces to $\delta ~\appropto~ a^{3.3}$. The initial void strength must then be $\approx 13\times$ larger ($\alpha_{\mathrm{void}} = 4.98\times 10^{-4}$) to compensate for the higher EFE in the past (the blue dot-dashed line in the right-hand panel of Figure~\ref{figure_void_density_profile_results}). Models with a time-dependent EFE are discussed further in Section~\ref{subsubsec:Structure formation in MOND}.

\subsubsection{Hubble diagram} \label{subsubsec:Hubble diagram}

\begin{figure}
    \includegraphics[width=\linewidth]{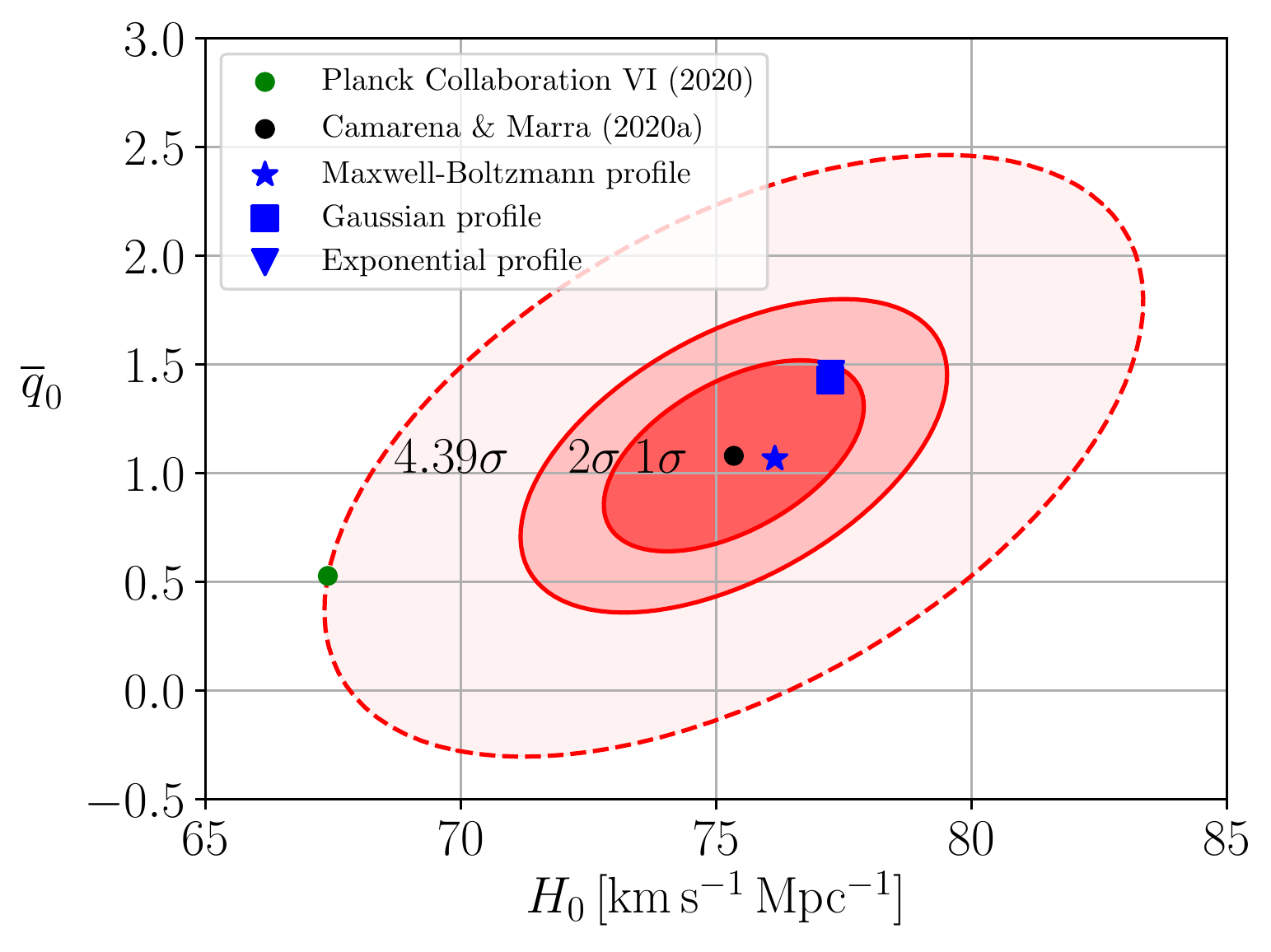}
    \caption{Combined inference on $H_{0}$ and $\overline{q}_{_0}$ (the black dot and red error ellipses) derived jointly from Pantheon SNe in the redshift range $0.023 \leq z \leq 0.15$ \citep{Camarena_2020}. The blue points show the results for the best-fitting MOND models with a Maxwell-Boltzmann (star), Gaussian (square), and exponential (triangle) void profiles. All three models are consistent with the observations at the $1\sigma$ confidence level. $H_{0}$ and $\overline{q}_{_0}$ derived from a Gaussian and an exponential void profile are in both models almost the same and cannot be distinguished in the plot (the triangle and the square coincide). Note that the indicated tension with Planck results (the green dot) differs from the $4.54 \sigma$ reported by \citet{Camarena_2020} because we have not included the correlation coefficient between $H_0$ and $\overline{q}_0$ for Planck.}
    \label{figure_H0_q0_plane}
\end{figure}

Our fiducial model yields $H_{0}^{\mathrm{model}} = 76.15 \, \rm{km\,s^{-1}\,Mpc^{-1}}$ and $\overline{q}_{_0}^{\mathrm{model}} = 1.07$. This is consistent with the observations of \citet{Camarena_2020} at the $84.20\%$ confidence level (only $0.20 \sigma$ tension). The combined inference on both parameters is shown in Figure~\ref{figure_H0_q0_plane}, which demonstrates that the best-fitting models with Maxwell-Boltzmann, Gaussian, and exponential initial profiles are all consistent with these observations within $1\sigma$. Thus, we show for the first time that the Hubble tension can be resolved in MOND. Note that the Planck parameters (the green dot) are in $\approx 4.39 \sigma$ tension with these local observations.

\begin{figure}
    \includegraphics[width=\linewidth]{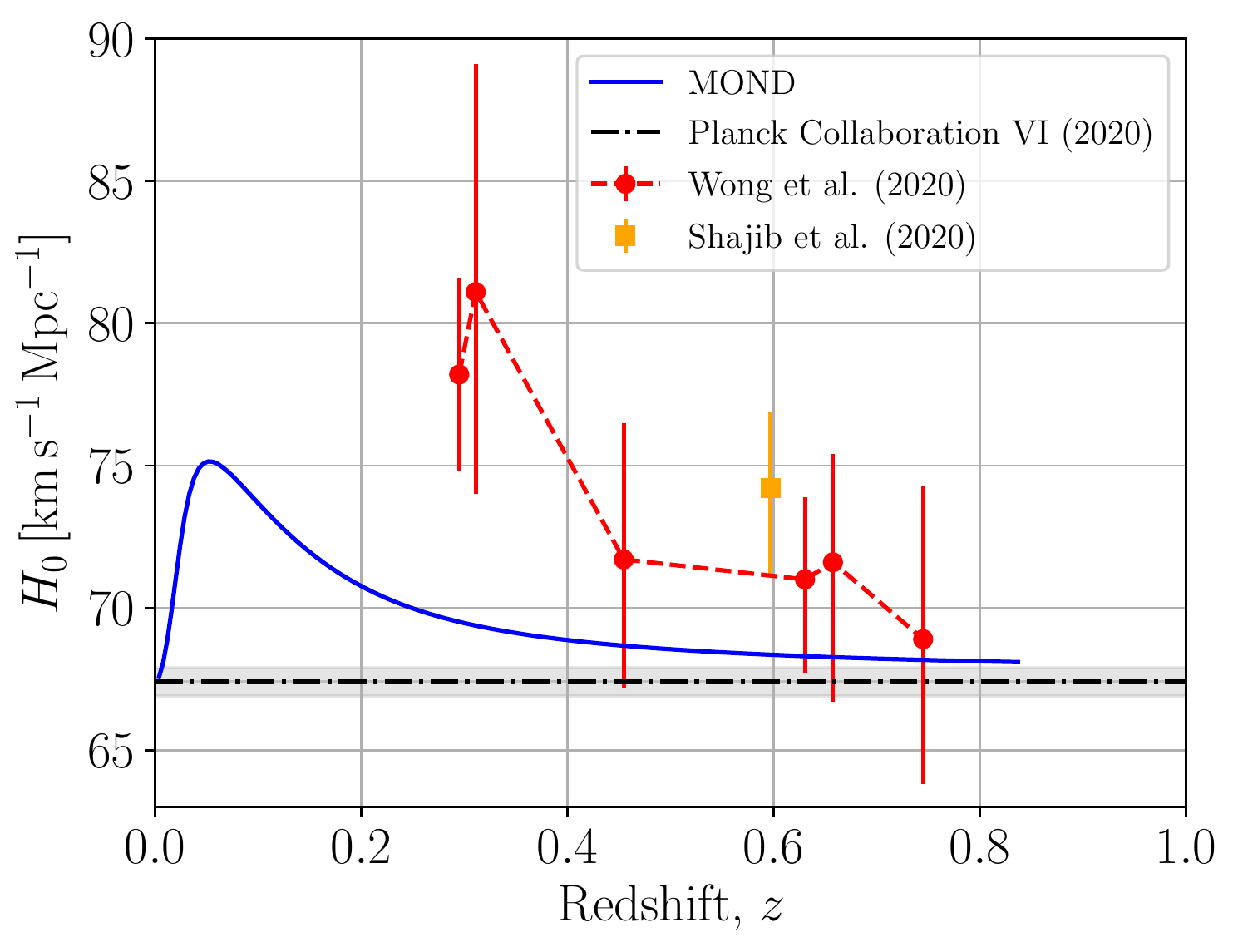}
    \caption{Hubble constant in dependence of redshift for the fiducial MOND model. The red data points are measurements of $H_{0}$ from lensed quasars by \citet{Wong_2020} and \citet{Shajib_2020} assuming a flat $\Lambda$CDM cosmology \citep[see also our Table~\ref{tab:observed_H0_strong_lensing} and figure~A1 in][]{Wong_2020}. The black horizontal dot-dashed line and its surrounding grey band marks $H_{0}^{\mathrm{global}} = 67.4 \pm 0.5 \, \rm{km\,s^{-1}\,Mpc^{-1}}$ \citep{Planck_2018}.}
    \label{figure:best_fitting_model_Hubble_vs_redshift}
\end{figure}

Time delays from strong gravitational lenses also provide an important constraint on our model. We use Figure~\ref{figure:best_fitting_model_Hubble_vs_redshift} to show $H_{0,\mathrm{model}}^{\mathrm{lensing}}$ in dependence of redshift, allowing a comparison with measurements from seven lens systems (Section~\ref{subsubsec:Hubble constant from strong lensing}). Interestingly, our model systematically underestimates $H_{0}$ especially at low redshifts, causing a $2.05 \sigma$ tension with the observations of \citet{Wong_2020} and \citet{Shajib_2020}. We expect that this discrepancy is partly caused by void motion due to the EFE, though there is also some internal inconsistency between the void profile of \citet{Keenan_2013} and the lensing Hubble data. In general, the latter are difficult to produce in any void model if the background $H_{0}^{\mathrm{global}} = 67.4 \, \rm{km\,s^{-1}\,Mpc^{-1}}$, since it is difficult to imagine a local void having substantial effects at $z = 0.5$. We discuss these issues in more detail in Section~\ref{subsec:Assessing the tension for MOND}. Although it is expected that strong lensing in MOND occurs similarly to standard cosmology (Section~\ref{subsubsec:Hubble constant from strong lensing}), we redo our analysis without constraints from lensing-based $H_0$ measurements in Section~\ref{subsubsec:Excluding strong lensing time delays}.

\subsubsection{LG peculiar velocity}
\label{subsubsec:LG_vpec}

\begin{figure}
    \includegraphics[width=\linewidth]{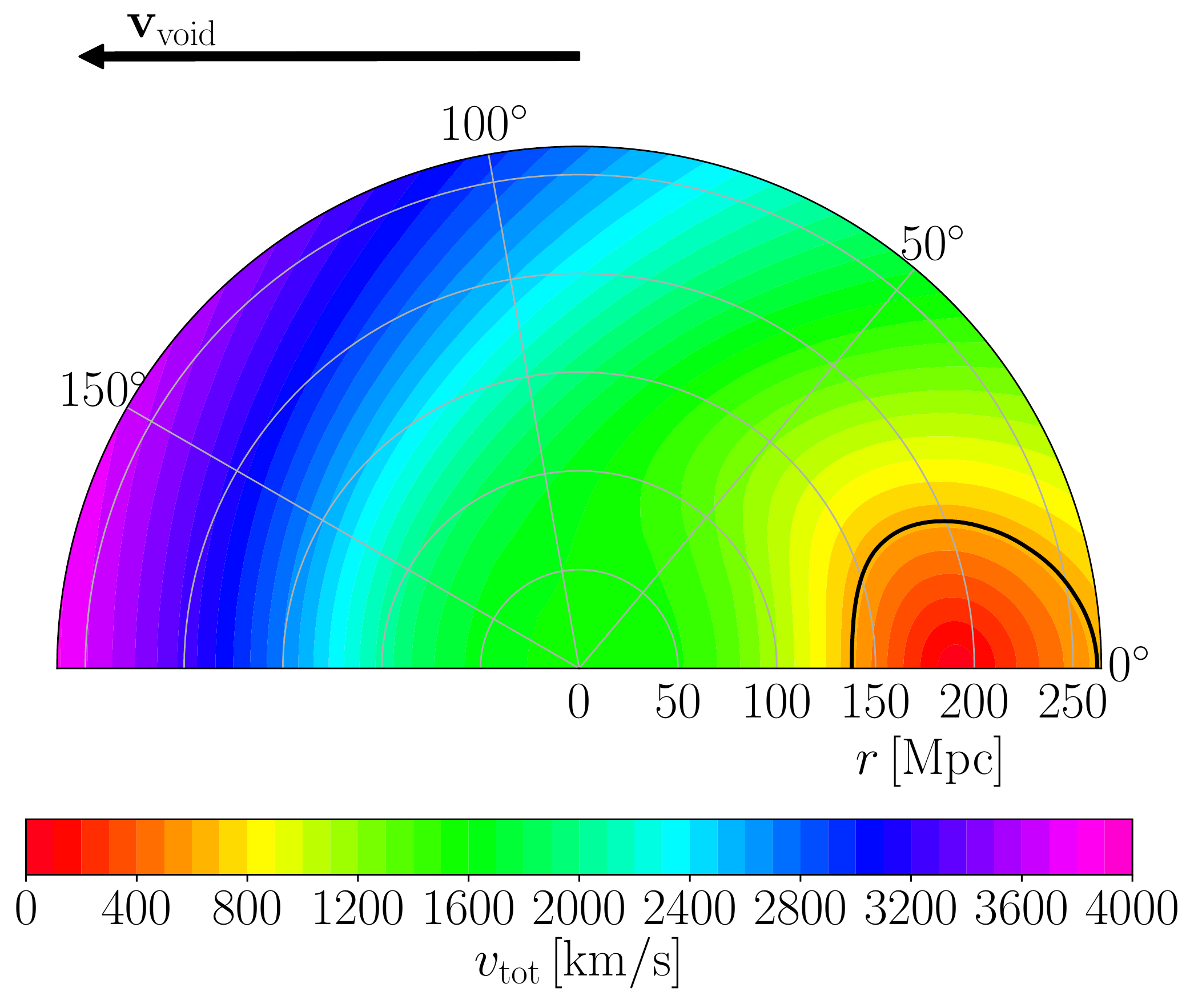}
    \caption{Total peculiar velocity (Equation~\ref{eq:total_peculiar_velocity}) map wrt. the CMB for the fiducial MOND model. The black arrow indicates the direction of the void motion $v_{\mathrm{void}}$. The black solid curve marks the region within which $v_{\mathrm{tot}} \leq 627 \, \rm{km\,s^{-1}}$. The LG is probably near the right end of this curve because the observed radio dipole (Section~\ref{subsec:KBC void}) indicates that we are currently moving away from the void centre in the CMB frame. We show only half of the velocity map because $v_{\mathrm{tot}}$ is axisymmetric about $v_{\mathrm{void}}$. The here shown total peculiar velocities are $v_{\mathrm{tot}} \la 0.01 c$, justifying the use of non-relativistic equations for the void gravitational field (Section~\ref{subsec:Governing equations}).}
    \label{figure_peculiar_velocity_results}
\end{figure}

Our model yields the total peculiar velocity wrt. the CMB in different parts of the void, as mapped in Figure~\ref{figure_peculiar_velocity_results}. The entire void moves in the direction indicated by the arrow, which arises from the EFE (Section~\ref{subsubsec:External field history}). Interestingly, the model allows for very high total and internal peculiar velocities, especially towards the void edge. We can also get partial or total cancellation between internal motions within the void and that of the void as a whole, creating a rather large region in which $v_{\mathrm{tot}} \leq v_{\mathrm{LG}} = 627 \, \rm{km\,s^{-1}}$ \citep{Kogut_1993}. This region is at a distance of $\approx \left( 150 - 270 \right) \, \rm{Mpc}$ from the void centre, implying that the LG must be slightly off-centre. Applying Equation~\ref{eq:P_value_motion_void} to find the fraction this region represents of the whole void, we estimate that the observed $v_{\mathrm{LG}}$ represents a $2.34 \sigma$ outlier to the simulated $v_{\mathrm{tot}}$ distribution, which causes therewith the highest tension amongst the here used observational constraints.

\begin{table*}
    \caption{Comparison of individual local Universe observables with our fiducial MOND model (Maxwell-Boltzmann profile, $g_{\mathrm{ext}} = 0.055 \, a_{_0}, r_{\mathrm{void}} = 228.2 \, \rm{cMpc}$, $\alpha_{\mathrm{void}} = 3.76 \times 10^{-5}$, $v_{\mathrm{void}} = 1586 \, \rm{km\,s^{-1}}$, $r_{\mathrm{void}}^{\mathrm{rms}} = 528.7 \, \rm{Mpc}$, $n_{\mathrm{EFE}} = 0$). The last row shows the probability of a higher $\chi^2$ given the number of degrees of freedom. We express this as the equivalent number of standard deviations for a 1D Gaussian (inverting Equation~\ref{eq:f_chi_to_P}). For $v_{\mathrm{LG}}$, we show the proportion of the void volume where $v_{\mathrm{tot}} \leq v_{\mathrm{LG}}$. Results for other void profiles are shown in Appendix~\ref{Appendix:Joint probabilities for different void profiles}.}
    \label{table:comparison_bestfittingmodel_with_observations}
    \begin{tabular}{lllllll} \hline 
    Parameter & $H_{0}^{\mathrm{local}} \, [\rm{km\,s^{-1}\,Mpc^{-1}}]$ & $\overline{q}_{_0}^{\mathrm{local}}$  & $H_{0}^{\mathrm{lensing}} \, [\rm{km\,s^{-1}\,Mpc^{-1}}]$ & $v_{\mathrm{LG}} \, [\rm{km\,s^{-1}}]$ & $\delta_{\mathrm{in}}$  & $\delta_{\mathrm{out}}$ \\ \hline 
    Observations & $75.35 \pm 1.68$ & $1.08 \pm 0.29$ & $--$ &  $627$ & $0.254 \pm 0.083$ & $-0.052 \pm 0.105$ \\  
MOND model & $76.15$ & $1.07$ & See Figure~\ref{figure:best_fitting_model_Hubble_vs_redshift} & See Figure~\ref{figure_peculiar_velocity_results} & $0.172$ & $0.050$  \\
$\chi^{2}$ & \multicolumn{2}{c}{$0.34$} & $14.66$ &$--$ & $0.99$ & $0.94$ \\
Degrees of freedom & \multicolumn{2}{c}{$2$}  & $7$ & $--$ & $1$ & $1$ \\
$\chi$ (1D Gaussian equivalent) & \multicolumn{2}{c}{$0.20$}  & $2.05$ &  $2.34$ &  $0.99$ & $0.97$ \\ \hline
    \end{tabular}
\end{table*}

\subsubsection{Overall agreement with observations}

Finally, we quantify the combined tension of our fiducial MOND model with local observations. As discussed earlier, most observables can be treated using a standard $\chi^2$ approach, but additional care is needed for $v_{\mathrm{LG}}$. Thus, we quantify the likelihood of different $\left(\chi^2, v_{\mathrm{tot}} \right)$ combinations according to our fiducial model, with $\chi^2$ found using Equation~\ref{eq:P_value_total}. We can then quantify the extent to which the actually observed combination is unlikely.

In our model universe, the joint probability that the observables can be summarized by some $\left(\chi^2, v_{\mathrm{tot}} \right)$ combination is
\begin{eqnarray}
    P \left( \text{Observations} | \text{Best-fitting model} \right) = P \left( \chi^{2} \right) \cdot P_{\mathrm{motion}} \left( v_{\mathrm{tot}} \right) \, , 
    \label{eq:P_value_total_bestfit}
\end{eqnarray}
where $P(\chi^{2})$ is the probability density function for a $\chi^{2}$ distribution with $8$ degrees of freedom (i.e. $11$ observational constraints and three model parameters). $P_{\mathrm{motion}}\left( v_{\mathrm{tot}} \right)$ is estimated from our simulation by splitting the volume into cells (Section~\ref{subsubsec:Motion of the void}) and assigning the volume in each cell to different bins in $v_{\mathrm{tot}}$, thereby building up a discretized picture of its distribution. This procedure does not assume that $v_{\mathrm{tot}}$ follows a Gaussian.

If our fiducial model is correct, $\chi^2$ must arise solely from measurement errors, while the observed $v_{\mathrm{LG}}$ reflects our position within the void. This causes that $\chi^2$ and $v_{\mathrm{tot}}$ have independent distributions, allowing them to be multiplied. We neglect the $22 \, \rm{km\,s^{-1}}$ uncertainty in $v_{\mathrm{LG}}$ \citep{Kogut_1993} because this is much smaller than the $\approx 4000 \, \rm{km\,s^{-1}}$ range in $v_{\mathrm{tot}}$ allowed by our model (Figure~\ref{figure_peculiar_velocity_results}).

\begin{figure}
	\includegraphics[width=\linewidth]{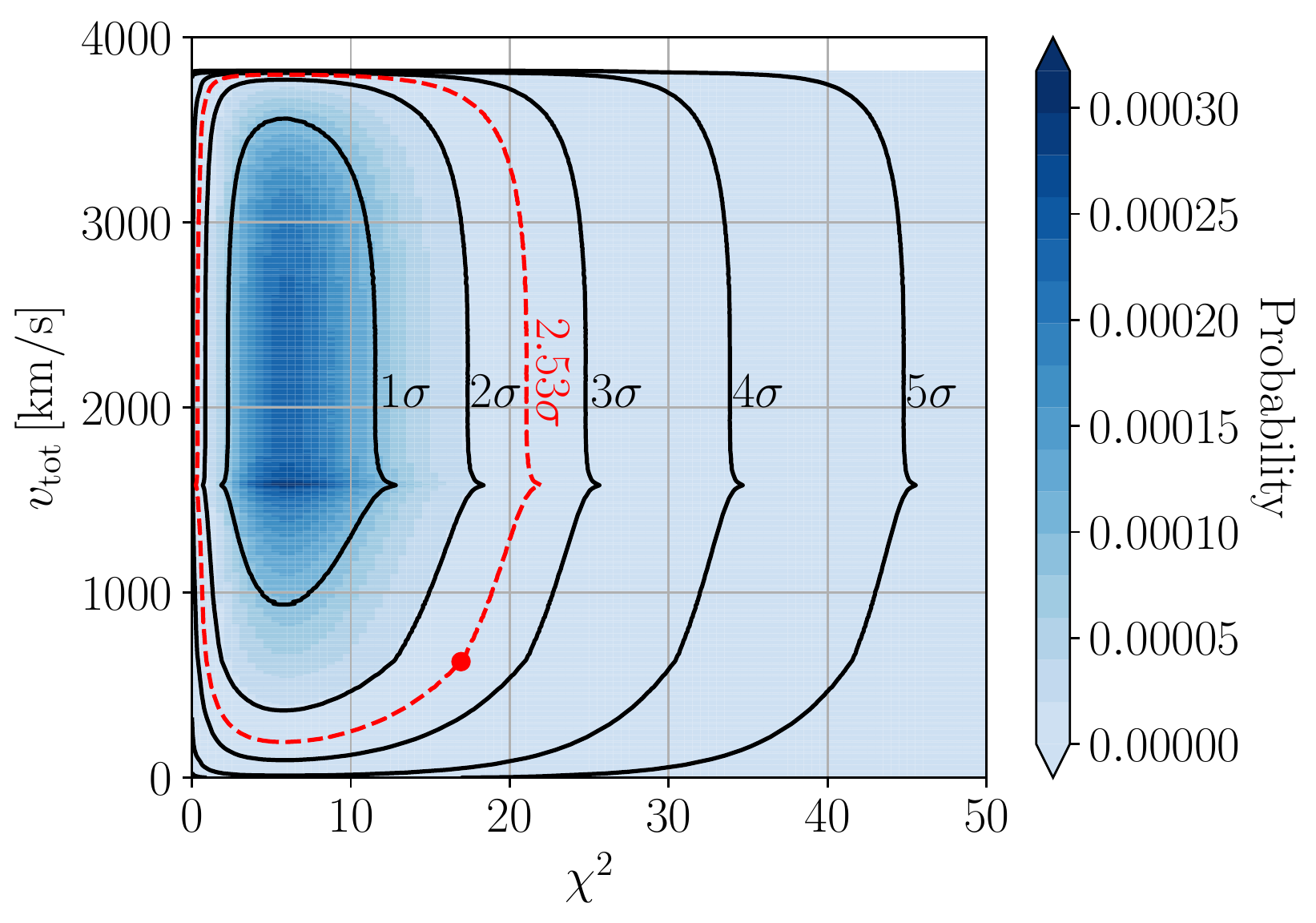}
	\caption{Joint probability of $\chi^{2}$ and $v_{\mathrm{tot}}$, the total peculiar velocity wrt. the CMB, in the fiducial MOND model obtained in Section~\ref{subsec:Joint probabilities}. The black contours show the indicated confidence levels. The red dot marks the total $\chi^{2}$ of the fiducial model and $v_{\mathrm{tot}} = v_{\mathrm{LG}} = 627 \, \rm{km\,s^{-1}}$. This is consistent with the model at the $1.14\%$ confidence level (the red dashed line), representing a $2.53 \sigma$ tension. The sharp feature in each contour occurs when $v_{\mathrm{tot}} = v_{\mathrm{void}} = 1586 \, \rm{km\,s^{-1}}$ (Section~\ref{subsubsec:Motion of the void}).}
	\label{figure_tension_with_observations}
\end{figure}

We use Figure~\ref{figure_tension_with_observations} to show the joint $\left(\chi^2, v_{\mathrm{tot}} \right)$ distribution based on our fiducial model. This explains the local observations at the $1.14\%$ confidence level ($2.53 \sigma$ tension). Individual observational constraints are summarized and compared with observations in Table~\ref{table:comparison_bestfittingmodel_with_observations}.

\begin{figure}
	\includegraphics[width=\linewidth]{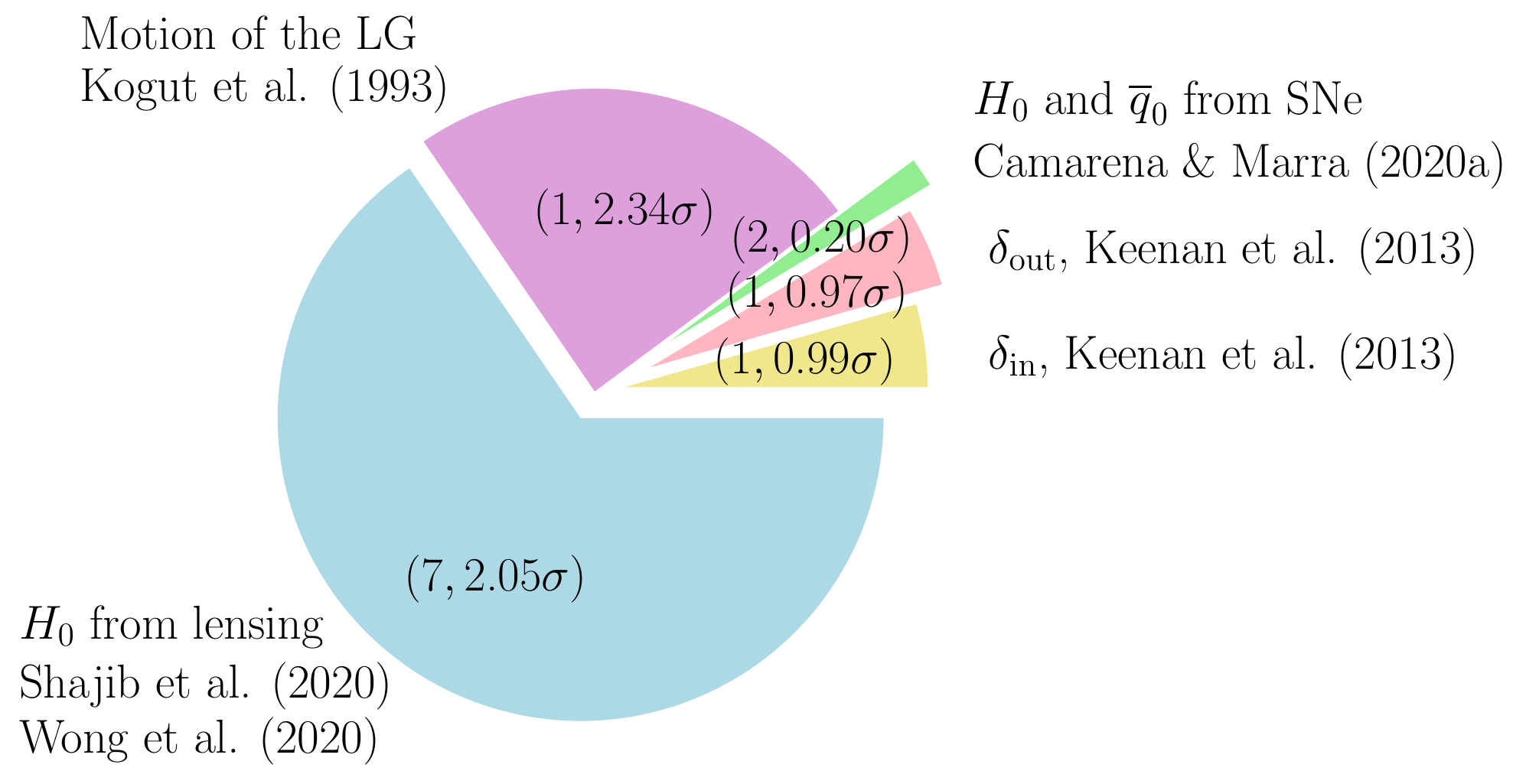}
	\caption{Pie chart showing the $\chi^{2}$ contributions from different constraints (Table~\ref{table:comparison_bestfittingmodel_with_observations}). The bracketed numbers show the number of degrees of freedom and corresponding level of tension for each constraint. The value for the motion of the LG is estimated based on the fraction of the void volume for which $v_{\mathrm{tot}} \leq v_{\mathrm{LG}} = 627 \, \rm{km\,s^{-1}}$.}
	\label{figure_Pichart_bestfitting}
\end{figure}

The $\chi^{2}$ contributions from different constraints are visualized in Figure~\ref{figure_Pichart_bestfitting} as a pie chart, which also shows the number of data points each constraint represents, and the corresponding level of tension. To facilitate a comparison with the other constraints, we use our previous estimate that $v_{\mathrm{LG}}$ is a $2.34 \sigma$ outlier to the simulated $v_{\mathrm{tot}}$ distribution (Section~\ref{subsubsec:LG_vpec}). Therefore, we assign a $\chi^2$ contribution of $2.34^2$ to this constraint.

The best-fitting MOND models with a Gaussian and an exponential void profile agree with the local observations at the $0.45\%$ ($2.84\sigma$) and $0.34\%$ ($2.93 \sigma$) confidence level, respectively (see also Appendix~\ref{Appendix:Joint probabilities for different void profiles}). Thus, our best-fitting void model with a constant EFE cannot be rejected regardless of the initial density profile. The implications of our results are discussed in Section~\ref{subsec:Assessing the tension for MOND}, which also looks at the overall picture of the $\nu$HDM model and the theoretical uncertainties of the here applied MOND approach (Section~\ref{subsubsec:Theoretical_assumptions}).

\section{Discussion} \label{sec:Discussion}

We discuss what our results in Sections~\ref{sec:LCDM framework} and \ref{sec:Results of MOND simulation} imply for $\Lambda$CDM and MOND cosmologies. This is followed by a consideration of commonly proposed arguments claiming that a local underdensity cannot solve the Hubble tension (Section \ref{subsec:Claimed_problems}).

\subsection{Assessing the tension for \texorpdfstring{$\Lambda$CDM}{LCDM}} \label{subsec:Assessing the tension for LCDM}

\citet{Keenan_2013} measured the $K$-band luminosity density as a function of redshift and found evidence for an underdensity around the LG with a radial extent of $\approx 300 \, \rm{Mpc}$ (see their figs.~9 and 10). They used the 2M++ catalogue \citep{Lavaux_2011}, which covers $\approx 90\%$ of the sky based on photometric data from the 2MASS-XSC catalogue and redshift data from the 2MRS, 6DFGRS, and SDSS catalogues. From the $K_{s} < 13.36$ luminosity density, \citet{Keenan_2013} estimated a relative density contrast of $\delta \equiv 1 - \rho/\rho_{0} \approx 0.5$ in the redshift range $0.0025 < z < 0.067$ (pink down-pointing triangle in their figure~11). Probing the luminosity density slightly deeper ($K_{s} < 14.36$) but only in the SDSS and 6DFGRS regions, they derived a slightly lower density contrast of $\delta = 0.46 \pm 0.06$ in the redshift range $0.01 < z < 0.07$ \citep[the light blue dot in figure~11 of][]{Keenan_2013}. We used this value for the inner density contrast of the KBC void in order to minimize tension with the $\Lambda$CDM framework.

Using the MXXL simulation \citep{Angulo_2012}, we calculated the density contrast for spheres with an outer radius of $300 \, \rm{Mpc}$ and an inner hole of radius $40 \, \rm{Mpc}$ around $10^{6}$ vantage points at $z = 0$. We also took into account the sky coverage of the 2M++ survey by generating at each vantage point a random observing direction from which $90\%$ of the mock sky is covered (Section~\ref{subsec:MXXL simulation}). The so-selected density fluctuations have an rms amplitude of $3.2\%$. This is consistent with the prediction of the Harrison-Zeldovich spectrum \citep{Harrison_1970, Zeldovich_1972} in combination with the early universe normalisation of $\sigma_{8} = 0.811 \pm 0.006$ \citep{Planck_2018}. Since \citet{Keenan_2013} used a fixed distance-redshift relation (see their section~4.7), we applied an RSD correction to the simulated density fluctuations. The rms fluctuation then became $4.8\%$, with the individual values closely following a Gaussian of this width (Appendix~\ref{Appendix_normality_tests}). Thus, the observational uncertainty of 6\% is larger than the expected cosmic variance.

Based on our analysis, the observed KBC void is in $6.04\sigma$ tension with standard $\Lambda$CDM cosmology and cannot be explained with cosmic variance (Section~\ref{KBC void in LCDM}). 
This contrasts with \citet{Sahlen_2016}, who concluded that supervoids such as the KBC void are consistent with standard theory. However, for this they used a top-hat galaxy density radius of $210 h^{-1} \, \rm{Mpc}$ and a DM density contrast of $\delta = 0.15-0.2$. Their assumed $\delta$ describes a much less pronounced void than the observed density contrast derived from the 2M++ survey \citep{Keenan_2013}.\footnote{In $\Lambda$CDM, the RSD effect implies $\delta \approx \frac{2}{3}\delta_{\mathrm{obs}}$ (Equation~\ref{eq:density_RSDcorrection}). Thus, $\delta_{\mathrm{obs}} = 0.46$ does not correspond to a true underdensity of $\delta = 0.2$.} Moreover, even a 15\% true underdensity on a 300 Mpc scale is very difficult to reconcile with $\Lambda$CDM due to the expected variance being only 3.2\% (Section~\ref{KBC void in LCDM}). While it may be possible for such large voids to exist somewhere in the Universe, it would be unlikely for us to live inside one $-$ unless they are more common.

\subsubsection{Hubble tension} \label{subsubsec:Hubble tension discussion}

Several studies have already discussed a potential connection between the local void and the Hubble tension \citep[e.g.][]{Keenan_2013, Enea_2018, Shanks_2019b, Kenworthy_2019}. Indeed, if mass conservation is assumed, a large underdensity in the local number density of galaxies should also show up in the velocity field. However, given the expected cosmic variance in $\Lambda$CDM, Figure~\ref{figure_LCDM_results_cosmic_variance} indicates that a $\approx 10 \sigma$ density fluctuation would be necessary to explain the locally observed expansion rate within its $2 \sigma$ confidence region \citep[$H_{0}^{\mathrm{local}} = 73.8 \pm 1.1 \, \rm{km\,s^{-1}\,Mpc^{-1}}$,][]{Riess_2019, Wong_2020}. The maximum plausible $5 \sigma$ density fluctuation is still not enough to explain $H_{0}^{\mathrm{local}}$ at the $5 \sigma$ level.

These findings are broadly consistent with \citet{Wu_2017}, who concluded that a void with $\delta = 0.8$ and a radius of $120 h^{-1} \, \rm{Mpc}$ could resolve the Hubble tension. Such a void would be in $\approx 20 \sigma$ tension with $\Lambda$CDM \citep{Kenworthy_2019}. While this by itself does not constitute an argument against such a large local underdensity, \citet{Wu_2017} stated that observations disfavour it. Section~\ref{subsec:Claimed_problems} of our work contains a more detailed discussion of claimed problems with a local void solution to the Hubble tension.

Combining the mutually consistent SH0ES and H0LiCOW results, \citet{Wong_2020} showed that the Hubble tension has now reached the $5.3\sigma$ level. Thus, both the KBC void and the Hubble tension falsify $\Lambda$CDM at $>5 \sigma$ significance. The most likely explanation in the context of standard theory is that both are caused purely by measurement errors, since there is less cosmic variance than observational uncertainty in both the KBC data and the $H_0$ measurement. This is especially true for the latter $-$ Equation~\ref{eq:Hubbe_RSDcorrection} shows that density fluctuations of 3.2\% would impact $H_{0}^{\mathrm{local}}$ by only $0.4 \, \rm{km\,s^{-1}\,Mpc^{-1}}$ \citep[see also][]{Wojtak_2014}. Since galaxy counting and measurements of the local Hubble constant face rather different observational challenges, the measurement errors would be independent, yielding a combined tension of $7.75 \sigma$. Using a more rigorous estimation that allows for cosmic variance results in a $7.09 \sigma$ falsification of the $\Lambda$CDM paradigm (Section~\ref{Combined implications for LCDM}).

Importantly, \emph{any} cosmological model which solves the Hubble tension without addressing the void (or vice versa) would violate the assumption of mass conservation in the Universe. Thus, early dark energy models \citep[e.g.][]{Hill_2020, Khoraminezhad_2020} which simply increase $H_{0}$ at the background level by $\approx 10 \%$ would overestimate the local $H_0$ by about this much once the observed KBC void is taken into account (Section~\ref{subsec:KBC void}). Perhaps the most important implication of the KBC void is that the Planck value of $H_0$ is probably correct at the background level, since any attempt to substantially change it would likely cause the void-corrected local value to disagree with local observations if mass is conserved in the Universe.

\begin{figure}
    \includegraphics[width=\linewidth]{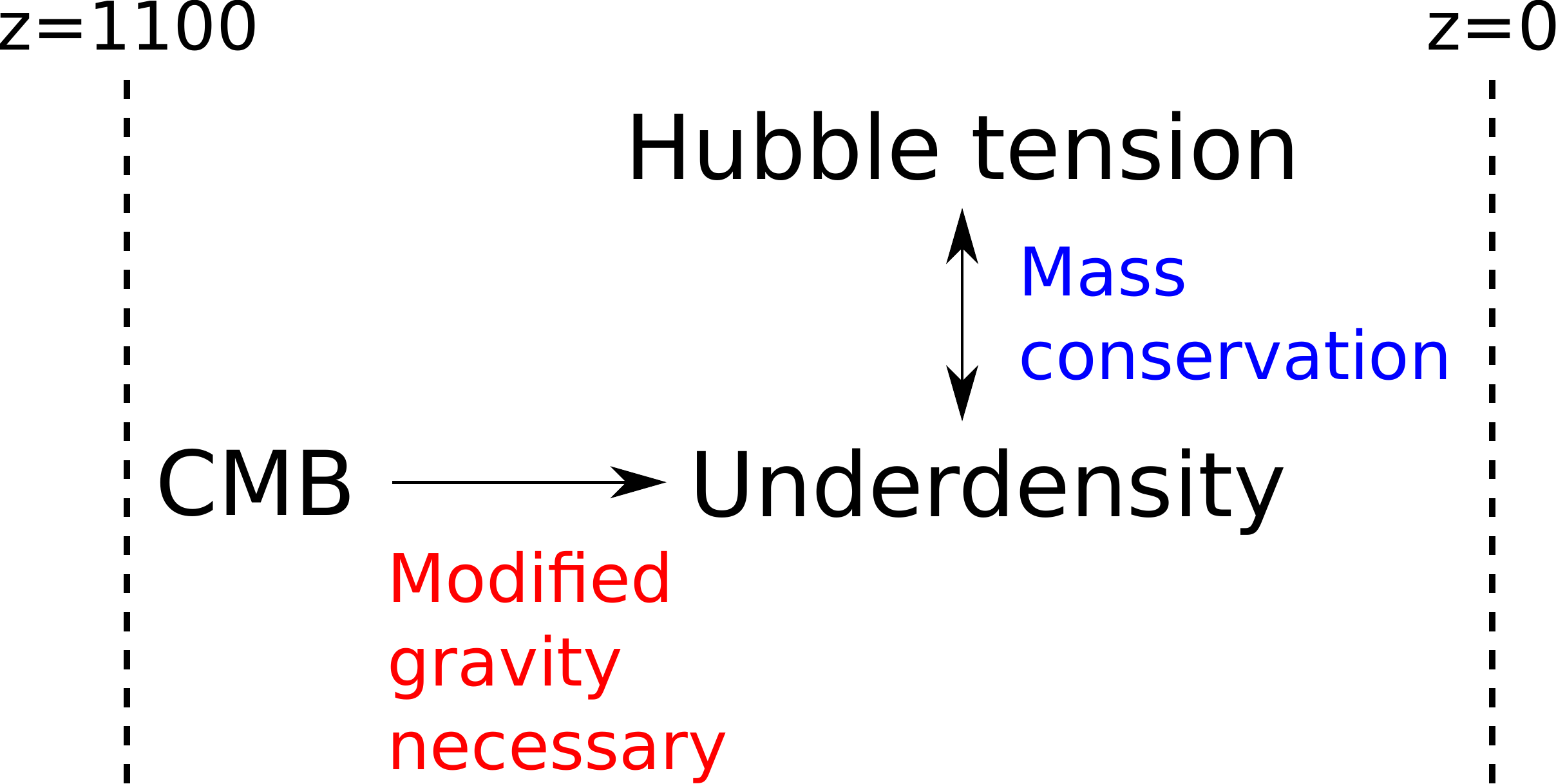}
    \caption{The KBC void in context. This large local underdensity and the Hubble tension can both be reconciled if mass is conserved in the Universe. However, such a large void cannot form out of the initial conditions of the CMB if the initial scale-invariant power spectrum \citep{Harrison_1970, Zeldovich_1972} is preserved. In particular, the existence of the KBC void within the $\Lambda$CDM framework is ruled out at $6.04 \sigma$, as demonstrated in Section~\ref{subsec:Comparison with observations}. Thus, modified gravity is required to explain low redshift observables (Section~\ref{subsec:Results The best-fitting model}) if the initial conditions at $z = 1100$ are indeed set by the CMB (Section \ref{subsubsec:CMB contamination by intergalactic dust}). Importantly, \emph{any} solution must simultaneously solve both the Hubble and underdensity tensions in order to conserve mass.}
    \label{figure_KKBC_void_context}
\end{figure}

Therefore, the KBC void is a plausible explanation for the Hubble tension if we can preserve a Planck background cosmology but enhance the cosmic variance. A schematic that considers the KBC void in a broader context is presented in Figure~\ref{figure_KKBC_void_context}. Starting from the initial conditions of the CMB at $z = 1100$ and assuming the scale-invariant Harrison-Zeldovich power spectrum to be valid at that time, we have shown that the existence of a KBC-like void at present is virtually impossible in a standard context. This indicates that a scale-invariant power spectrum is violated today $-$ order unity fluctuations on a $10~\,\rm{Mpc}$ scale \citep[e.g.][]{Mantz_2015} do not give way to 3\% fluctuations on a 300 Mpc scale. \emph{This is a very strong hint that the gravitational inverse square law has to break down.} Thus, the spatial distribution of matter on both an $8 \, \rm{Mpc}$ scale \citep{Peebles_2010} and on an $\approx 1 \, \rm{Gpc}$ scale (this work) suggest a long-range enhancement to gravity.

\subsection{Assessing the tension for MOND} \label{subsec:Assessing the tension for MOND}

MOND \citep{Milgrom_1983} is a low-acceleration modification to gravity originally designed to explain galaxy rotation curves without CDM. It has enjoyed a great deal of predictive success in this regard (Section~\ref{subsec: Milgromian dynamics}). Therefore, extrapolating MOND from kpc to Gpc scales could be a promising way to address large-scale challenges for standard cosmology such as massive high-redshift galaxy clusters \citep[e.g. El Gordo,][]{Katz_2013} and supervoids.

In this context, we study the possible origin of the KBC void from small initial density fluctuations, and its impact on the local Hubble constant. Unfortunately, we do not presently have a large enough cosmological $N$-body or hydrodynamical MOND simulation to quantify the likelihood of the KBC void, as done for the $\Lambda$CDM framework with the MXXL simulation (Section~\ref{sec:LCDM framework}).

We therefore used the $\nu$HDM framework (Section~\ref{subsec:nuHDM cosmological model}) to develop a semi-analytical MOND simulation in which the expansion history is a standard flat background cosmology consistent with the latest Planck data \citep{Planck_2018} $-$ which should be a good approximation also in MOND (Section \ref{subsubsec:Background cosmology}). We applied MOND only to the density deviations from the cosmic mean (Section \ref{subsubsec:Large-scale structure}). In Section \ref{subsubsec:Theoretical_assumptions}, we discuss the possibility of a non-trivial coupling between density perturbations and the background.

Our main MOND models assume an initial Maxwell-Boltzmann density profile motivated by the radial density distribution of the Local Volume. \citet{Karachentsev_2018} showed that the matter density within a sphere of $r = 40 \, \rm{Mpc}$ ($r = 135 \, \rm{Mpc}$) around the LG is only $\Omega_{\mathrm{m,loc}} = 0.09 - 0.14$ ($\Omega_{\mathrm{m,loc}} = 0.05 - 0.16$). This is $\approx {2-3\times}$ lower than the cosmic mean density measured by \citet{Planck_2018}, confirming the existence of a large local underdensity (Section~\ref{subsec:KBC void}). \citet{Karachentsev_2018} also showed that the density increases inwards for heliocentric distances $\la 40 \, \rm{Mpc}$ (see their figure~3), justifying our choice of a Maxwell-Boltzmann void profile (Section~\ref{subsubsec:Void initial profiles}). In addition, we also run void models initialized with a Gaussian and an exponential profile (Appendix~\ref{Appendix: KBC void mass profile}). In all cases, the void profiles are parametrized by an initial void size and strength at $z = 9$. The initial void strengths range from $\alpha_{\mathrm{void}} = 10^{-5} - 10^{-2}$, with the lower limit based on the observed density fluctuations in the CMB. By the time that $z = 9$, we expect significantly larger perturbations. Our upper limit on $\alpha_{\mathrm{void}}$ is sufficient to capture the range of models preferred by our analysis (Section~\ref{subsubsec:Structure formation in MOND}).

The EFE is strongly constrained in our models because it affects the formation of cosmic structure and thus internal velocities within the void, in addition to the void's motion as a whole (Section~\ref{subsubsec:Motion of the void}). Models with a very small EFE create extremely deep and extended voids, which disagrees with the density contrast of the KBC void $-$ especially for its outer region. This also results in a much larger local Hubble constant than observed. Increasing the EFE leads to ${v_{\mathrm{void}} \gg v_{\mathrm{LG}}}$, so the observed $v_{\mathrm{LG}}$ can only be explained by nearly complete cancellation with a large $v_{\mathrm{int}}$. However, a strong EFE makes the MONDian system more Newtonian and suppresses therewith the growth of structure. Consequently, models with a very high EFE produce very shallow voids and a local Hubble constant similar to its global value, causing that $v_{\mathrm{int}}$ is not large enough to cancel $v_{\mathrm{void}}$.

Our analysis for the Maxwell-Boltzmann profile restricts $g_{\mathrm{ext}}$ to the range $\left( 0.030 - 0.053 \right) a_{_0}$ at the $1 \sigma$ confidence level. Models with a Gaussian and an exponential void profile prefer a slightly larger EFE, i.e. $\left( 0.054 - 0.094  \right) a_{_0}$ (Figure~\ref{figure_MOND_GAUSSIAN_results}) and $\left( 0.054 - 0.092 \right) a_{_0}$ (Figure~\ref{figure_MOND_EXP_results}) at the $1 \sigma$ confidence level, respectively. This is because the Maxwell-Boltzmann profile reduces $\delta$ near the void centre and therewith slows down the internal peculiar velocities of individual particles within the void. Thus, a lower EFE is required to achieve $v_{\mathrm{tot}} \leq v_{\mathrm{LG}} = 627 \, \rm{km\,s^{-1}}$ over a large part of the void.

Our analysis rules out models without an EFE, which is in any case a logical consequence of the non-linearity inherent to Milgrom's law of gravity \citep{Milgrom_1986}. Observationally, MOND without the EFE is strongly disfavoured by the velocity distribution of wide binary stars in the Solar neighbourhood \citep{Pittordis_2019}. The EFE is also necessary to explain the internal velocity dispersions of dwarf galaxies (Section~\ref{subsec: Milgromian dynamics}). We discuss the time-dependence of the EFE in more detail in Section~\ref{subsubsec:Structure formation in MOND}.

In contrast to the EFE, Figure~\ref{figure:marginalization_MBprofile} indicates that the initial void size and strength are not strongly constrained by observations. Thus, other initial void parameters could in principle also yield reasonable results at the present time. In particular, our analysis of Maxwell-Boltzmann voids yields ${1\sigma}$ confidence intervals on $r_{\mathrm{void}}$ and $\alpha_{\mathrm{void}}$ of $\left(173.9 - 636.9\right) \, \rm{cMpc}$ and $\left(1.07 - 8.12 \right) \times 10^{-5}$, respectively. Models with Gaussian or exponential initial profiles allow for larger voids, but with a similar void strength (Appendix~\ref{Appendix:Joint probabilities for different void profiles}). There are two main reasons why both void parameters are only weakly constrained by local observations. First of all, $H_{0}$ and $\overline{q}_{_0}$ are derived from data in the redshift range $0.023 \leq z \leq 0.15$ and constrain therewith only the inner and not the outer part of the void. Secondly, the uncertainties of $H_{0}$ measured from strong lens systems are relatively large, which allows for a wide range of possible void behaviours in the outskirts.

We found that our best-fitting Maxwell-Boltzmann MOND model requires an EFE of $g_{\mathrm{ext}} = 0.055 \, a_{_0}$, an initial void size of $r_{\mathrm{void}} = 228.2 \, \rm{cMpc}$, and an initial void strength of $\alpha_{\mathrm{void}} = 3.76 \times 10^{-5}$ at $z = 9$. The EFE causes a bulk flow of $v_{\mathrm{void}} = 1586 \, \rm{km\,s^{-1}}$ at $z = 0$. Our fiducial model explains the local observations listed in Table~\ref{tab:comparison_bestfittingmodel_with_observations} at the $1.14\%$ confidence level ($2.53 \sigma$ tension). Figure~\ref{figure:marginalization_MBprofile} shows that models with somewhat different initial conditions also provide reasonable results.

The rms density fluctuation in the total matter field at the CMB ($z = 1100$) is $\delta_{\mathrm{rms}} \approx 10^{-4}$ (Section~\ref{subsubsec:Radiation dominanted era and CMB}), which implies $\delta_{\mathrm{rms}} \approx 10^{-2}$ at $z = 9$ for a $\Lambda$CDM cosmology. Thus, $\alpha_{\mathrm{void}}$ of our best-fitting model is much lower than the expected cosmic variance in $\Lambda$CDM. This could make KBC-like voids very common in the universe, potentially conflicting with the observed foreground lensing of the CMB (Section~\ref{Other_voids}). This problem could be alleviated if the EFE was stronger in the past (Section~\ref{subsubsec:Structure formation in MOND}), or if the peculiar accelerations and the Hubble flow are coupled (Section~\ref{subsubsec:Theoretical_assumptions}) $-$ both would slow down the growth of structure. It would be highly interesting to quantify the existence of KBC-like voids in a large cosmological MONDian $N$-body simulation, especially if it accounts for the HFE in some way. 

As already shown in several previous studies \citep[e.g.][and references therein]{Sanders_1998,Famaey_2012}, we affirm that structure formation is much more efficient in MOND compared to the Newtonian case (right-hand panel of Figure~\ref{figure_void_density_profile_results}). Applying an RSD correction based on the best-fitting model, the observed underdensity is $25.4\pm8.3 \%$ ($-5.2\pm10.5 \%$) in the inner (outer) part of the KBC void. The enhanced growth of structure allows our fiducial model to match these constraints at the $0.99 \sigma$ ($0.97 \sigma$) confidence level. In contrast, the KBC void rules out the $\Lambda$CDM framework at $6.04 \sigma$ (Section~\ref{KBC void in LCDM}).

In MOND, the long-range modification to gravity causes a very shallow decrease of the density contrast with distance, causing our model to systematically underestimate the density at the outer part of the KBC void as derived from the $K$-band luminosity data of \citet{Keenan_2013}. However, observational uncertainties on the density contrast there are still relatively large. Future surveys would be necessary to more precisely measure the density profile beyond $\approx 400 \, \rm{Mpc}$. This may provide an important test of our model because the radial density profile should be sensitive to the underlying growth rate.

Interestingly, \citet{Angus_2011} found some evidence for large voids with a diameter of $250 h^{-1} \, \rm{Mpc}$ in their $512 h^{-1} \, \rm{cMpc}$ $N$-body cosmological MOND simulation with massive neutrinos. Although both large voids and massive galaxy clusters are expected in a MOND cosmology \citep[e.g.][]{Sanders_1998}, it is not fully clear if those were formed artificially due to the low particle resolution \citep{Angus_2013}. Their simulations also assume no coupling between peculiar accelerations and the Hubble flow \citep[i.e. $\beta = 0$ in equation~8 of][]{Sanders_2001}. As discussed further in Section~\ref{subsubsec:Theoretical_assumptions}, a coupling to the Hubble flow would suppress the formation of massive voids and clusters on scales $\ga 100$~Mpc. Therefore, it would be very valuable to revisit their cosmological simulations with a higher particle resolution and an AMR grid code such as \textsc{phantom} of \textsc{ramses} \citep{Teyssier_2002,Lueghausen_2015}. Such MOND simulations would require very large box sizes in order to include large-scale modes and the resulting EFE on smaller regions (Section~\ref{subsubsec:Large-scale structure}). For very long modes, light travel time effects could be important such that a relativistic code is required. This could be based on the model of \citet{Skordis_2019}.

A unique characteristic of our void model is the prediction of very high total peculiar velocities, especially towards the void edge in the direction parallel to $\bm{g}_{\mathrm{ext}}$. In the best-fitting model, the void as a whole moves with a peculiar velocity of $v_{\mathrm{void}} = 1586 \, \rm{km\,s^{-1}}$ due to the EFE from source(s) beyond the void (i.e. at $z \ga 0.15$). Thus, our model predicts a sphere centred on the LG should have a large bulk flow of $\approx 1000 \, \rm{km\,s^{-1}}$ in a similar direction to $\bm{g}_{\mathrm{ext}}$. This is qualitatively similar to the results of previous $\nu$HDM simulations \citep{Katz_2013}. Interestingly, some evidence for a large bulk flow has been found \citep{Kashlinsky_2008, Kashlinsky_2011}. We discuss this further in Section~\ref{subsubsec:KSZ effect}.

Partial cancellation between the void's motion and internal motions within it leads to a region $\approx \left(150 - 270 \right) \, \rm{Mpc}$ from the void centre in which $v_{\mathrm{tot}} \leq v_{\mathrm{LG}} = 627 \, \rm{km\,s^{-1}}$. The fraction that this volume represents of the whole void corresponds to a $2.34\sigma$ event, implying that the LG is statistically not at a special position in the void. Note that the LG motion causes the highest tension amongst our constraints (Table~\ref{table:comparison_bestfittingmodel_with_observations} and Figure~\ref{figure_Pichart_bestfitting}).

\begin{figure}
    \includegraphics[width=86mm]{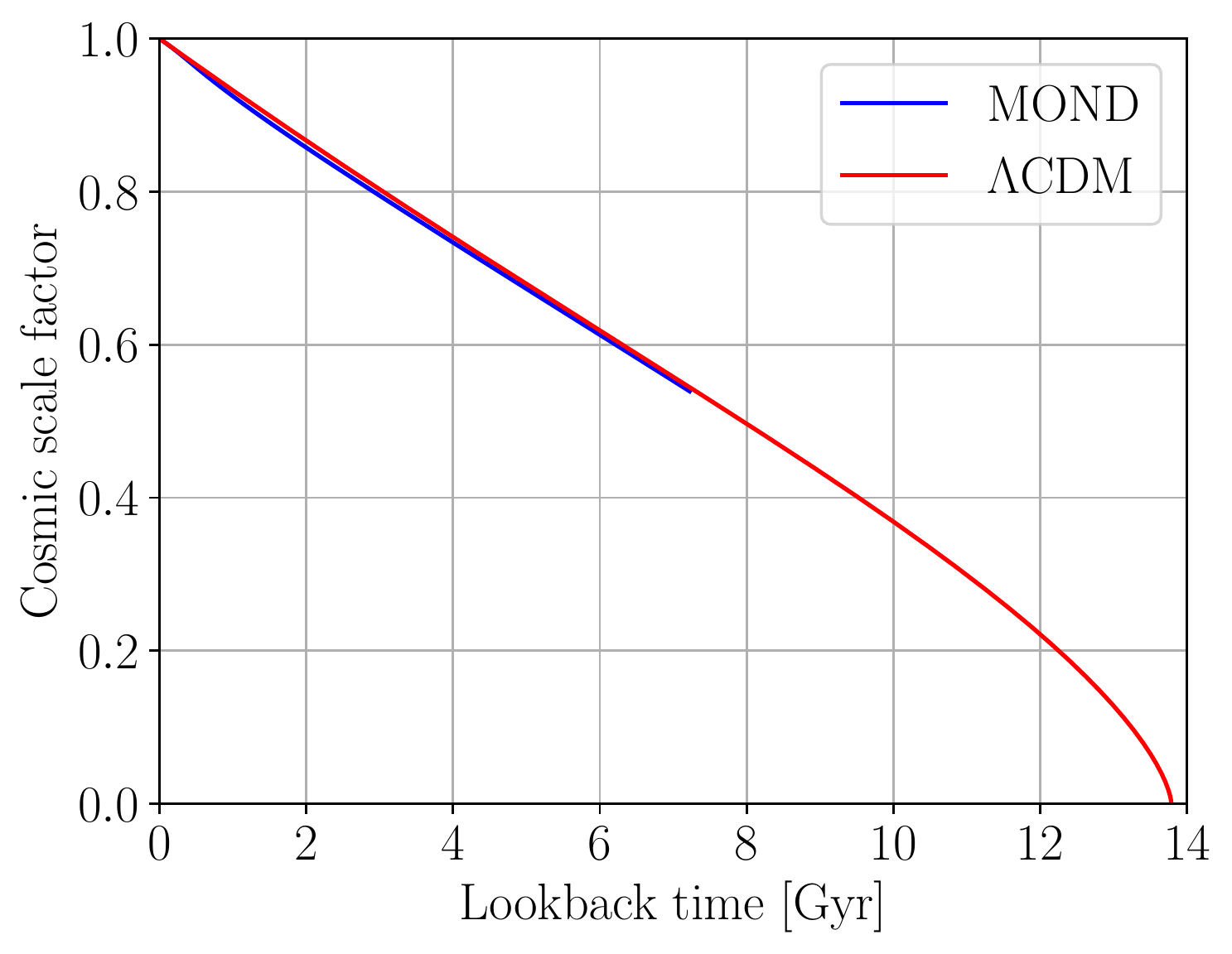}
    \caption{Time dependence of the cosmic scale factor in $\Lambda$CDM (red) and the apparent scale factor in our fiducial MOND model (blue). This model has a Maxwell-Boltzmann initial profile with $r_{\mathrm{void}} = 228.2 \, \rm{cMpc}$ and $\alpha_{\mathrm{void}} = 3.76 \times 10^{-5}$ embedded in a constant ($n_{\mathrm{EFE}} = 0$) external field of strength $g_{\mathrm{ext}} = 0.055 \, a_{_0}$.}
    \label{figure_best_fitting_model_scalefactor}
\end{figure}

Our fiducial model gives an apparent expansion history very close to $\Lambda$CDM (Figure~\ref{figure_best_fitting_model_scalefactor}), but with local Hubble constant $H_{0}^{\mathrm{model}} = 76.15 \, \rm{km\,s^{-1}\,Mpc^{-1}}$ and acceleration parameter $\overline{q}_{_0}^{\mathrm{model}} = 1.07$ in the redshift range $0.023 \leq z \leq 0.15$. This agrees with the observations of \citet{Camarena_2020} at the $84.20\%$ confidence level ($0.20 \sigma$ tension). The best-fitting models with a Gaussian and an exponential void profile agree at the $0.83\sigma$ and $0.89\sigma$ level, respectively (Figure~\ref{figure_H0_q0_plane}). \emph{Thus, we showed for the first time that the KBC void can arise in MOND and solve therewith the Hubble tension.}

The locally observed acceleration parameter $\overline{q}_{_0} = 1.08 \pm 0.29$ \citep{Camarena_2020} disagrees with the $\Lambda$CDM expectation of $\overline{q}_{0,\mathrm{\Lambda CDM}} = 0.53$ \citep{Planck_2018} at the $1.9\sigma$ level. In combination with the $H_{0}$ discrepancy between these studies, this would falsify $\Lambda$CDM at $4.54\sigma$ confidence (see also Figure~\ref{figure_H0_q0_plane}). Interestingly, $\overline{q}_{_0} > 1$ is not possible for a standard background cosmology. The locally observed high Hubble constant and acceleration parameter provide compelling evidence that the Hubble tension is caused by a local effect like the KBC void. This addresses the concern of \citet{Kenworthy_2019} that the KBC void is not evident in the SNe distance-redshift relation (Section~\ref{subsubsec:Fixing the deceleration parameter}) $-$ both the first and second derivatives of the distance-redshift relation very much point to a local void. Observationally, a discrepancy could also appear as a third order effect in the jerk parameter $j \equiv a^{2} \dddot{a}/{\dot{a}}^{3}$, but given the already large uncertainty of $\overline{q}_{_0}$, it would be difficult to measure $j_{_0}$ precisely.

As discussed in Section~\ref{subsubsec:Hubble constant from strong lensing}, strong lensing does not occur in the MOND regime and so should be similar to in $\Lambda$CDM cosmology \citep{Sanders_1999}. Thus, our main analysis includes $H_{0}$ constraints from seven strongly lensed quasars as obtained by \citet{Shajib_2020} and \citet{Wong_2020}. The latter work applied a blinded analysis (described in their section~3.6) and found that $H_{0}$ decreases as a function of lens redshift at $1.9 \sigma$ significance (see their appendix~A). $H_0$ becomes consistent with Planck expectations at $z \ga 0.5$, well beyond the void. This is again a very strong indication that the Hubble tension is driven by a local environmental effect such as the KBC void. A decrease of the inferred $H_{0}$ with redshift is also apparent in our MOND model (Figure~\ref{figure:best_fitting_model_Hubble_vs_redshift}) and is a generic consequence of any local resolution to the Hubble tension. The redshift dependence of $H_{0,\mathrm{lensing}}^{\mathrm{model}}$ depends mainly on the density profile of the void. For our fiducial model, the combined tension with all seven lensing-based $H_{0}$ measurements is $\chi^{2} = 14.66$, which represents $2.05\sigma$ tension for $7$ degrees of freedom. In the case of a Gaussian and an exponential void profile, this would reduce to $1.76\sigma$ and $1.83\sigma$, respectively, because the best-fitting models have a larger void (Appendix \ref{Appendix:Joint probabilities for different void profiles}). This discrepancy with the lensing data is mainly caused by a systematic underestimation of $H_{0}$ by our models, especially for the two lowest redshift lenses RXJ1131$-$1231 at $z = 0.295$ and PG 1115+080 at $z = 0.311$ (Section~\ref{subsubsec:Excluding strong lensing time delays}).

The high values of $H_{0}$ at $z \ga 0.4$ are also conspicuous because according to \citet{Keenan_2013} the density should have reached the cosmic mean already at $z \approx 0.2$ (see their figure~11). Consequently, we would expect that $H_{0}$ obtained from lenses located at $z \ga 0.4$ must be very similar to the Planck prediction. The $H_{0}$ values from the four highest redshift lenses of \citet{Wong_2020} give a median (mean) of $71.3 \,\rm{km\,s^{-1}\,Mpc^{-1}}$ ($70.8 \,\rm{km\,s^{-1}\,Mpc^{-1}}$), which is $3.9\,\rm{km\,s^{-1}\,Mpc^{-1}}$ ($3.4\,\rm{km\,s^{-1}\,Mpc^{-1}}$) higher than the Planck prediction. This systematic offset can be reduced if the background $H_{0}$ is underestimated due to errors in the Planck measurements caused by intergalactic dust (\citealt{Yershov_2020}; see also \citealt{Vavrycuk_2018, Vavrycuk_2019}). We discuss this issue further in Section~\ref{subsubsec:CMB contamination by intergalactic dust}. It is also possible that there is some systematic offset in $H_0$ measurements using strong lensing time delays \citep{Kochanek_2020}. Since we assume mass conservation in our models, this discrepancy cannot be fully resolved in our analysis $-$ the algorithm searches for a compromise between the KBC and lensing data. It seems that there is some internal inconsistency between them. A strong test of our model would be to infer $H_{0}$ very accurately from nearby and high-redshift lens systems. The model predicts that $H_{0}$ should be almost identical to the Planck prediction at $z \ga 0.9$. However, a substantial anomaly is expected for a lens at $z = 0.1$. A measurement here would nicely complement the SNe results of \citet{Camarena_2020}, which go out to $z = 0.15$.

\subsubsection{Excluding strong lensing time delays} \label{subsubsec:Excluding strong lensing time delays}

Although strong lensing in MOND should be similar to the $\Lambda$CDM framework \citep{Sanders_1999}, the $H_0$ measurements from strong lensing could be affected by the EFE on the void. This requires a better understanding of its origin. We have assumed that the EFE in our simulations affects the void as a whole, implying that it must be sourced by something beyond the void, i.e. at $z \ga 0.15$. This approach would be valid for deriving $H_{0}$ and $\overline{q}_{_0}$ from SNe data in the range $0.023 \leq z \leq 0.15$ since the EFE would similarly move everything in this region. However, this may not be true for even the lowest redshift lens as its $z = 0.295$ (Table~\ref{tab:observed_H0_strong_lensing}). If the EFE is sourced by something at lower $z$, it would move the void $-$ but not the lens.

In this context, we consider in more detail the geometry of the void and the sky positions of the lenses. The observed CMB dipole shows that the LG moves with a velocity of $627 \, \rm{km\,s^{-1}}$ wrt. the CMB towards Galactic coordinates $\left( 276^{\circ} \pm 3^{\circ}, 30^{\circ} \pm 3^{\circ} \right)$, which roughly matches the direction of the radio dipole (Section~\ref{subsec:KBC void}). Thus, the LG motion wrt. the CMB frame is probably directed away from the void centre (see also Figure~\ref{figure_peculiar_velocity_results}). Interestingly, the two lowest redshift lens systems (RXJ1131$-$1231 and PG 1115+080) are located at Galactic coordinates $\left(274.4^{\circ}, +45.9^{\circ}\right)$ and $\left(249.9^{\circ}, +60.6^{\circ} \right)$, respectively, which roughly coincides with the directions of the CMB and radio dipoles. Thus, both lenses are also most likely located on the opposite side to the void centre. Assuming these lenses are not affected by the EFE on the void, this would cause an extra redshift. Consequently, a larger Hubble anomaly would be expected than calculated thus far.

Relative to the CMB dipole, the direction to each lens subtends an angle $\theta$, where $\cos \theta = 0.96$ and $\cos \theta = 0.82$ for RXJ1131$-$1231 and PG 1115+080, respectively. The impact on the measured $H_{0}$ from these two low-$z$ lenses can be estimated as
\begin{eqnarray}
	\Delta H_{0} &=& \frac{\Delta a}{\Delta t} \, , \quad \mathrm{with} \\
	\Delta a &=& \frac{a v_{\mathrm{void,r}}}{c} \, , \\
	v_{\mathrm{void,r}} &\equiv& v_{\mathrm{void}} \cos{\theta} \, ,
	\label{eq:strong_lensing_discussion}
\end{eqnarray}
where $a$ is the scale factor at which the lens is observed, and $\Delta t$ is the corresponding lookback time. Thus, assuming the lenses are unaffected by the EFE, $H_{0}$ derived from the lens systems RXJ1131$-$1231 and PG 1115+080 would be overestimated by $1.10$ and $0.88 \, \rm{km\,s^{-1}\,Mpc^{-1}}$, respectively. This would only slightly reduce the tension with our best-fitting models. Therefore, the sharp rise in $H_{0}$ values for the two lowest redshift lenses cannot be fully accounted for with the EFE and is still hard to explain. Additional lenses are needed to confirm this feature.

We address the possible impact of the EFE on the lensing-based $H_{0}$ measurements by redoing the analysis for our Maxwell-Boltzmann model without the constraints from strong lensing time delays. In this case, the best-fitting model has $g_{\mathrm{ext}} = 0.050 \, a_{_0}$ causing $v_{\mathrm{void}} = 1442 \, \rm{km\,s^{-1}}$, $r_{\mathrm{void}} = 208.4 \, \rm{cMpc}$, and $\alpha_{\mathrm{void}} = 2.15 \times 10^{-5}$. This model yields $H_{0}^{\mathrm{model}} = 76.47 \, \rm{km\,s^{-1}\,Mpc^{-1}}$ and $\overline{q}_{_0}^{\mathrm{model}} = 1.21$ ($0.26\sigma$ combined tension), $\delta_{\mathrm{in}} = 0.167$ ($\delta_{\mathrm{obs, corr}}^{\mathrm{in}} = 0.258 \pm 0.082$; $1.11\sigma$), and $\delta_{\mathrm{out}} = 0.037$ ($\delta_{\mathrm{obs, corr}}^{\mathrm{out}} = -0.038 \pm 0.104$; $0.73\sigma$). The fraction of the void with $v_{\mathrm{tot}} \leq v_{\mathrm{LG}}$ represents a $2.25 \sigma$ tension. The model explains all these local observations at the $1.96 \sigma$ ($5.0\%$) confidence level. Thus, excluding the lensing data allows for a somewhat better overall fit, but has little effect on the preferred model parameters.

\subsubsection{Structure formation and external field history in MOND} \label{subsubsec:Structure formation in MOND}

\begin{figure}
	\includegraphics[width=\linewidth]{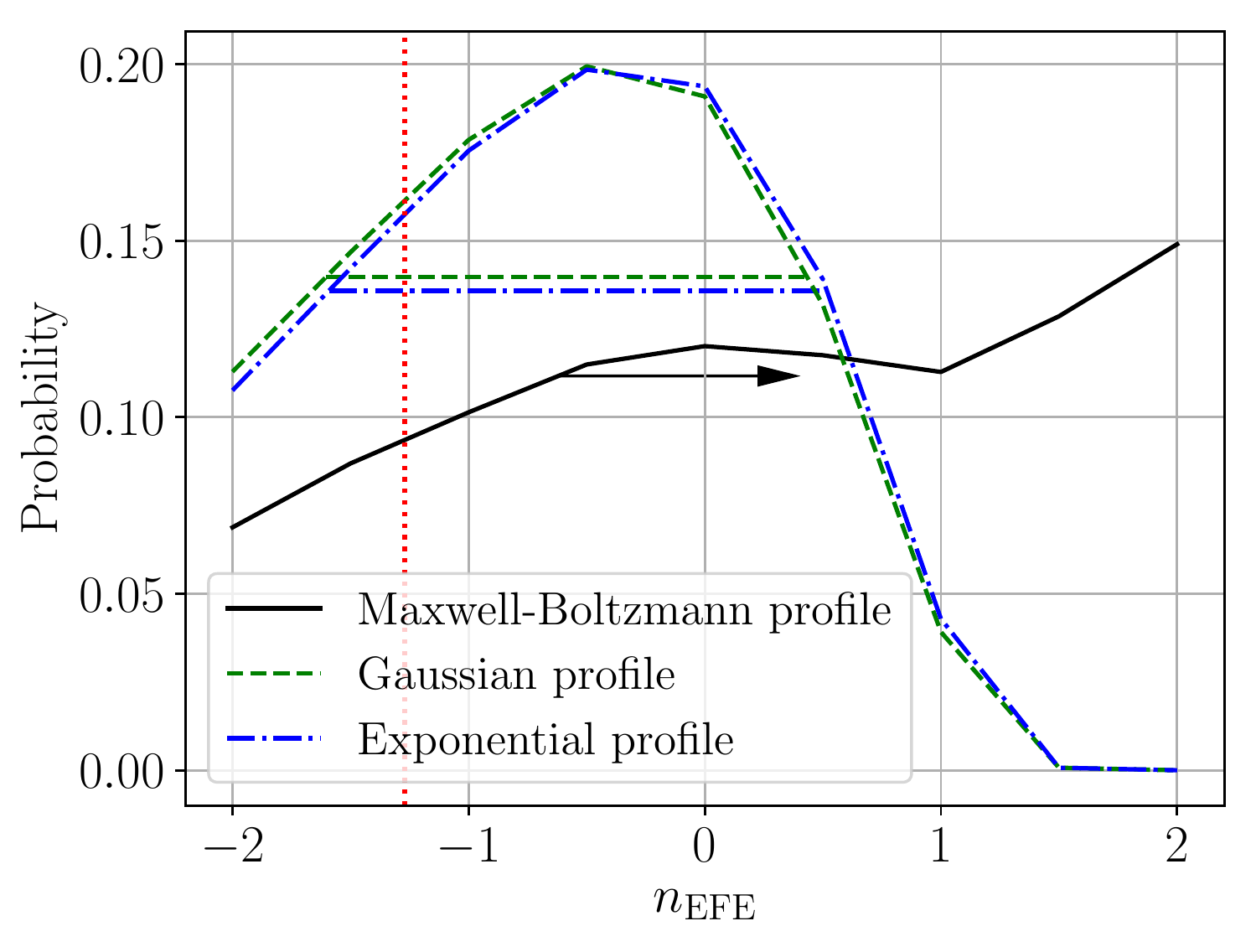}
	\caption{Marginalized posterior distribution of the time dependence of the external field ($n_{\mathrm{EFE}}$ in Equation~\ref{eq:time_dependent_external_field_history}), based on $9 \times 10^{6}$ MOND models where the initial void profile is Maxwell-Boltzmann (solid black), Gaussian (dashed green), or exponential (dot-dashed blue). The black arrow and the horizontal lines indicate the $1 \sigma$ confidence interval for each profile. The red-dotted vertical line marks the expected $n_{\mathrm{EFE}} = -1.27$ (Equation~\ref{eq:n_EFE_history_expected}), which we derive from the $g_{\mathrm{ext}}$ of our fiducial model and the expected gravitational acceleration at recombination.}
	\label{figure_EFE_history_results}
\end{figure}

Since the EFE acting on a MONDian subsystem depends on surrounding structure and therewith on the scale factor, we made in Section~\ref{subsubsec:External field history} the ansatz $g_{_\mathrm{N,ext}} \left( t \right) = g_{_\mathrm{N,ext}} \left( t_{_0} \right)a^{n_{\mathrm{EFE}}}(t)$. So far, we have restricted attention to the case ${n_{\mathrm{EFE}} = 0}$. Letting $n_{\mathrm{EFE}}$ vary in the range $\left(-2, 2 \right)$, Figure~\ref{figure_EFE_history_results} shows its marginalized posterior based on $9 \times 10^{6}$ different models. In the case of a Maxwell-Boltzmann profile, the analysis yields $n_{\mathrm{EFE}} > -0.62$ at the $1\sigma$ confidence level. Gaussian and exponential initial profiles allow for $(-1.60,+0.43)$ and $(-1.59,+0.52)$ at the $1\sigma$ level, respectively. Thus, only the Maxwell-Boltzmann profile prefers a weaker EFE in the past, while the other profiles prefer the opposite. A time-independent EFE ($n_{\mathrm{EFE}} = 0$) $-$ assumed for all our models thus far $-$ lies within the $1\sigma$ range for all three considered void profiles. This justifies our assumption of a time-independent EFE in our main analysis.

\begin{figure*}
	\includegraphics[width=58mm]{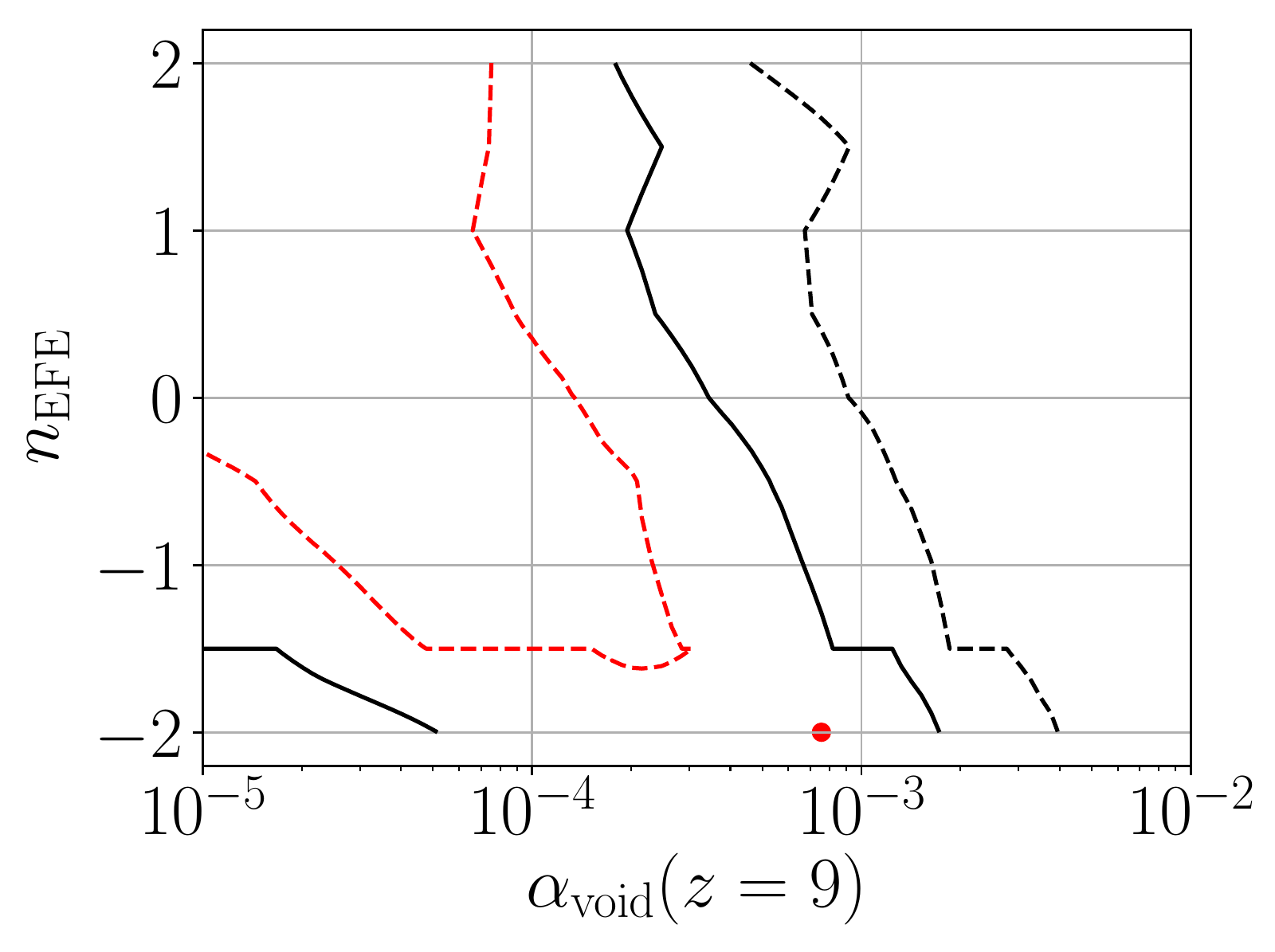}
	\includegraphics[width=58mm]{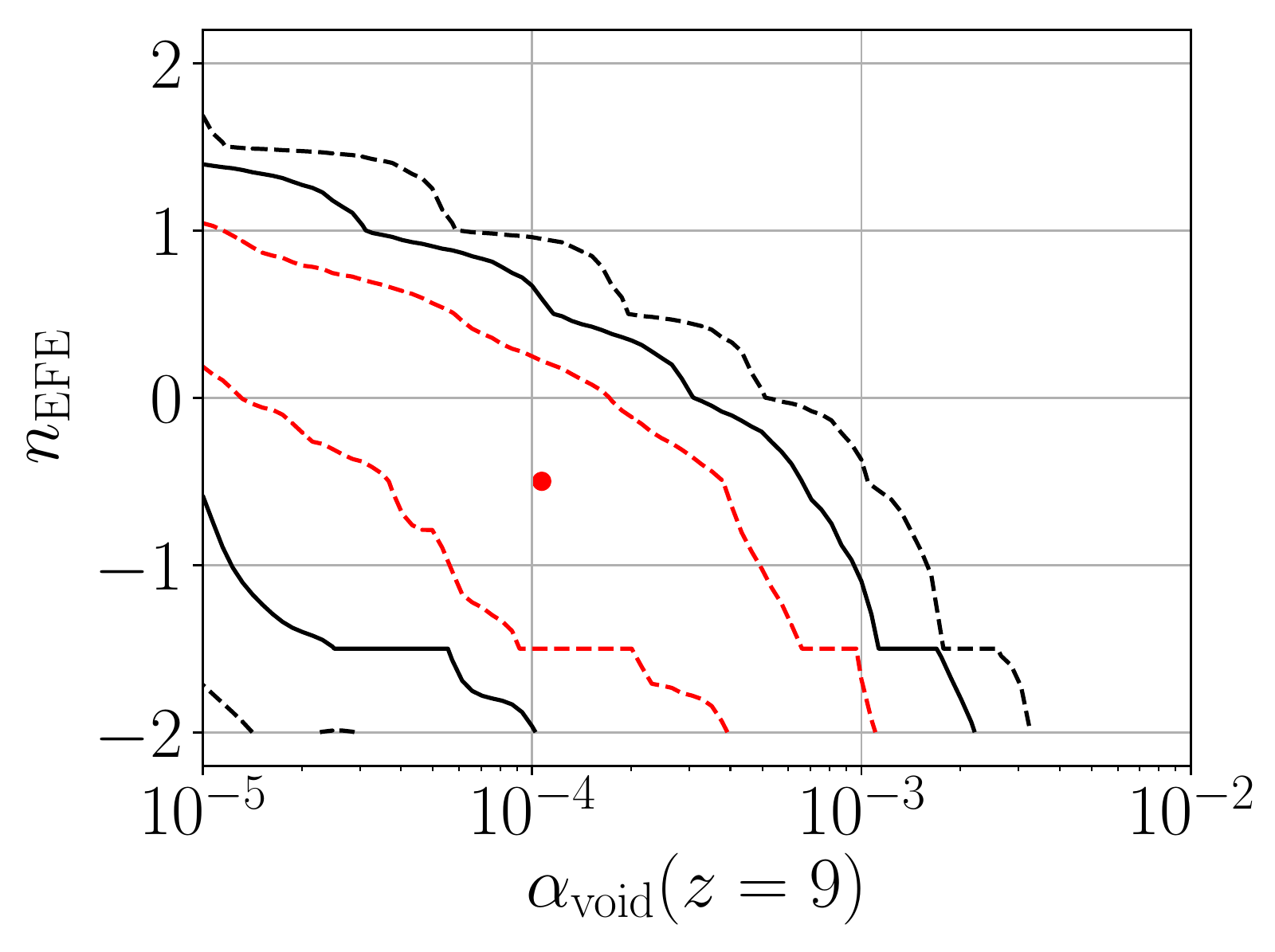}
	\includegraphics[width=58mm]{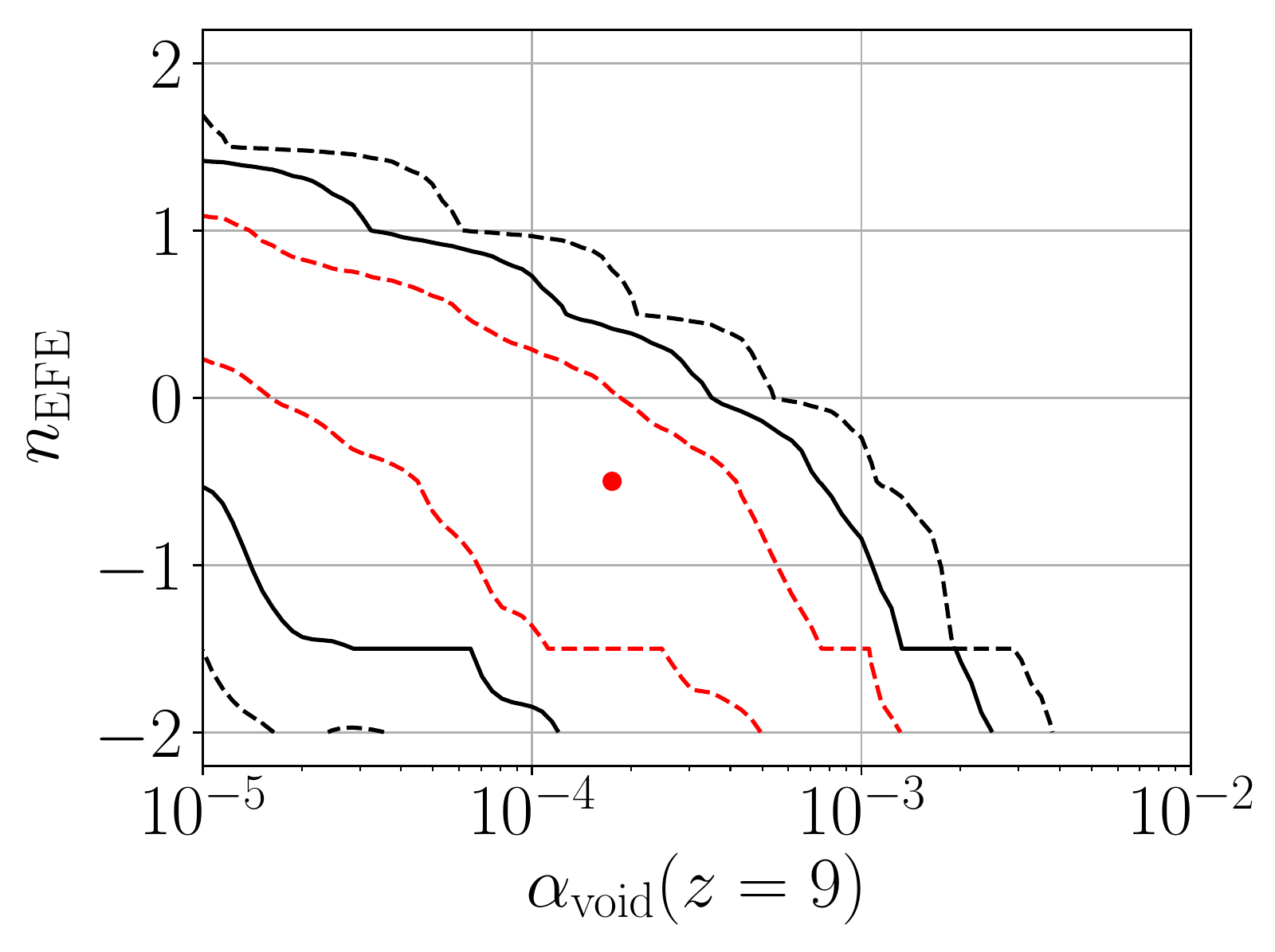}
	\caption{Marginalized posterior distribution of the indicated model parameters based on $9 \times 10^{6}$ MOND models for a Maxwell-Boltzmann (left), Gaussian (middle), and exponential (right) initial profile. The red dashed, black solid, and black dashed lines mark the $1\sigma$, $2\sigma$, and $3\sigma$ confidence levels, respectively. A stronger EFE in the past ($n_{\mathrm{EFE}} < 0$) requires a stronger initial void strength at $z = 9$. The red dots mark the best-fitting models: $g_{\mathrm{ext}} = 0.030 \, a_{_0}$, $r_{\mathrm{void}} = 218.3 \, \rm{cMpc}$, $\alpha_{\mathrm{void}} = 7.56 \times 10^{-4}$, $n_{\mathrm{EFE}} = -2$ (Maxwell-Boltzmann profile, left-hand panel); $g_{\mathrm{ext}} = 0.065 \, a_{_0}$, $r_{\mathrm{void}} = 1030.0 \, \rm{cMpc}$, $\alpha_{\mathrm{void}} = 1.07 \times 10^{-4}$, $n_{\mathrm{EFE}} = -0.5$ (Gaussian profile, middle panel); and $g_{\mathrm{ext}} = 0.070 \, a_{_0}$, $r_{\mathrm{void}} = 1030.0 \, \rm{cMpc}$, $\alpha_{\mathrm{void}} = 1.75 \times 10^{-4}$, $n_{\mathrm{EFE}} = -0.5$ (exponential profile, right-hand panel).}
	\label{figure_EFE_history_results_alpha_n}
\end{figure*}

The marginalized posterior distributions for $n_{\mathrm{EFE}}$ and the initial $\alpha_{\mathrm{void}}$ are shown in Figure~\ref{figure_EFE_history_results_alpha_n}. As expected, a stronger EFE in the past ($n_{\mathrm{EFE}} < 0$) requires a deeper initial void. In particular, values of $\alpha_{\mathrm{void}}$ up to $\approx 10^{-3}$ are now allowed, contrary to the case where we fix $n_{\mathrm{EFE}} = 0$ (Figure~\ref{figure:marginalization_MBprofile}). This would be closer to the expected density fluctuations in $\Lambda$CDM when $a = 0.1$, since the $\approx 10^{-5}$ density fluctuations in the CMB should have grown $\approx 100 \times$. This is only mildly disfavoured by our analysis.

We can estimate the external field history based on the parameters of our fiducial model and the gravitational acceleration at recombination:
\begin{eqnarray}
	n_{\mathrm{EFE}} ~\approx~ \frac{\ln g_{_\mathrm{N}} \left( t_{_0} \right) - \ln g_{_\mathrm{N}} \left( t_{_\mathrm{CMB}} \right)}{\ln a \left( t_{_0} \right) - \ln a \left( t_{_\mathrm{CMB}} \right)} \, ,
	\label{eq:n_EFE_history_expected}
\end{eqnarray}
where $t_{_\mathrm{CMB}} = 380 \, \rm{kyr}$, and $t_{_0} = 13.8 \, \rm{Gyr}$. As explained in Section~\ref{subsubsec:Radiation dominanted era and CMB}, $g_{_\mathrm{N}} \left( t_{_\mathrm{CMB}} \right) \approx 21 \, a_{_0}$, causing MOND to have only a very small effect before recombination. Since our fiducial model prefers an external field of $0.055 \, a_{_0}$, we use this as our estimate of $g \left( t_{_0} \right)$, which via Equation~\ref{eq:MOND_simple_interpolation_function} yields $g_{_\mathrm{N}} \left( t_{_0} \right) = 0.0029 \, a_{_0}$. Comparing the expected large-scale Newtonian gravitational fields at these times gives $n_{\mathrm{EFE}} = -1.27$ (the red vertical line in Figure~\ref{figure_EFE_history_results}). This is consistent with the marginalized posterior distribution for the Gaussian and exponential profile models at the $1\sigma$ level and for the Maxwell-Boltzmann models at the $2\sigma$ level. The Gaussian and exponential models prefer $n_{\mathrm{EFE}} \approx -0.5$.

Equation~\ref{eq:n_EFE_history_expected} implies $n_{\mathrm{EFE}} = -1.27$, which is a faster decline than the $n_{\mathrm{EFE}} = -1$ suggested by Equation~\ref{eq:MOND_scaling_relations_9} for linear perturbations in the matter-dominated era (Section~\ref{subsubsec:Radiation dominanted era and CMB}). This could be due to the effect of dark energy, which is not taken into account in Equation~\ref{eq:MOND_scaling_relations_9} as it only dominates at $z \la 0.8$. Since dark energy slows down the growth of structure, $n_{\mathrm{EFE}}$ would be shifted to more negative values $-$ if the fractional density perturbations are frozen in co-moving coordinates, we would get $n_{\mathrm{EFE}} = -2$.

Though it is beyond the scope of our semi-analytic study, we mention briefly that as structure grows in a MONDian universe, it imposes an external field on surrounding structures, thereby hampering their growth. This leads to structure formation in different regions becoming mutually correlated. In particular, since MOND gravity declines as $1/r$ whereas Newtonian tides scale as $1/r^3$ (there being no EFE), any structure in a Milgromian universe is affected by much more distant structures compared to the Newtonian case. This makes it difficult to conduct a MOND simulation with sufficiently large volume to satisfy the CP.

\subsubsection{Theoretical uncertainties in the MOND approach \& outlook for further studies}
\label{subsubsec:Theoretical_assumptions}

At present, it is not known if MOND is related to a fundamental (quantum) theory \citep[i.e. `FUNDAMOND',][]{Milgrom_2020b, Milgrom_2020}. As a result, we do not have a completely secure understanding of cosmology and structure formation in MOND. In fact, it is likely that the implications on these scales are not uniquely derivable from the RAR in disc galaxies. Although the relativistic MOND theory of \citet{Skordis_2019} seems quite promising, its consequences for cosmology are not yet established. Therefore, the here applied cosmological model required us to make some ansatzes, whose uncertainties will be summarized and discussed in the following (see also Section~\ref{subsec:nuHDM cosmological model}).

Motivated by previous theoretical studies, we assumed that the background cosmology in a Milgromian universe obeys the same Friedmann equation as in $\Lambda$CDM (Section~\ref{subsubsec:Background cosmology}). While this is not necessarily true in MOND,  the observations of \citet{Joudaki_2018} suggest that this works well empirically. Moreover, constraints from BBN imply rather small deviations from the standard $a \left( t \right)$ during the radiation-dominated era. In a MOND context, this forces the expansion history to obey the standard Friedmann equation to sub-per cent precision in the matter and $\Lambda$-dominated eras \citep{Skordis_2006}. Moreover, the source term for the Friedmann equation remains the same if CDM is replaced with the same density in sterile neutrinos with $m_{\nu_{s}} = 11 \, \rm{eV}/c^2$ since both would be non-relativistic up to very high $z \gg z_{\mathrm{eq}}$. Consequently, there is very good reason to suppose that the background $a \left( t \right)$ is very nearly the same as in $\Lambda$CDM.

Sterile neutrinos would only slightly affect primordial nucleosynthesis and plasma physics prior to recombination \citep[Sections~\ref{section:BBN} and \ref{subsubsec:Radiation dominanted era and CMB}, respectively, see also figure~1 in][]{Angus_2009}. Given also a very nearly standard expansion history and high peculiar gravitational accelerations at that time, we expect the $\nu$HDM framework to yield the same CMB power spectrum as $\Lambda$CDM. The angular diameter distance to the CMB would also be the same in both frameworks, causing both to suffer from the Hubble tension if $H_{0}^{\mathrm{local}}$ is little affected by cosmic variance. Our main argument is that this last assumption holds in $\Lambda$CDM but not MOND.

To simulate structure formation in MOND, we made the usual ansatz that MOND should be applied only to the density deviations from the cosmic mean \citep[e.g.][]{Llinares_2008, Angus_2011, Angus_2013, Katz_2013, Candlish_2016}. We justified this in Section~\ref{subsubsec:Large-scale structure} based on the fact that \citet{Sanders_2001} showed in his section~2 that this approach \citep[elaborated further in][]{Sanders_2002} is valid in a non-relativistic Lagrangian that has the MOND behaviour. This so-called Jeans Swindle \citep{Binney_1987_book} is one of the strongest assumptions of current cosmological MOND models. \citet{Falco_2013} has formally shown that it can be justified in an expanding General Relativistic universe, but this needs to be mathematically demonstrated for a MONDian framework in which the Poisson equation is non-linear. Despite this uncertainty, it seems inevitable that structure formation would be significantly faster in MOND compared to $\Lambda$CDM on a 100~Mpc scale. This is because 100~Mpc is much larger than the free streaming length of both sterile neutrinos and CDM, so the only major difference between the $\Lambda$CDM and $\nu$HDM frameworks is a different gravity law. Since the accelerations are much smaller than $a_{_0}$ (Section \ref{subsec: Milgromian dynamics}), we expect any MOND theory to yield a significant enhancement to the gravity generated by density perturbations. As a result, we argue that MOND models naturally possess the ability to explain the KBC void and Hubble tension.

We now discuss whether this conclusion remains valid if the Jeans swindle approach is not applicable because of the HFE, a coupling between the background cosmology and structures within it. As mentioned in Section~\ref{subsubsec:Large-scale structure}, \citet{Sanders_2001} developed a non-relativistic two-field Lagrangian-based theory of MOND that couples the Hubble flow and the peculiar accelerations from inhomogeneities. This coupling is described by the adjustable parameter $\beta$ in his equation~8, and is elaborated further in \citet{Sanders_2002}. If $\beta = 0$, the coupling between these fields vanishes, which is equivalent to the above-mentioned Jeans swindle. In the case $\beta \neq 0$, the background cosmology would remain intact, but the Hubble flow acceleration ($\bm{g}_{\mathrm{Hubble}}$ in Equation~\ref{eq:governing_equation_Force_1}) would appear as an additional source of gravity that suppresses the $\nu$ factor in Equation~\ref{eq:MOND_basic}.

\begin{figure}
    \includegraphics[width=\linewidth]{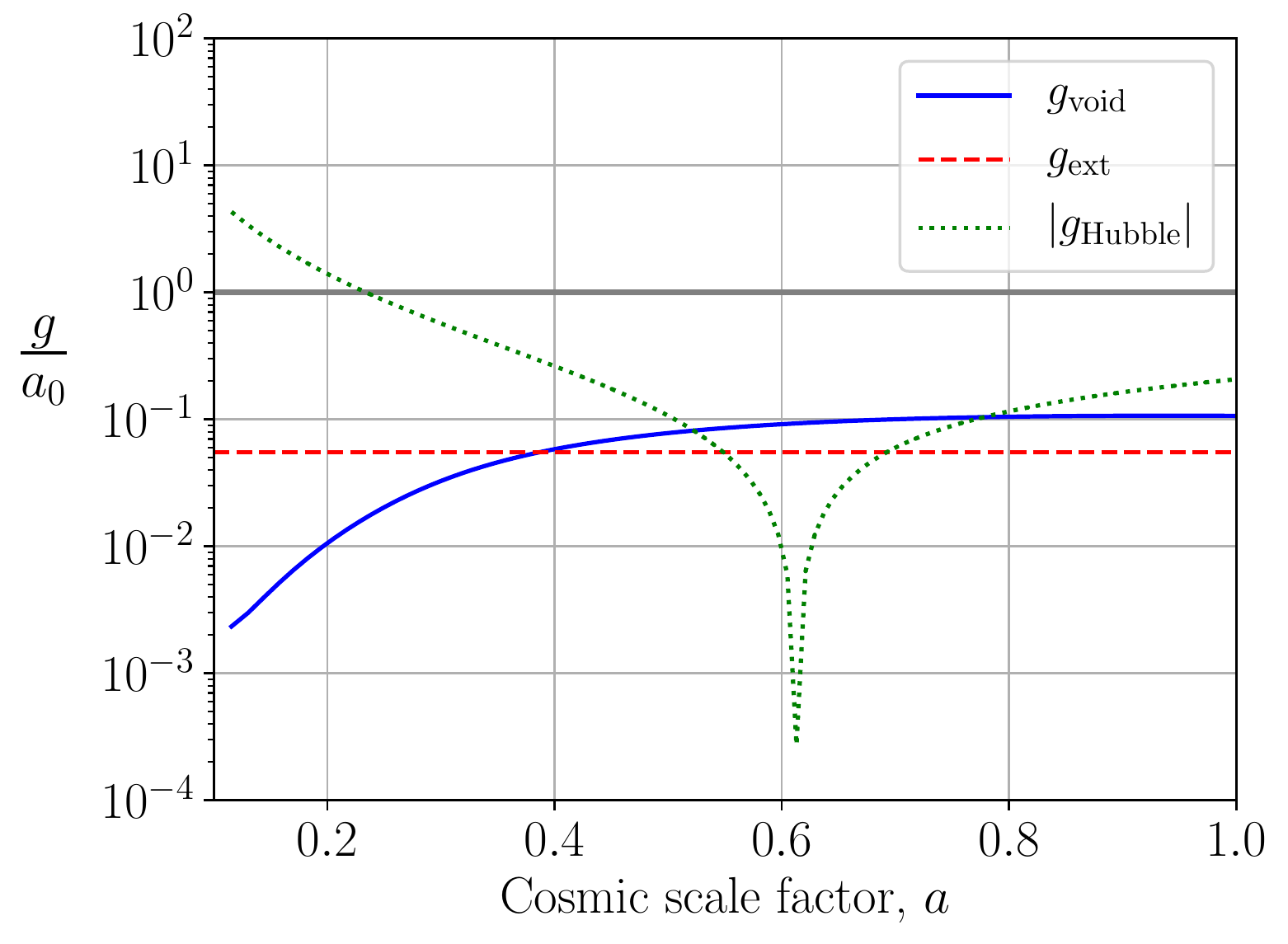}
    \caption{Accelerations along the trajectory of a particle ending up 300~Mpc from the void centre in our best-fitting model. We show the time evolution of $g_{\mathrm{void}}$ (the blue solid line), $g_{\mathrm{ext}}= 0.055 \, a_{_0} $ (the red horizontal dashed line), and $\lvert g_{\mathrm{Hubble}} \rvert$ for a standard background cosmology (the green-dotted line). The thick horizontal grey line refers to $a_{_0}$. Note that $g_{\mathrm{Hubble}}$ changes sign when $a = 0.61$ ($z = 0.63$) $-$ it is positive at later times and negative earlier on. At present, $g_{\mathrm{void}} \approx 0.1 \, a_{_0}$ and $g_{\mathrm{Hubble}} \approx 0.2 \, a_{_0}$, so the latter would limit the MOND boost to gravity at the void edge in case of a strong HFE.}
    \label{figure_MOND_gravity_time}
\end{figure}

As argued in Section~\ref{subsub:Implications for the Hubble tension}, the existence of the KBC void and the Hubble tension can be simultaneously reconciled in the $\Lambda$CDM framework due to mass conservation, but only for a $10\sigma$ density fluctuation (Figure~\ref{figure_LCDM_results_cosmic_variance}). We demonstrated that in the absence of any HFE, our best-fitting model can explain both the KBC void and the Hubble tension because of an enhanced cosmic variance compared to $\Lambda$CDM (Section~\ref{subsec:Results The best-fitting model}). A coupling to $g_{\mathrm{Hubble}}$ would reduce the cosmic variance in our MOND model, and therewith also the frequency of KBC-like voids. To estimate the possible impact, we use Figure~\ref{figure_MOND_gravity_time} to plot the various accelerations entering Equation~\ref{eq:governing_equation_Force_1} over time for the same test particle presently 300~Mpc from the void centre. MOND should boost structure formation somewhat at all epochs with $a \ga 0.2$ since $\lvert g_{\mathrm{Hubble}} \rvert < a_{_0}$ and the other acceleration terms are even smaller. However, in order to obtain a lower limit on the cosmic variance expected in the MOND framework, we assume the most conservative scenario in which structures grow only as fast as in the $\Lambda$CDM framework $\left( \delta \appropto a^{0.8} \right)$ at all epochs when $g_{\mathrm{Hubble}}$ is dominant $-$ it is after all unclear exactly what would happen then. As shown in the right-hand panel of Figure~\ref{figure_void_density_profile_results}, density fluctuations in our best-fitting model grow as $\delta \appropto a^{3.0}$ for the period $0.5 \leq a \leq 0.8$. During this period, our previous approach should be valid because $\lvert g_{\mathrm{Hubble}} \rvert < g_{\mathrm{void}}$ (Figure~\ref{figure_MOND_gravity_time}). Thus, the $3.2\%$ standard $\Lambda$CDM cosmic variance is increased by a factor of at least $\left(0.8/0.5\right)^{3.0-0.8} \approx 2.8$, implying a $9.0\%$ cosmic variance. Since our best-fitting MOND model yields a present underdensity of $\delta_{\mathrm{in}} = 0.172$ on a 300 Mpc scale\footnote{This is consistent with the observed $0.46 \pm 0.06$ \citep{Keenan_2013} due to the RSD correction (Table~\ref{table:comparison_bestfittingmodel_with_observations}).}, it requires an $\approx 1.9\sigma$ fluctuation. In a MONDian model with $n_{\mathrm{EFE}} = -1$ (Equation~\ref{eq:time_dependent_external_field_history}), density fluctuations grow slower as $\delta \appropto a^{2.8}$ for $0.5 \leq a \leq 0.8$, implying an $8.2\%$ cosmic variance and thus a $2.1 \sigma$ fluctuation. The latter scenario may be more realistic in light of our discussion in Section \ref{subsubsec:Structure formation in MOND}. Since the present value of $\delta$ is essentially fixed by the observations of \citet{Keenan_2013}, the slower structure growth in such a model implies a deeper void in the past, increasing $g_{\mathrm{void}}$ relative to our fiducial model. This would increase the timespan during which $g_{\mathrm{Hubble}}$ is sub-dominant, thereby allowing MOND to enhance structure growth to a greater extent. We conclude that even in the case of a strong HFE and making very conservative assumptions, our MOND approach still succeeds in explaining the KBC void, which by mass conservation also resolves the Hubble tension.

This is mainly because $g_{\mathrm{Hubble}}$ was much smaller a few Gyr ago $-$ it is $\propto \ddot{a}$ (Equation~\ref{eq:governing_equation_Force_1}), which crossed zero at a lookback time of $\approx 6 \, \rm{Gyr}$ (Figure~\ref{figure_MOND_gravity_time}). $r$ was also smaller in the past if we consider the same co-moving scale. While $g_{\mathrm{Hubble}}$ can undoubtedly suppress the growth of structure, it was sub-dominant for an extended period $-$ during which it should be appropriate to neglect the HFE. Indeed, a much more rapid growth of perturbations around the time $\ddot{a} = 0$ is also evident in figure~1 of \citet{Sanders_2001}, who considered MOND with a strong HFE. Thus, the presence of dark energy can actually promote the growth of structure in a MOND context by reducing $\lvert g_{\mathrm{Hubble}} \rvert$. Observations of structure growth on a $\ga~100$~Mpc scale at $a \approx 0.61$ might reveal evidence for this growth spurt, which would occur at a redshift beyond the extent of the KBC void.

\citet{Sanders_2002} showed that since $g_{\mathrm{Hubble}}$ scales directly with the size of a system, perturbations on smaller length-scales would be shielded from the HFE, allowing them to grow much faster in case of a strong coupling (see their figure~12). For instance, $g_{\mathrm{Hubble}} \approx 7 \times 10^{-3} a_{_0}$ for a particle at a scale of $10\,\rm{Mpc}$, so $g_{\mathrm{Hubble}}$ would be very sub-dominant compared to typical external fields of a few percent of $a_{_0}$ \citep[e.g.][]{Famaey_2007}. This justifies the Jeans swindle approach used in numerical simulations if the goal is to address problems on this scale.

The initial $\alpha_{\mathrm{void}}$ required by our fiducial model is lower than the expected rms density fluctuations at $z = 9$ for a $\Lambda$CDM cosmology, which we can estimate by scaling the present value of 0.032 by $0.1^{0.8}$ to obtain $\approx 0.005$. Our results in Figure~\ref{figure_EFE_history_results_alpha_n} show that this remains true even with a stronger EFE in the past ($n_{\mathrm{EFE}} < 0$), implying that KBC-like voids would be quite common in MOND. Such a void could explain the Cold Spot in the CMB, which is often interpreted as a huge underdensity \citep{Nadathur_2014}. However, such structures are quite rare, suggesting that large voids are not very frequent in the Universe (Section~\ref{subsec:Assessing the tension for MOND}). Moreover, a high frequency of KBC-like voids could cause too much foreground lensing of the CMB $-$ though this is far from clear (Section~\ref{Other_voids}).

In principle, the implications of MOND on large scales depend on an adjustable parameter analogous to $\beta$ in equation~8 of \citet{Sanders_2001}, which can be used to alter the frequency of KBC-like voids or massive galaxy clusters such as El Gordo at high redshift. A strong HFE would reduce the frequency of such structures. We apply Occam's Razor and assume $\beta = 0$ since there is no compelling observational or theoretical evidence for $\beta \neq 0$. In particular, it is not yet clear if covariant theories \citep[such as that proposed recently by][]{Skordis_2019} have any flexibility in the coupling between peculiar and Hubble flow accelerations, at least when we impose other constraints, e.g. that gravitational waves travel at $c$. Since $g_{\mathrm{Hubble}}$ acts to suppress the void gravity and increases with distance, the outer density profile of the KBC void could empirically constrain the coupling. Our best-fitting model implies that at present, $g_{\mathrm{ext}} \ll g_{\mathrm{void}} \ll a_{_0}$ at $300\,\rm{Mpc}$, implying that $g_{\mathrm{void}} \appropto r^{-1}$ in the outer part of the void. Since $g_{\mathrm{Hubble}} > g_{\mathrm{void}}$, a strong background coupling would make the system more Newtonian, causing therewith a steeper decline of gravity with distance. However, current measurements of the KBC void's outer density profile are not sufficiently precise to strongly constrain the HFE.

If a particular cosmological MOND model predicts that structure formation on 300~cMpc scales is very similar to standard cosmology, such a model would not be able to account for the KBC void $-$ and would have to be rejected for similar reasons to $\Lambda$CDM~(Section \ref{sec:LCDM framework}). Any viable covariant formalism of MOND has to describe the density and velocity field of the local Universe. Interestingly, the here applied approach can reproduce the KBC void, which solves therewith the Hubble tension due to mass conservation. Therefore, our model can serve to guide further theoretical development of MOND in a cosmological context.

The $g_{\mathrm{Hubble}}$ term would be much larger in the CMB era. For the sound horizon scale at recombination \citep[$147.09\pm0.26\,\rm{cMpc}$,][]{Planck_2018}, $\lvert g_{\mathrm{Hubble}} \rvert \approx 35000 \, a_{_0}$. Thus, even a slight background-perturbation coupling would completely eliminate any MOND effects, making the pre-CMB universe purely Newtonian. This is also true in the absence of any such coupling, since the gravity from inhomogeneities is $\approx 21 \, a_{_0}$ (Section \ref{subsubsec:Radiation dominanted era and CMB}). Thus, our conclusions regarding the CMB are not affected by a possible HFE.

This is also true for the model of \citet{Zhao_2008}, which implies that the MOND acceleration scale $a_\dagger$ is time-dependent, with
\begin{eqnarray}
	a_{_\dagger} ~=~ a_{_0} a^{-3/2} \, .
	\label{eq:a_0_time_dependent}
\end{eqnarray}
At the time of recombination, $a_\dagger = 36500 \, a_{_0}$, which exceeds $\lvert g_{\mathrm{Hubble}} \rvert$ at that epoch. Thus, even in the presence of a very strong coupling of perturbations to $g_{\mathrm{Hubble}}$, his model implies significant MOND effects in the CMB, making it very difficult to fit its power spectrum. Furthermore, \citet{Milgrom_2017} showed that the model of \citet{Zhao_2008} is in tension with the rotation curves of galaxies at high redshift.

In this contribution, we assumed that Milgrom's constant $a_{_0}$ is constant in space and cosmic time. While the former is expected in a fundamental theory, a time-varying $a_{_0}$ is in principle quite possible. Although this is observationally not supported at the moment \citep[e.g.][]{Milgrom_2017}, $a_{_0} \approx c H_{0}/\left( 2\mathrm{\pi} \right)$, which could be a hint that MOND is fundamentally related to cosmology. If this relation is true, $a_{_0}$ would decrease over cosmic time, implying that the early universe was more MONDian than assumed in our models. Thus, postulating that $a_{_0} \propto H$ would cause strong MOND effects in the CMB, arguing against the idea (Section \ref{subsubsec:Radiation dominanted era and CMB}). On the other hand, a much lower $a_{_0}$ in the past would significantly raise the MOND timing argument mass of the LG, which for a constant $a_{_0}$ ends up rather similar to its baryonic mass \citep{Banik_2018a}. Since HDM should not significantly cluster on such a small scale to avoid disrupting MOND fits to galaxy rotation curves, it appears that very strong time evolution of $a_{_0}$ in either sense is ruled out empirically if not theoretically.

MOND as currently understood cannot explain the CMB power spectrum and the dynamics of galaxy clusters without an extra matter component. Therefore, we follow \citet{Angus_2009} in postulating the existence of HDM. This is another strong assumption of our model $-$ but not more hypothetical than the existence of CDM. If anything, sterile neutrinos have been described as ``almost part of the standard model'' of particle physics \citep{Merle_2017}, while CDM particles are generally thought to require supersymmetry. In future, it will be very important to directly search for sterile neutrinos in terrestrial experiments.

In addition, precise measurements of the CMB power spectrum at $\ell > 4900$ could put strong constraints on our model. This is because small shifts to the CMB power spectrum may arise when applying MOND supplemented with sterile neutrinos rather than Newtonian gravity with CDM. We have assumed that the effects are either not detectable or can be compensated through small adjustments to the cosmological parameters.

Furthermore, we modelled the gravitational field of the void using non-relativistic equations. This should be quite accurate because the total peculiar velocities are $v_{\mathrm{tot}} \la 0.01 c$ within $\approx 250\,\rm{Mpc}$ of the void centre (Figure~\ref{figure_peculiar_velocity_results}). Moreover, the void is much smaller than the cosmic horizon. Gravity travel time effects should thus not be too significant if gravitational waves travel at $c$, as occurs in the model of \citet{Skordis_2019}.

The exact density profile of the KBC void is not fully known, so further assumptions are required when modelling it. Motivated by observations of an increasing density as one goes inwards for distances $\la 40\,\rm{Mpc}$ \citep{Karachentsev_2018}, we assumed an initial Maxwell-Boltzmann profile (Section~\ref{sec:Results of MOND simulation}). We demonstrated the robustness of our results by also implementing Gaussian and exponential void profiles (Appendix~\ref{Appendix:Joint probabilities for different void profiles}). The best-fitting models with those profiles yield a slightly larger overall tension compared to our main analysis, with the best parameters shifting to a stronger EF and larger initial void with comparable depth (Table~\ref{table:comparison_bestfittingmodels_with_observations}). Thus, other void profiles could yield reasonable results with adjusted EF and void parameters. This issue could be constrained with better knowledge regarding the exact density profile of the KBC void. 

Although the here presented MOND approach suffers from theoretical uncertainties especially with regards to the HFE, the encouraging results of our best-fitting models (Section~\ref{subsec:Results The best-fitting model} and Appendix~\ref{Appendix:Joint probabilities for different void profiles}) suggest that our assumptions are reasonable. Furthermore, our models allow a wide range of possible void parameters (Figs.~\ref{figure:marginalization_MBprofile}, \ref{figure_MOND_GAUSSIAN_results}, and \ref{figure_MOND_EXP_results}), so adjusting these could in principle compensate theoretical uncertainties. In particular, a stronger HFE would require an initially deeper void $-$ though we argued that the required depth would be reasonable even under conservative assumptions.

Once it is clear which covariant MONDian framework should be applied, the role of the HFE (if any) would become apparent. It would then be valuable to statistically quantify the existence of the KBC void within a large numerical MONDian cosmological simulation, enabling a comparison with our analysis of its likelihood in $\Lambda$CDM (Section~\ref{sec:LCDM framework}). Such a simulation would also deliver a better understanding of void profiles at low redshift, and on how the growth of structure is regulated by the EFE from surrounding structures (Section~\ref{subsubsec:Structure formation in MOND}).

\subsection{Claimed problems for a local void solution to the Hubble tension} \label{subsec:Claimed_problems}

In the following, we address some commonly used arguments for why a void model cannot resolve the Hubble tension.

\subsubsection{Other anomalies in large-scale structure} \label{Other_voids}

If the growth of structure is much more rapid than predicted by standard theory, large underdensities such as the KBC void should also exist at higher redshift. Large voids are not evident in the galaxy two-point correlation function, suggesting that large-scale structure seems to be consistent with the $\Lambda$CDM paradigm. However, it must be borne in mind that the underlying matter density field is not measured directly $-$ it is estimated from the distribution of galaxies. At large distances, only the brightest galaxies can be observed, so one has to assume the so-called bias factor:
\begin{eqnarray}
	b ~\equiv~ \frac{\delta_{\mathrm{galaxy}}}{\delta} \, ,
	\label{eq:bias_factor}
\end{eqnarray}
which relates the galaxy density contrast $\delta_{\mathrm{galaxy}}$ to that of the underlying matter distribution. This bias factor is typically chosen to match the $\Lambda$CDM expectation for cosmic variance at the relevant scale. Thus, an accurate model-independent estimation of the density contrast can only be achieved for low-mass galaxies observed in the NIR, for which $b \approx 1$ on 100 Mpc scales in any cosmological model. This makes it very difficult to perform a similarly detailed analysis to \citet{Keenan_2013} for $z \ga 0.5$. Even the 2MASS survey they used only covers $57-75 \%$ of the luminosity function (see their figure~9). This fraction would be much lower for higher redshift galaxy samples.

We are also faced with the problem that galaxy positions are in general unknown $-$ in the distant Universe, only redshifts are available. Even the redshifts are often not measured directly but are estimated photometrically. This leads to a significant smearing effect along the line of sight, making it rather difficult to identify distant supervoids \citep{Kovacs_2019}. The situation is reminiscent of the LG satellite planes $-$ due to distance uncertainties, it is difficult to know if the satellites of a distant galaxy are distributed anisotropically. In both cases, if similar anomalies had not been reported at larger distances, this would not tell us whether such anomalies exist.\footnote{A satellite plane has recently been discovered around Cen A \citep{Mueller_2018}.}

However, supervoids identified in the Dark Energy Survey do seem to show an enhanced ISW effect \citep{Kovacs_2019}. Their stacked analysis of 87 supervoids found that the effect has an amplitude of ${5.2 \pm 1.6}$ times the conventional expectation when combined with the earlier results of \citet{Kovacs_2018}.

Another possibly related anomaly is that the lensing amplitude implied by the CMB power spectrum is stronger than predicted \citep{DiValentino_2019}. They suggested that this problem could instead be an indication that the Universe has a positive curvature. This would have serious implications for our entire understanding of the Universe and require a completely different cosmological model. For instance, a closed universe would imply a very low $H_0$ of $54^{+3.3}_{-4.0} \, \rm{km\,s^{-1}\,Mpc^{-1}}$, which is completely inconsistent with local measurements (see their figure~7). However, the enhanced lensing amplitude $-$ evident also in \citet{DiValentino_2020} $-$ could be the imprint of unexpectedly large density fluctuations caused by more rapid growth of structure. In this scenario, a supervoid would be a more likely explanation for the CMB Cold Spot \citep{Nadathur_2014}. Indeed, their suggested void profile has a central underdensity of 0.25 and characteristic size of 280 Mpc, rather similar to the KBC void (see their equation~1). They concluded that such a void is highly unlikely in $\Lambda$CDM, as also implied by our results in Figure~\ref{figure_LCDM_results}. It would be very valuable to empirically determine the actual frequency of such voids.

\subsubsection{High peculiar velocities} \label{subsubsec:KSZ effect}

Our fiducial model implies the existence of void regions with total peculiar velocity $v_{\mathrm{tot}} \leq v_{\mathrm{LG}} = 627 \, \rm{km\,s^{-1}}$ relative to the surface of last scattering. Such low velocities are unlikely but allowed at the $2.34\sigma$ level.

Interestingly, our model predicts $v_{\mathrm{tot}}$ of up to $\approx 4000 \, \rm{km\,s^{-1}}$ for objects $\approx 250 \, \rm{Mpc}$ from the void centre in the direction of its motion (Figure~\ref{figure_peculiar_velocity_results}). Such high peculiar velocities are potentially detectable with the kinematic Sunyaev-Zeldovich (kSZ) effect of galaxy clusters \citep{Sunyaev_1980}. However, this is difficult to disentangle from the thermal Sunyaev-Zeldovich (tSZ) effect because our model predicts similar peculiar velocities to the internal velocity dispersions of galaxy clusters. A large local underdensity also reduces the number of clusters available for kSZ studies, increasing the uncertainty.

\citet{Hoscheit_2018} concluded that the KBC void is consistent with the linear kSZ effect (see their figure~6). Similar results were obtained by \citet{Ding_2020}. Some evidence for a bulk flow of $\approx 1000 \, \rm{km\,s^{-1}}$ has been found \citep{Kashlinsky_2008, Kashlinsky_2011}. This is broadly consistent with the expected motion of the whole void due to the EFE ($v_{\mathrm{void}} = 1586 \, \rm{km\,s^{-1}}$), though the bulk flow of a smaller region will depend on our exact location within the void and the survey volume. Bulk flows of $\approx 1000 \, \rm{km\,s^{-1}}$ are not possible on a 100 Mpc scale in a $\Lambda$CDM universe, but would be expected in MOND \citep{Katz_2013}.

\subsubsection{Gravitational redshifting of the CMB monopole} \label{subsubsec:CMB_monopole}

A large local underdensity like the KBC void should also affect the mean temperature (monopole) of the CMB. This is because the height of the potential at our location causes a gravitational redshift. Using Equation~\ref{eq:governing_equation_Lambda_ratio_GR}, the general relativistic redshift for a photon travelling uphill from distance $r$ to the centre becomes
\begin{eqnarray}
	1 + z_{\mathrm{GR}} ~=~ \exp \left( \frac{1}{c^2} \int g \, dr \right) \, .
	\label{eq:gravitational_redshift}
\end{eqnarray}
In the best-fitting MOND model, we obtain $z_{\mathrm{GR}} = 8.4\times 10^{-3}$ for the most distant test particle, which is $700\times$ larger than the $1\sigma$ rms fluctuations of $z_{\mathrm{GR}} = 1.2 \times 10^{-5}$ assumed in the study of \citet{Yoo_2019}. Their figure~2 shows that the impact of such a gravitational redshift on the inferred cosmological parameters is very small, even with an extra factor of 700. Moreover, we expect that the actual gravitational redshift at our position in the void should be much smaller. This is because we are not exactly at the centre of the void, and thus not at the highest part of its gravitational potential hill (Figure~\ref{figure_peculiar_velocity_results}). Redshifting from the void's gravity would also be partially counteracted by the EFE, which is required in order to explain the rather slow motion of the LG wrt. the CMB (Section~\ref{subsubsec:LG_vpec}).

None the less, it is possible that gravitational redshifting of the CMB would change the best-fitting HDM and baryon fractions by a few times their official uncertainties. Since these are nowadays rather small \citep{Planck_2018}, we conclude that this effect has only a small impact on the CMB power spectrum, which moreover could probably be compensated through slight adjustments to the cosmological parameters. Of particular relevance for the Hubble tension is that gravitational redshifting of the CMB has a negligible impact on the precisely measured angular scale of the first acoustic peak \citep[figure~2 of][]{Yoo_2019}.

\subsubsection{Assumption of Newtonian gravity for the void dynamics} \label{subsubsec:Assumption of Newtonian gravity for the void dynamics}

Applying Newtonian gravity to the dynamics of any void would lead to sharp gradients in the predicted density and velocity profiles due to the steep inverse square law. \citet{Kenworthy_2019} found no evidence for such a sharp edge in the SN luminosity-distance relation, which would $-$ according to them $-$ rule out the existence of a large local void with $\delta > 0.2$ at the $4\sigma-5\sigma$ confidence level. Also, \citet{Hoscheit_2018} applied the large scale void radial profile of \citet{Keenan_2013} to show that $H_{0}$ is $1.27 \pm 0.59 \, \rm{km\,s^{-1}\,Mpc^{-1}}$ higher in the redshift range $0.0233 < z< 0.07$ compared to $0.07 < z < 0.15$. This only modestly reduces the Hubble tension, e.g. with the SNe data of \citet{Riess_2016}. 

However, as shown in Section~\ref{subsec:Comparison with observations}, an inverse square law is too weak to produce a deep and extended underdensity like the KBC void. Therefore, the assumption of Newtonian gravity for the void dynamics is not sustainable. In MOND, the long-range modification to gravity would cause a much more gradual return from the void-induced peculiar velocities to the background cosmology, as demonstrated in Figure~\ref{figure_best_fitting_model_scalefactor}. Therefore, sharp features in the density profile and Hubble diagram are not expected in a MONDian model. This holds especially for $H_{0}$ derived from SNe data because in order to constrain the cosmological model, one has to consider many individual SNe. Consequently, the inferred $H_0$ only gradually declines towards the Planck prediction as SNe beyond the void are included in the analysis \citep[as e.g. done by][]{Colgain_2019}.

\subsubsection{Restrictive upper limit on the void size} \label{subsubsec:Restrictive upper limit on the void size}

The present void size can be treated as a model parameter independently of the applied gravity theory. Adopting a very low upper limit on the allowed void size would unavoidably cause sharp features in the density and velocity profiles in any framework. Moreover, the Hubble tension cannot be resolved by a small void unless we postulate that it is extremely deep. This issue affected the analysis of \citet{Wu_2017}, who assumed a void size of $180 \, \rm{Mpc}$. They noticed that since the SNe data go out much further, it is difficult for such a small void to resolve the Hubble tension. None the less, they did not consider a larger void, opting instead for a very large density contrast of $\delta = 0.8$. This led to poor agreement with direct measurements of the density field. However, a larger and shallower void would have provided much better agreement with observations, as shown in this work.

Using the high-resolution $\Lambda$CDM $N$-body cosmological simulation called Millennium-II, \citet{Xie_2014} obtained that $\approx 14 \%$ of LG-like systems are located in a region that resembles the observed local void. Thus, they concluded that ``the emptiness of the Local Void is indeed a success of the standard $\Lambda$CDM theory.'' However, by ``Local Void'', they meant a sphere of radius $\approx 8 \, \rm{Mpc}$, which is much smaller than the KBC void. Thus, their work cannot be used as an argument that the local void observed by \citet{Keenan_2013} is consistent with $\Lambda$CDM cosmology, as is done in section~5 of \citet{Sahlen_2016}.

\subsubsection{Fixing the acceleration parameter} \label{subsubsec:Fixing the deceleration parameter}

The acceleration parameter $\overline{q}_{_0}$ describes the second time derivative of the scale factor (Equation~\ref{eq:q0_new}). It is therefore a measure of the void's gravity. As discussed in Section~\ref{subsubsec:Assumption of Newtonian gravity for the void dynamics}, \citet{Kenworthy_2019} concluded that the KBC void is not evident in the SN luminosity-distance relation. In addition to assuming Newtonian gravity, they fixed the acceleration parameter to the $\Lambda$CDM prediction of $\overline{q}_{_0} = 0.55$. In general, $\overline{q}_{_0}$ would have a higher value if there is a large local void. To allow for this possibility, $\overline{q}_{_0}$ must be treated as a free parameter when using the apparent expansion rate history to constrain the properties of a local void.

Fortunately, \citet{Camarena_2020} address this shortcoming by deriving $\overline{q}_{_0}$ and $H_{0}^{\mathrm{local}}$ jointly from SNe data without a restrictive choice of prior. Their analysis yields $\overline{q}_{_0} = 1.08 \pm 0.29$, much higher than in the Planck cosmology. A high $\overline{q}_{_0}$ is also evident when using BAO data or treating the SNe Ia absolute magnitude as a free parameter \citep{Camarena_2020b}. This is a strong hint for the existence of a local void independently of the galaxy luminosity density \citep[e.g.][]{Keenan_2013}. Indeed, \citet{Colgain_2019} inferred a local underdensity at $z \la 0.15$ using SNe data alone. Both the local Hubble constant and acceleration parameter can be explained in our best-fitting MOND models (Figure~\ref{figure_H0_q0_plane}).

\subsubsection{Effect of the void at high redshift} \label{subsubsec:Effect of the void at high redshift}

\begin{figure}
    \includegraphics[width=86mm]{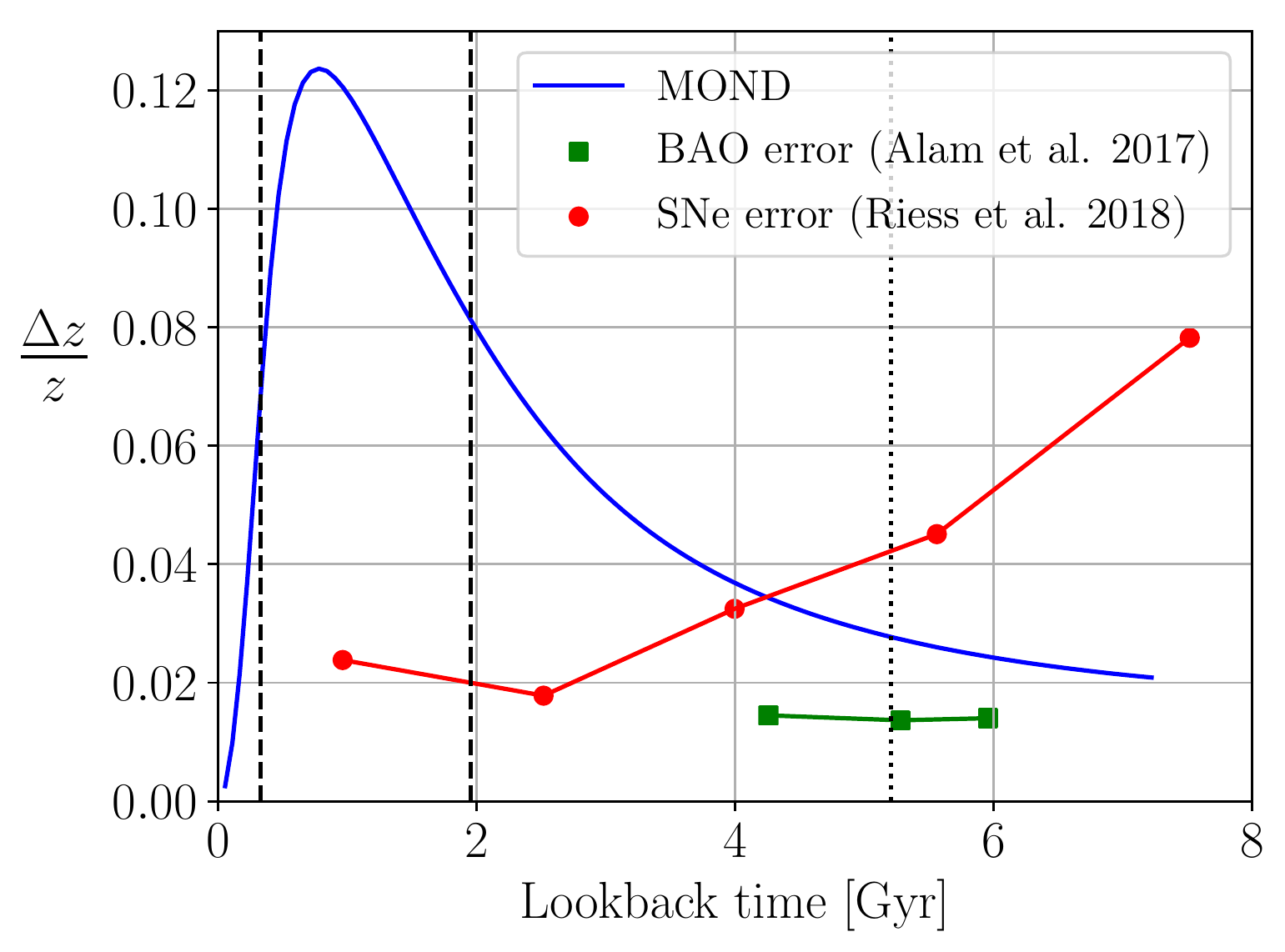}
    \caption{Difference in the redshift between our best-fitting model and a standard void-free cosmology (latter subtracted), shown as a function of lookback time. The first two dashed vertical lines show the range $0.023 \leq z \leq 0.15$ covered by \citet{Camarena_2020}. The dotted vertical line marks $z = 0.5$. For comparison, the red points and green squares show fractional distance uncertainties using SNe \citep[table~6 of][]{Riess_2018} and BAO \citep[table~8 of][]{Alam_2017}, respectively. Since fractional redshift errors are generally rather small, this gives an estimate of the uncertainty in the inferred Hubble constant, which has similar sensitivity to redshift and distance.}
    \label{figure_best_fitting_model_redshift_uncertainty}
\end{figure}

The apparent expansion rate history in our fiducial MOND model is very similar to the Planck cosmology (Figure~\ref{figure_best_fitting_model_scalefactor}). None the less, the fractional difference in $z$ between void and void-free cosmologies (i.e. $\Delta z/z$) reaches the 12\% level and is sufficient to solve the Hubble tension (Figure~\ref{figure_best_fitting_model_redshift_uncertainty}). Observations at higher $z$ could distinguish our void model from other possible solutions, e.g. miscalibrated SNe, early dark energy, etc. This is because a local void predicts that the inferred $H_{0}$ decreases with the redshift of the data set used, and asymptotically approaches the Planck prediction (Figure~\ref{figure:best_fitting_model_Hubble_vs_redshift}).

Unfortunately, Figure~\ref{figure_best_fitting_model_redshift_uncertainty} shows that high-redshift SNe currently do not pose strong constraints on our model. This is because the simulated $\Delta z/z$ decreases with redshift, while the uncertainty of binned SNe distance measurements to fixed $z$ increases for $z \ga 0.2$. It crosses the simulated $\Delta z/z$ curve at $z \approx 0.38$, corresponding to a lookback time of $\approx 4.2\,\rm{Gyr}$. At $z \approx 0.5$ (the dotted vertical line in Figure~\ref{figure_best_fitting_model_redshift_uncertainty}), the predicted $\Delta z/z \approx 3\%$, but the observational uncertainty is much larger \citep[see also][]{Cuceu_2019, Macaulay_2019}. As a result, even the 9\% Hubble tension could not be reliably detected in SNe at these redshifts, which is reasonable given that the uncertainty of $H_{0}$ derived from SNe at $0.023 \leq z \leq 0.15$ is already $\approx 2\%$ \citep{Camarena_2020}. Moreover, there are much fewer observed SNe at high redshifts, which could increase systematic errors.

In contrast to high-redshift SNe, the current BAO precision \citep{Alam_2017} lies slightly below the predicted $\Delta z/z$. However, the uncertainties are still too large to distinguish our model from a void-free Planck cosmology at high significance. We note that BAO-based $H_{0}$ measurements \citep{Alam_2017, Zhang_2019} are very close to the Planck prediction (Section~\ref{subsec:Hubble tension}), which is consistent with our void model. The small excess it predicts can only be confirmed or ruled out with more precise observations.

In conclusion, it is currently difficult to distinguish void and void-free models with data only at $z \ga 0.5$. Data at lower $z$ are more useful in this regard. In particular, the redshift range $0.023 \leq z \leq 0.15$ covered by \citet{Camarena_2020} brackets the peak of the simulated $\Delta z/z$ curve and poses therewith a strong test of our model. In this redshift range, our void model differs from $\Lambda$CDM by $\Delta z/z = 7-12\%$, which is quite consistent with the $9.5\%$ difference between local and early universe measures of $H_0$. Note that the position of the peak in $\Delta z/z$ depends on the underlying void profile $-$ it would occur at the void centre for a Gaussian or an exponential profile.

\subsubsection{CMB contamination by intergalactic dust} \label{subsubsec:CMB contamination by intergalactic dust}

Finally, we consider the possibility that the CMB is contaminated by intergalactic dust, which in turn would affect the $H_{0}^{\mathrm{global}}$ required by \citet{Planck_2018}. Some distant foreground emission can increase $H_{0}^{\mathrm{global}}$ \citep{Yershov_2020}, which would slightly reduce the mild tension between our model and the strong lensing data (Figure~\ref{figure:best_fitting_model_Hubble_vs_redshift}). This is because the $z > 0.4$ lenses all give $H_0$ systematically above the Planck prediction by a similar extent. But changing the Planck $H_0$ would not explain the high inferred $H_0$ from the two lowest $z$ lenses, which would continue to hint at a local void. 

It is also possible that the CMB is more substantially affected by dust. In contrast to the $\Lambda$CDM model in which the CMB is explained as relic radiation from the early Universe at $z~\approx~1100$ \citep[e.g.][]{Bennett_2003, Planck_2018}, an alternative model is that the entire CMB is thermal radiation from intergalactic dust particles heated up by starlight \citep{Vavrycuk_2018}. 
Assuming the observed intergalactic dust is in thermal equilibrium with the radiation field from galaxies, the model of \citet{Vavrycuk_2018} implies a dust temperature of $T_{\mathrm{D}} = 2.776\,\rm{K}$. This is only slightly higher than the measured $T_{\mathrm{CMB}} = 2.72548 \pm 0.00057\,\rm{K}$ \citep{Fixen_2009}. The exact value of $T_{\mathrm{D}}$ depends on the amount of intergalactic dust and the intergalactic opacity ratios, which are both poorly known observationally. In future, it would be important to study which dust parameters are necessary to match the observed $T_{\mathrm{CMB}}$. In other words, it would be important to quantify the uncertainty on $T_{\mathrm{D}}$, which was not explicitly addressed by \citet{Vavrycuk_2018}. Furthermore, it needs to be demonstrated that the model can yield the observed perfect black body spectrum within rather small uncertainties \citep{Planck_2013}, and also yield nearly Gaussian temperature fluctuations \citep{Planck_2014_XXIII}. It is therefore possible that intergalactic dust significantly affects the CMB, but a detailed consideration of such a scenario is beyond the scope of our work. Interestingly, \citet{Vavrycuk_2019} showed in a subsequent study that the anomalous dimming of SNe Ia can in principle be explained by light extinction due to intergalactic dust.

In addition to heating by starlight, dust grains would also be heated by the primordial CMB, especially at high $z$. This may have caused rethermalization of the CMB by dust from the first stars at ${z \approx 15}$. In this scenario, the angular scale evident in BAO measurements corresponds to a different co-moving length than the sound horizon at the time of last scattering. However, agreement can be recovered if we assume a non-standard background cosmology where $a \propto t$ \citep{Melia_2020}. In fact, it is not possible that $a \propto t$ without such a late rethermalization of the primordial CMB \citep{Fujii_2020}. While the late-time expansion history is indeed approximately of this form (Figure \ref{figure_best_fitting_model_scalefactor}), the model does not yet explain the nature of the acoustic oscillations in the CMB power spectrum. Moreover, BBN would be modified to a very substantial extent, making it difficult to explain the observed light element abundances \citep{Lewis_2016}.

\section{Conclusions} \label{sec:Conclusion}

Cosmic structure $-$ and therewith the distribution of galaxies $-$ provide strong constraints on the underlying cosmological model. In this context, we used the framework of the standard $\Lambda$CDM theory and MOND \citep{Milgrom_1983} to investigate the KBC void, a large underdensity with a relative density contrast of $\delta \equiv 1 - \rho/\rho_{0} = 0.46 \pm 0.06$ between $z = 0.01$ and $z = 0.07$ \citep{Keenan_2013}. A large local underdensity is evident throughout the whole electromagnetic spectrum (Section~\ref{subsec:KBC void}).

Using the MXXL simulation \citep{Angulo_2012}, we showed that the KBC void is in $6.04 \sigma$ tension with standard cosmology (Section~\ref{KBC void in LCDM}). In principle, if mass conservation is assumed, such an immense void should also show up in the velocity field, and would approximately solve the Hubble tension (Equation~\ref{eq:H_0_impact}). This tension nowadays exceeds the $5 \sigma$ threshold based on numerous independent techniques (Section \ref{subsec:Hubble tension}). However, we demonstrated that a $10 \sigma$ density fluctuation would be necessary to solve the Hubble tension at the $2 \sigma$ level (Figure~\ref{figure_LCDM_results_cosmic_variance}). This is due to the very small expected cosmic variance in $\Lambda$CDM \citep[e.g.][]{Macpherson_2018}. The most likely explanation in this framework is that both the KBC void and Hubble tension are caused by measurement errors. However, the measurements rely on very different observational techniques. For instance, a zero-point error in SNe Ia fluxes would change the inferred $H_0$ but would not affect the KBC void. Thus, both phenomena would independently falsify $\Lambda$CDM at more than $5 \sigma$ confidence, yielding a combined tension of $7.75 \sigma$. Taking into account the cosmic variance expected in $\Lambda$CDM, both tensions are not completely independent, reducing the combined tension with standard cosmology to $7.09 \sigma$ (Section~\ref{Combined implications for LCDM}). The $\Lambda$CDM model (or any dark-matter-based Einsteinian/Newtonian cosmological model) is thus rigorously ruled out by the data on kpc, Mpc, and Gpc scales \citep[see also][]{Kroupa_2012, Kroupa_2015}.

As discussed in Section~\ref{subsubsec:Hubble tension discussion}, an early change in the expansion history is unlikely to solve the Hubble tension, and would in any case not explain the KBC void. Importantly, we argued that the locally measured Hubble constant is very similar to the Planck prediction in $\Lambda$CDM once the KBC void is accounted for (Equation~\ref{eq:H_0_impact}). Our results thus support the Planck cosmology at the background level and in the early Universe.

However, a deep and large void such as the KBC void implies that the growth of structure must be more rapid than predicted by standard theory. This would also fit into the picture obtained by \citet{Peebles_2010}, who concluded that the structure of the Local Volume with its void and sheet on an $8 \, \rm{Mpc}$ scale points to a faster growth rate of cosmic structure. Since gravity is the dominant force on these scales, it is very likely that gravity has to be enhanced at long range (Figure~\ref{figure_KKBC_void_context}).

Consequently, we aimed to study the KBC void and its velocity field in MOND, an acceleration-dependent modification of Newtonian gravity. MOND was originally designed to explain the dynamical discrepancies in galaxies without the need of CDM \citep{Milgrom_1983}. Unfortunately, there is currently no $N$-body or hydrodynamical cosmological MOND simulation large enough to quantify the likelihood of a KBC void, as done for the $\Lambda$CDM framework. Therefore, we developed a semi-analytic approach based on the \citet{Angus_2009} cosmological model, which relies on MOND supplemented by sterile neutrinos with a mass of $m_{\nu_{s}} = 11 \, \rm{eV}/c^{2}$. We call this the $\nu$HDM framework, where $\nu$ refers to both the interpolating function in QUMOND \citep{Milgrom_2010} and sterile neutrinos as an HDM component. The energy budget would be similar to the $\Lambda$CDM cosmology, with a baryonic matter density of $\Omega_{\mathrm{b},0} \approx 0.05$, a sterile neutrino density of $\Omega_{\mathrm{\nu_s},0} \approx 0.25$, and a dark energy density of $\Omega_{\mathrm{\Lambda},0} \approx 0.7$ at the present time (Section~\ref{subsec:nuHDM cosmological model}).

This paradigm is mainly motivated by a sample of $30$ virialized galaxy groups and clusters which all reach the Tremaine-Gunn limit for sterile neutrinos with $m_{\nu_{s}} = 11 \, \rm{eV}/c^{2}$ \citep{Angus_2010}. Moreover, the model provides a viable explanation of the thermal history of the Universe. As discussed in Section~\ref{subsubsec:Radiation dominanted era and CMB}, an extra sterile neutrino species is consistent with the standard model of particle physics (which accommodates neutrino oscillations) and only slightly affects the nucleosynthesis era $-$ it would raise the primordial helium abundance from $Y_{\mathrm{p}} = 0.247$ to $Y_{\mathrm{p}} = 0.259$. Measurements of high-redshift metal-poor gas clouds backlit by quasars do not rule this out \citep{Aver_2012, Cooke_2018b}. At $z = 1100$, sterile neutrinos with $m_{\nu_{s}} = 11 \, \rm{eV}/c^{2}$ have a free-streaming length of $\approx 3.5 \, \rm{cMpc}$, implying that they would only affect multipoles ${\ell \ga 4900}$ in the CMB power spectrum, beyond the range of Planck. This is confirmed in section~6.4.3 of \citet{Planck_2015}, which states that sterile neutrinos with $m_{\nu_s} > 10 \, \rm{eV}/c^2$ are indistinguishable from CDM in Planck measurements of the power spectrum. Importantly, typical accelerations at the CMB would exceed $a_{_0}$, causing structure formation to be little affected by MOND until $z \la 50$ (Equation~\ref{eq:EFE_CMB_2}). Furthermore, the $\nu$HDM model closely recovers the standard expansion history \citep{Angus_2009}, which is currently favoured by observations \citep[e.g.][]{Joudaki_2018}. This is because not only the overall matter content but also the Friedmann equation should be very similar to $\Lambda$CDM \citep{Skordis_2006}.

Within this framework, we developed a semi-analytical model with the usual ansatz that density perturbations obey Milgrom's law of gravity \citep[e.g.][]{Angus_2013, Katz_2013, Candlish_2016}, but a standard background cosmology applies (Section~\ref{subsec:nuHDM cosmological model}). In particular, we adopted a background Hubble constant of $H_{0}^{\mathrm{global}} = 67.4 \, \rm{km\,s^{-1}\,Mpc^{-1}}$, $\Omega_{\mathrm{m},0} = 0.315$, and $\Omega_{\mathrm{\Lambda},0} = 0.685$ \citep{Planck_2018}. In this way, we extrapolated Milgrom's gravitational theory from sub-kpc to Gpc scales without further theoretical assumptions to specifically address the local density and velocity field (Section~\ref{subsec:Governing equations}). 

Our simulation starts at $z = 9$ with a void profile parametrized by an initial void size $r_{\mathrm{void}}$ ranging from $50 - 1030 \, \rm{cMpc}$ and an initial void strength $\alpha_{\mathrm{void}}$ ranging from $10^{-5}$ to $10^{-2}$. In our fiducial model, the void is described by a Maxwell-Boltzmann profile (Section~\ref{subsubsec:Void initial profiles}) motivated by the density profile of the Local Volume \citep{Karachentsev_2018}. We also run models with a Gaussian and an exponential initial profile (Appendices~\ref{Appendix: KBC void mass profile} and \ref{Appendix:Joint probabilities for different void profiles}). Furthermore, we vary the present EFE from $g_{\mathrm{ext}} = 0$ up to $0.5 \, a_{_0}$. For our main analysis, we assume that the EFE is constant over cosmic time, but models with different power-law dependencies on the scale factor are also considered (Section~\ref{subsubsec:Structure formation in MOND}). In total, we run $10^{6}$ MOND models for our main analysis (Maxwell-Boltzmann initial profile with time-independent $g_{\mathrm{ext}}$).

We constrain our models with observations of the local Universe, i.e. the inner ($0.01 < z < 0.07$) and outer ($600 - 800 \, \rm{Mpc}$) density contrast of the KBC void \citep[see also figure~11 and table~1 in][]{Keenan_2013}, the local Hubble constant and acceleration parameter derived jointly from SNe in the redshift range $0.023 \leq z \leq 0.15$ \citep{Camarena_2020}, $H_{0}$ measured from seven strong lenses \citep{Wong_2020, Shajib_2020}, and the motion of the LG wrt. the CMB \citep{Kogut_1993}.

Our fiducial MOND model has $g_{\mathrm{ext}} = 0.055 \, a_{_0}$ causing a bulk flow of $v_{\mathrm{void}} = 1586 \, \rm{km\,s^{-1}}$ at $z=0$, $r_{\mathrm{void}} = 228.2 \, \rm{cMpc}$ at $z = 9$, and $\alpha_{\mathrm{void}} = 3.76 \times 10^{-5}$ then. In this model, the density contrast in a $300\,\rm{cMpc}$ sphere grows as $\delta \appropto a^{3.8}$, which is much faster than in $\Lambda$CDM where $\delta \appropto a^{0.8}$ (Figure~\ref{figure_void_density_profile_results}). At the present epoch, our model yields $\delta_{\mathrm{in}} = 0.172$ and $\delta_{\mathrm{out}} = 0.050$, which explains the observed density contrasts after an RSD correction (Section~\ref{subsubsec:Density contrast}) at $0.99\sigma$ and $0.97\sigma$, respectively. The model yields a local Hubble constant of $H_{0}^{\mathrm{model}} = 76.15 \, \rm{km\,s^{-1}\,Mpc^{-1}}$ and an acceleration parameter of $\overline{q}_{_0}^{\mathrm{model}} = 1.07$ in the redshift range $0.023 \leq z \leq 0.15$, consistent with the observations of \citet{Camarena_2020} at the $84.20\%$ confidence level ($0.20 \sigma$ tension, Section~\ref{subsec:Results The best-fitting model}). Similar results are obtained for models initialized with a Gaussian or an exponential profile (Appendix~\ref{Appendix: KBC void mass profile}). Thus, we have shown for the first time that the Hubble tension can be solved in MOND. Several other tensions are also simultaneously resolved, notably the KBC void and that in $\overline{q}_{_0}$ \citep[see also][]{Camarena_2020b}.

While all our best-fitting models generally imply larger peculiar velocities than the observed $v_{\mathrm{LG}}$ of only $627 \, \rm{km\,s^{-1}}$, the possibility that $v_{\mathrm{tot}} \leq v_{\mathrm{LG}}$ cannot be excluded at the $99\%$ confidence level (Section~\ref{subsec:Results The best-fitting model} and Appendix~\ref{Appendix:Joint probabilities for different void profiles}). Thus, we do not require the LG to be at a special position within the KBC void in a statistically significant sense. Our results indicate that we should be $150 - 270$~Mpc from the void centre in roughly the opposite direction to the external field on the void (Figure~\ref{figure_peculiar_velocity_results}).

As we go beyond the void, all our models predict that the inferred $H_{0}$ decreases with redshift. Indeed, observations of strongly lensed quasars taken by the H0LICOW team have shown that $H_{0}$ decreases with the lens redshift at a significance level of $1.9 \sigma$ \citep{Wong_2020}. However, our best-fitting model systematically underestimates the lensing-inferred $H_{0}$, which may be related to the EFE sourced by a massive object beyond the void at $z \ga 0.15$ but closer than the lenses at ${z > 0.3}$ (Section~\ref{subsubsec:Excluding strong lensing time delays}). It could also be a sign of systematic errors \citep{Kochanek_2020}, but the sharp rise for the two nearest lenses is suggestive of a void-induced effect.

Taking into account all observational constraints (Section~\ref{subsec:Observational constraints}), our fiducial MOND model explains these local observations at the $1.14\%$ confidence level, representing $2.53 \sigma$ tension (Section~\ref{subsec:Results The best-fitting model}). The best-fitting MOND models with a Gaussian and an exponential void profile are consistent with observations at $0.45\%$ ($2.84 \sigma$) and $0.34\%$ ($2.93 \sigma$), respectively (Appendix~\ref{Appendix:Joint probabilities for different void profiles}).

Although strong lensing does not occur in the MOND regime \citep{Sanders_1999} and works similarly to standard cosmology (Section~\ref{subsubsec:Hubble constant from strong lensing}), we also redo our analysis without the $H_{0}$ constraints from this method. Our best-fitting model is then consistent with observations at the $5.0\%$ ($1.96 \sigma$) confidence level, with only small changes to the best-fitting parameters (Section~\ref{subsubsec:Excluding strong lensing time delays}). 

Our analysis strongly disfavours models without an EFE, consistent with results from wide binaries \citep{Pittordis_2019}. Furthermore, we showed that allowing time variation of the EFE has only a minor impact on our results because a constant EFE is well within uncertainties (Section~\ref{subsubsec:Structure formation in MOND}). The main effect of allowing a stronger EFE in the past is to raise the required void strength at $z = 9$, with values up to $\approx 10^{-3}$ becoming allowed at $1 \sigma$ (Figure~\ref{figure_EFE_history_results_alpha_n}). This is more in line with the expected cosmic variance at that epoch.

We also discussed structure formation and the implications for the KBC void in MOND if peculiar accelerations are coupled to the Hubble flow acceleration $\bm{g}_{\mathrm{Hubble}}$, as proposed by \citet{Sanders_2001}. Such a coupling (or HFE) would effectively add $g_{\mathrm{Hubble}}$ as an extra source of gravity when calculating the MOND boost to gravity, making the behaviour more nearly Newtonian (Sections~\ref{subsubsec:Large-scale structure} and \ref{subsubsec:Theoretical_assumptions}). However, even a strong HFE implies a significant enhancement to gravity and the formation of voids compared to the Newtonian case. This is because $g_{\mathrm{Hubble}} \approx 0.2 \, a_{_0}$ on a $300\,\rm{Mpc}$ scale, and completely vanished $6\,\rm{Gyr}$ ago (Figure \ref{figure_MOND_gravity_time}). As a result, we conservatively estimated that even with a strong HFE, the cosmic variance in MOND would still be at least $2.8\times$ that in standard $\Lambda$CDM on a 300~Mpc scale ($\approx 9.0\%$ instead of $3.2\%$). This would mean that whereas $\Lambda$CDM needs a $10\sigma$ density fluctuation to simultaneously explain the KBC void and Hubble tension (Figure~\ref{figure_LCDM_results_cosmic_variance}), a MOND cosmology would only need an $\approx 2 \sigma$ fluctuation (Section \ref{subsubsec:Theoretical_assumptions}). Thus, MOND can successfully describe the density and velocity field on a Gpc scale under a wide range of plausible theoretical assumptions on how density perturbations couple to the background cosmology. In principle, the strength of the coupling introduces additional degrees of freedom that could be used to match the observed frequency of KBC-like voids, the observed lensing of the CMB, and the ISW effect. However, it is not clear if a covariant version of MOND has this flexibility when other constraints are imposed, e.g. that gravitational waves should travel at $c$. These theoretical uncertainties should be addressed in future work.

While the MONDian framework provides a reasonable fit to the locally observed density and velocity field, we emphasize that other alternative cosmologies might do so as well. Our results suggest that a successful model should have an expansion history similar to $\Lambda$CDM, but yield significantly more cosmic variance on a $300\,\rm{Mpc}$ scale. Additionally, the model must also accurately describe the dynamics of galaxies in order to provide a holistic explanation of the observed Universe. In this regard, a modification to gravity at length-scales beyond e.g. 10~Mpc would not be sufficient as it would face the same issues as $\Lambda$CDM on galaxy scales.

There are still considerable theoretical uncertainties in the here developed cosmological MOND simulation (Sections~\ref{subsec:nuHDM cosmological model} and \ref{subsubsec:Theoretical_assumptions}) because we lack an understanding of the fundamental theory behind MOND \citep[i.e. FUNDAMOND, ][]{Milgrom_2020b, Milgrom_2020}. Nevertheless, a promising relativistic MOND version was recently developed in which gravitational waves travel at the speed of light \citep{Skordis_2019}. Its implications for cosmology should be explored, though a rather large box size would be required to reach the scale at which the CP holds in a Milgromian universe. This is because in MOND the EFE suppresses the growth of structure, causing structure formation in different regions to become correlated (Section~\ref{subsubsec:Structure formation in MOND}). Without such simulations and/or further analytic work, we cannot draw any strong conclusions on the expected time evolution of the EFE. We nonetheless expect our results to hold because a wide range of possible EFE histories yield reasonable results, and because other void parameters such as its initial size and strength could be adjusted to optimize the fit (Figure~\ref{figure_EFE_history_results_alpha_n}).

Any viable cosmological model has to explain both the local and global Universe. The KBC void is virtually impossible within the $\Lambda$CDM framework (Section~\ref{subsec:Comparison with observations}). Consequently, the $\Lambda$CDM model faces serious challenges on $\rm{Gpc}$ scales, as shown in this contribution $-$ the KBC void and Hubble tension falsify the $\Lambda$CDM paradigm at the $7.09 \sigma$ level, and point towards much more rapid growth of structure than predicted by standard cosmology. Moreover, \citet{DiValentino_2019} reported ``a possible crisis for cosmology'' based on the Planck power spectra, while \citet{DiValentino_2020arxiv} concluded that the $\Lambda$CDM paradigm has to be replaced. These large-scale issues should be addressed together with the severe problems faced by $\Lambda$CDM on galactic scales \citep[e.g. the satellite planes and the RAR, see also][and references therein]{Kroupa_2015}.

Previous studies have shown that MOND is successful on several astrophysical scales ranging from the equilibrium dynamics of galaxies \citep{Famaey_2012} and their formation out of gas clouds \citep{Wittenburg_2020}, to the equilibrium dynamics of virialized galaxy clusters \citep{Angus_2013}, and the formation of extreme clusters like El Gordo \citep[e.g.][]{Katz_2013}. The cluster-scale successes require the assumption of sterile neutrinos as HDM, which allows MOND to produce a standard expansion history and have very little effect on BBN and the high-acceleration CMB (Section~\ref{subsec:nuHDM cosmological model}). Consequently, there exist only very few (if any) scales at which the $\Lambda$CDM framework provides a unique explanation for the observations. Rather, observations of the local and global Universe strongly suggest that we should replace $\Lambda$CDM with the $\nu$HDM framework, which relies on MOND and sterile neutrinos. 

The encouraging results we obtained using this approach should be put on a more secure theoretical footing using a covariant framework such as that of \citet{Skordis_2019}. In particular, it is important to rigorously demonstrate that the background cosmology behaves like in $\Lambda$CDM at the sub-per cent level. A covariant framework would also clarify if there is any coupling between the Hubble flow acceleration and that sourced by inhomgeneities. If there is and if its strength is adjustable, the value could be found empirically using numerical simulations of large-scale structure. Calculating photon propagation through the resulting time-varying inhomogeneous gravitational field would then allow comparison with the observed lensing of the CMB by intervening structures, and the resulting ISW effect \citep{Buchert_2001, Wiltshire_2007}. Although these both appear to be underestimated in the $\Lambda$CDM framework (Section \ref{Other_voids}), they may be overestimated in $\nu$HDM. In this context, it is worth mentioning that the CMB Cold Spot could be caused by a KBC-like void \citep{Nadathur_2014}. The expected frequency of such voids should be quantified using numerical simulations, which would also account for more complicated effects such as non-sphericity of the void. This may lead to predictions for angular dependence of the apparent expansion rate, which could be contrasted with observations \citep[e.g. those of][]{Migkas_2020}.

We conclude that unlike $\Lambda$CDM as presently understood, MOND supplemented by HDM appears to be a promising way to explain observations across all astrophysical scales. In particular, we expect this $\nu$HDM model to yield an almost standard expansion history but with enhanced cosmic variance on a 300~Mpc scale, allowing it to explain the observed KBC void and therewith the Hubble tension. This scenario has to be investigated in an open-minded manner in future studies.

\section*{Data availability}

The data underlying this article are available in the article.

\section*{Acknowledgements}

 IB is supported by an Alexander von Humboldt Foundation postdoctoral research fellowship. We are grateful to Karl Menten for his support and helpful suggestions. We would also like to thank Raul Angulo for providing data from the MXXL simulation, and David Camarena for providing the observational correlation coefficient between their inferred $H_{0}$ and $\overline{q}_{_0}$. We thank the referee for her/his useful comments to improve especially the theoretical aspects of this publication.

\bibliographystyle{mnras}
\bibliography{KBC_bbl}

\begin{appendix}

\section{Gaussianity of the \texorpdfstring{$\Lambda$CDM}{LCDM} density fluctuations} \label{Appendix_normality_tests}

\begin{figure*}
    \includegraphics[width=58mm]{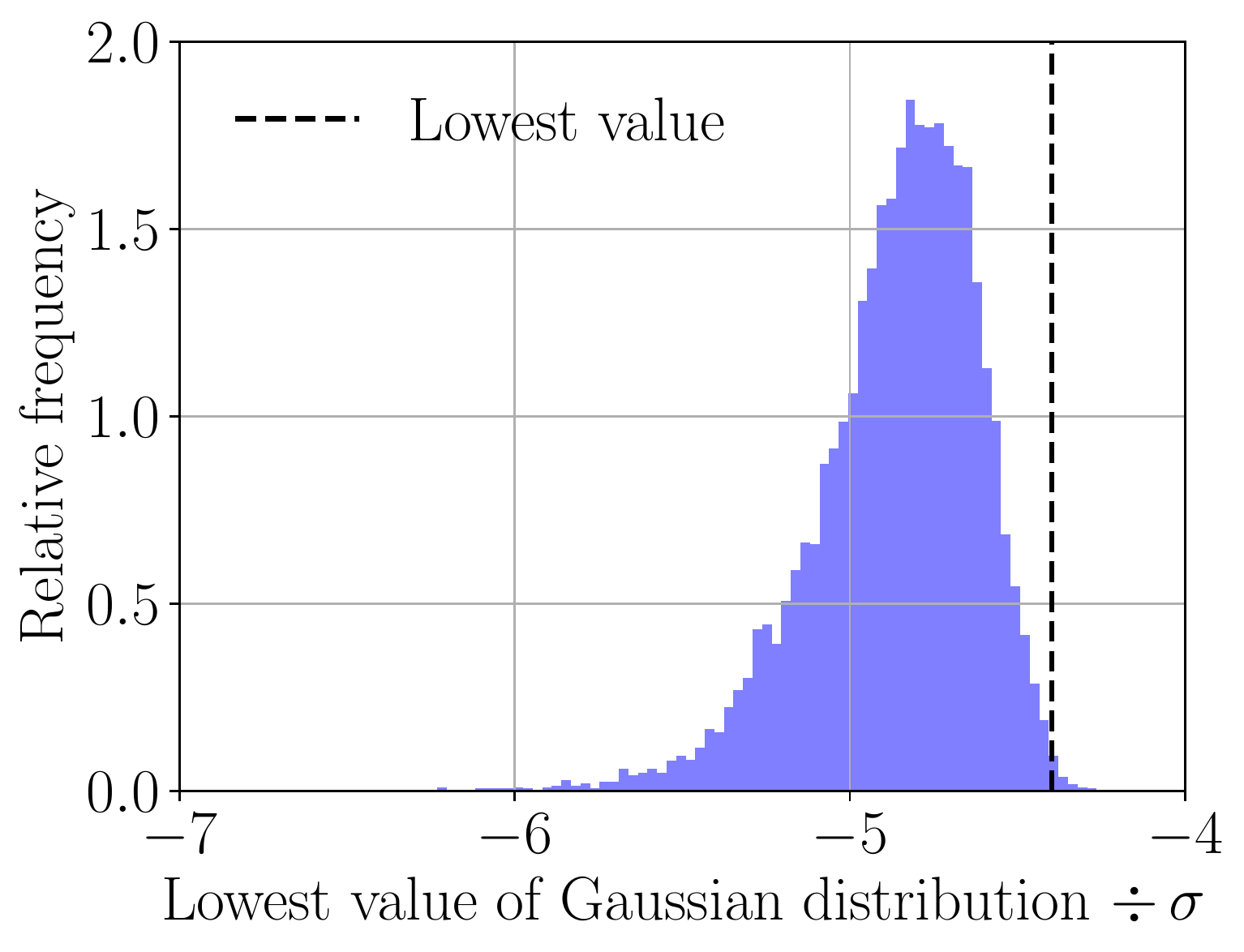}
    \includegraphics[width=59mm]{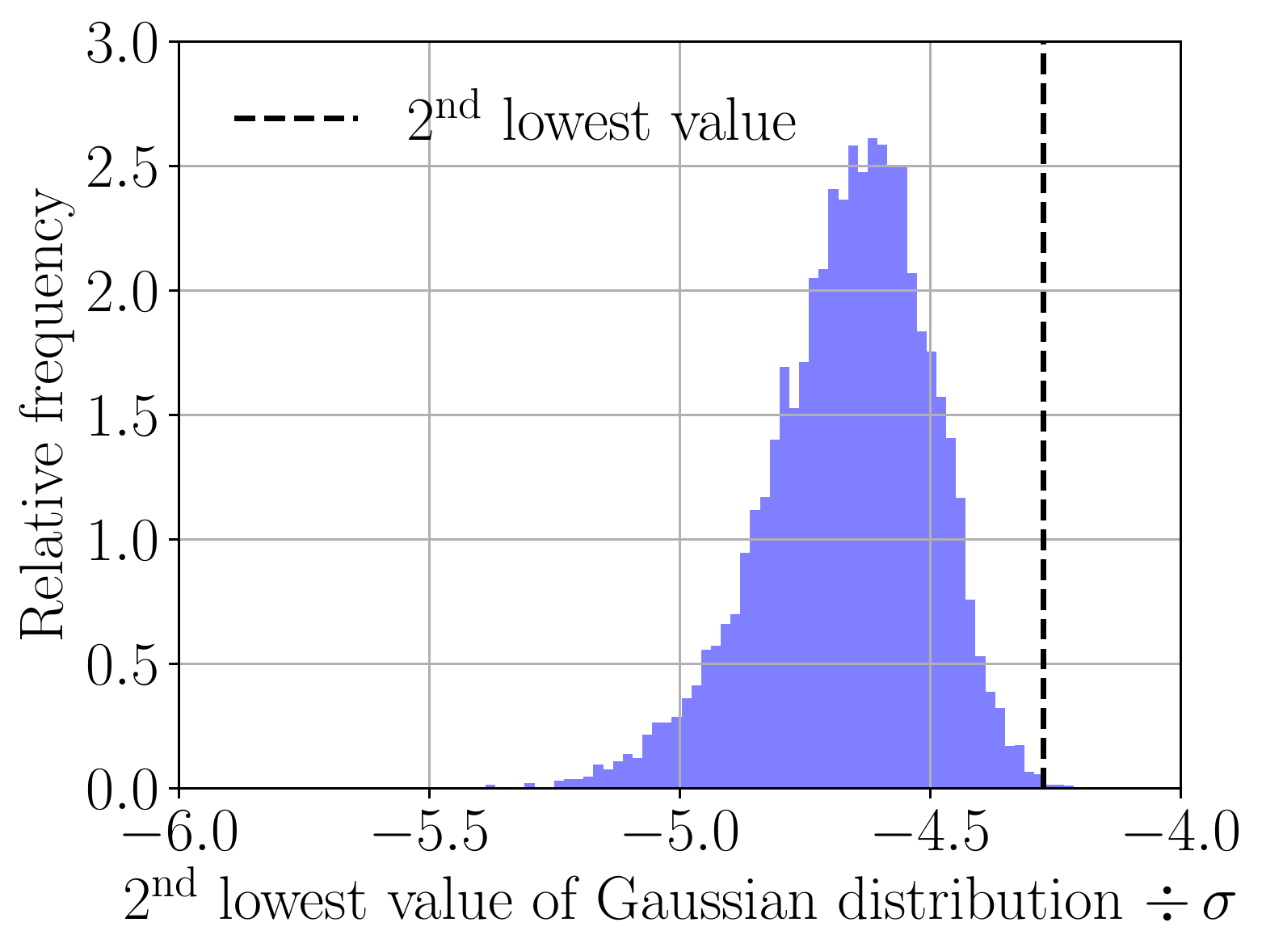}
    \includegraphics[width=56.5mm]{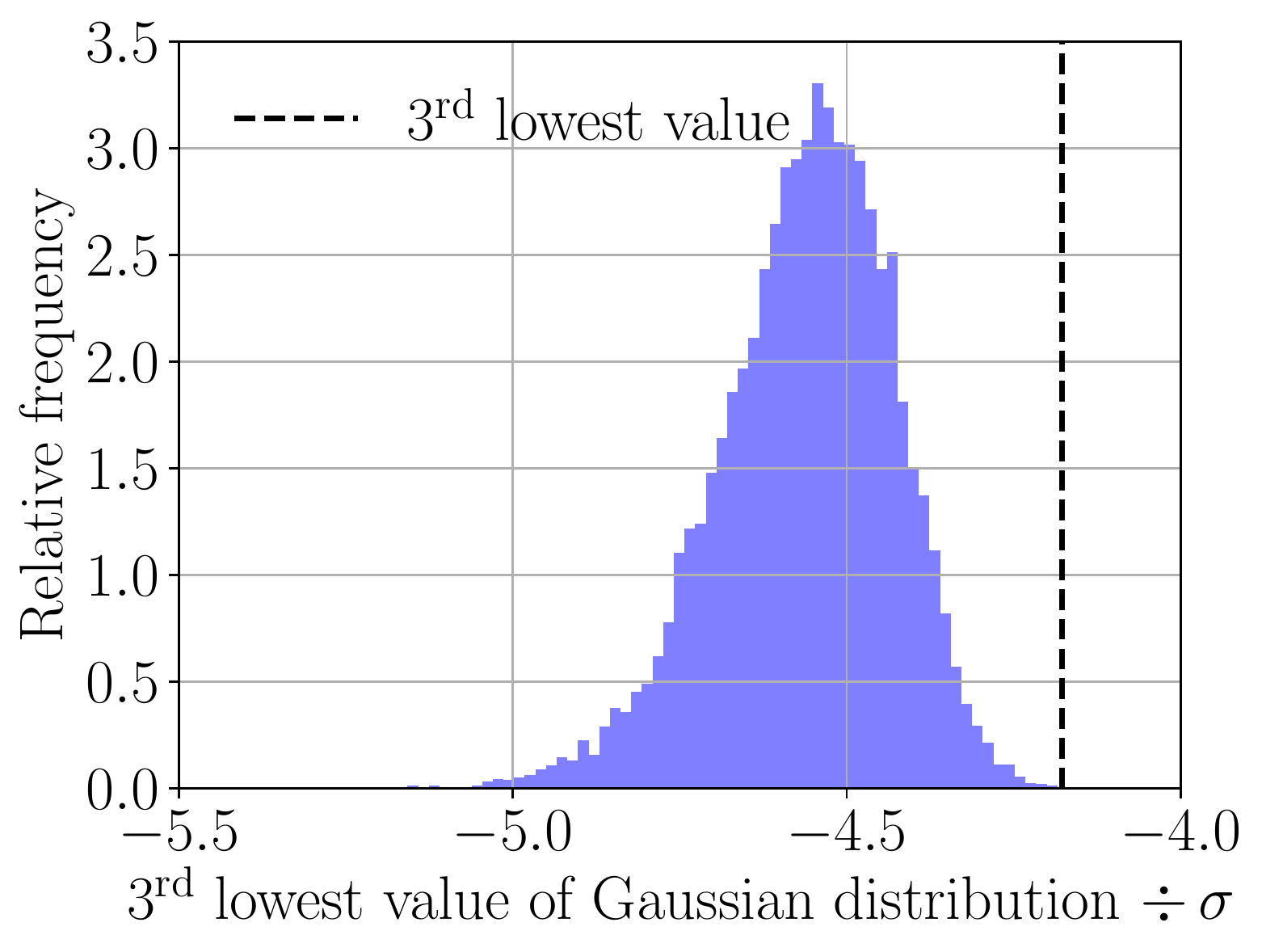}
\caption{Normality tests on the density fluctuations in the $\Lambda$CDM MXXL simulation within a spherical shell with an inner radius of $40 \, \rm{Mpc}$ and an outer radius of $300 \, \rm{Mpc}$ at redshift $z = 0$. The distributions (blue) show the lowest (left-hand panel), second (middle panel), and third (right-hand panel) lowest values generated using $10^{4}$ Monte Carlo trials, with each value shown based on $10^{6}$ Gaussian random numbers to mimic the number of vantage points used in MXXL. The dashed lines mark the lowest, second, and third lowest relative density contrast scaled by the rms fluctuations of the MXXL simulation (Section~\ref{subsec:MXXL simulation}).}
\label{figure_Gaussianity_check}
\end{figure*}

We perform a Gaussianity test to determine if the density fluctuations calculated in the MXXL simulation (Section~\ref{sec:LCDM framework}) follow a normal distribution. For this, we run $10^{4}$ Monte Carlo trials in which each time we select the three lowest values out of $10^{6}$ randomly generated Gaussian numbers. The left-hand panel of Figure~\ref{figure_Gaussianity_check} shows the distribution of the lowest value of each Monte Carlo trial compared with the lowest relative density contrast in MXXL scaled by the rms fluctuation. The same procedure is applied for the second and third lowest values in the middle and right-hand panels of Figure~\ref{figure_Gaussianity_check}, respectively. As expected, the lowest, second and third lowest values generated by the Monte Carlo trials cluster in a narrow region around $-5\sigma$. The three  most underdense regions in the MXXL simulation match roughly with the expected values from the Monte Carlo distributions, indicating that the MXXL density fluctuations closely follow a normal distribution.
 
\section{KBC void mass profiles} \label{Appendix: KBC void mass profile}

In addition to our fiducial MOND simulation based on a Maxwell-Boltzmann void density profile (Section~\ref{subsubsec:Void initial profiles}), we also model the void with a Gaussian and an exponential profile. The enclosed mass of the void within co-moving radius $r_{\mathrm{com}}$ for a Gaussian profile is 
\begin{align}
	M_{\mathrm{enc}} = 4 \mathrm{\pi} \rho_{0} r_{\mathrm{void}}^{3} \left( \frac{x^{3}}{3} - \alpha_{\mathrm{void}} \left[ \sqrt{\frac{\mathrm{\pi}}{2}} \mathrm{erf}\left( \frac{x}{\sqrt{2}} \right) - x \exp \left(-\frac{x^{2}}{2}\right) \right] \right) \, .
	\label{eq:governing_equation_Menc_Gauss}
\end{align}
As before, $x \equiv r_{\mathrm{com}}/r_{\mathrm{void}}$, $\alpha_{\mathrm{void}}$ is the initial void strength, and $r_{\mathrm{void}}$ is the initial co-moving void size at $z = 9$.

The corresponding result for an exponential profile is
\begin{align}
	M_{\mathrm{enc}} = 4 \mathrm{\pi} \rho_{0} r_{\mathrm{void}}^{3} \left( \frac{x^{3}}{3} - \alpha_{\mathrm{void}} \left[2 - \left( x^2 + 2x + 2 \right) \exp \left( -x \right) \right] \right).
	\label{eq:governing_equation_Menc_EXP}
\end{align}

In both cases, $\alpha_{\mathrm{void}}$ is the initial underdensity at the void centre. The results of using these void profiles are presented and compared with local observations in Appendix \ref{Appendix:Joint probabilities for different void profiles}.

\section{Results for different void profiles} \label{Appendix:Joint probabilities for different void profiles}

\begin{figure*}
    \includegraphics[width=\linewidth]{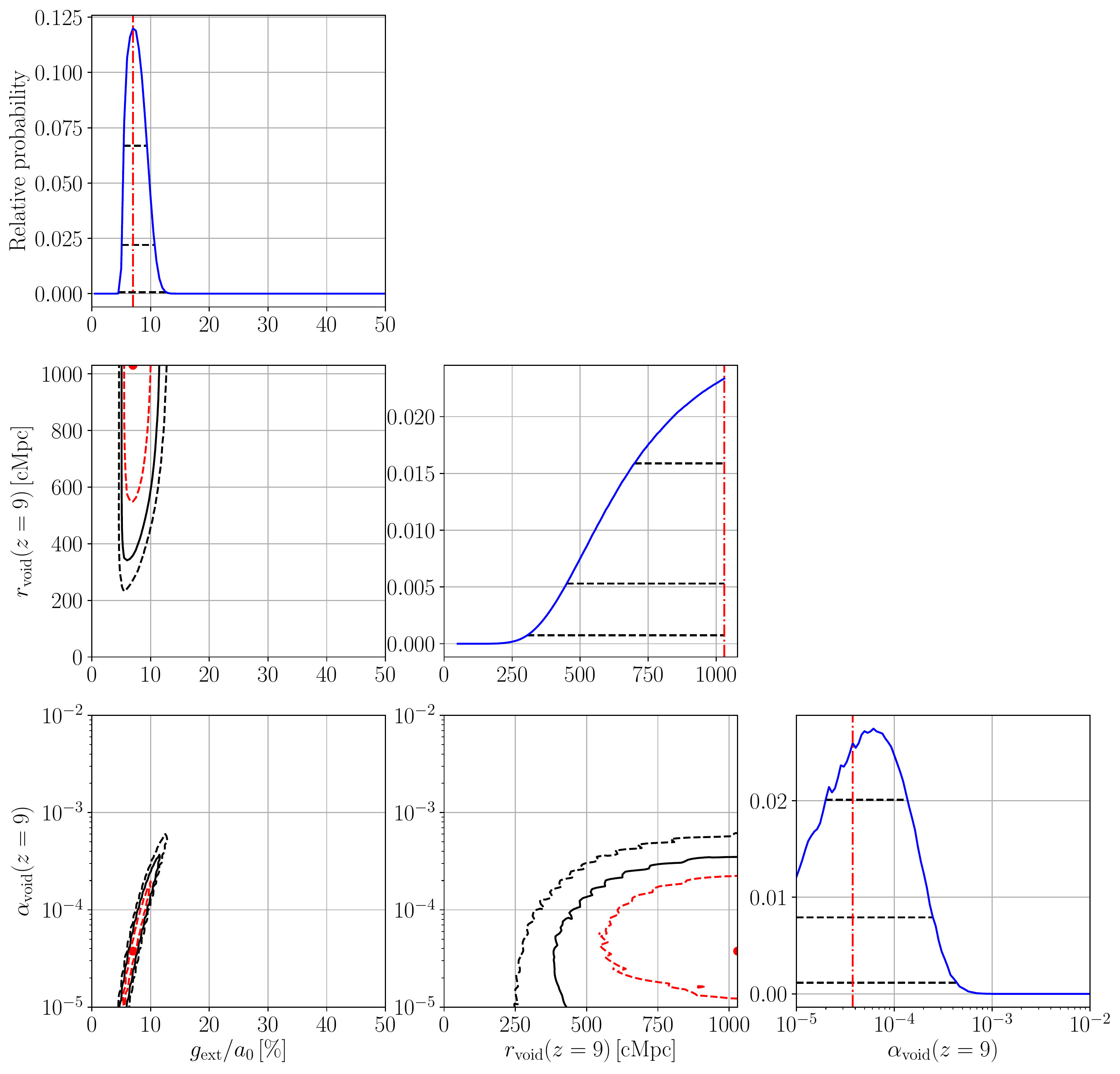}
    \caption{Similar to Figure~\ref{figure:marginalization_MBprofile}, but for a void modelled with a Gaussian profile (Equation~\ref{eq:governing_equation_Menc_Gauss}). The red dashed, black solid, and black dashed lines mark the $1 \sigma$, $2 \sigma$, and $3 \sigma$ confidence levels, respectively. For 1D posteriors, these are shown using horizontal black lines. The red dot or vertical line marks the best-fitting model with an external field strength of $g_\mathrm{ext} = 0.070 \, a_{_0}$, an initial void size of $r_{\mathrm{void}} = 1030 \, \rm{cMpc}$ (the upper limit of the allowed parameter range), and an initial void strength of $\alpha_{\mathrm{void}} = 3.76 \times 10^{-5}$ at $z = 9$.}
    \label{figure_MOND_GAUSSIAN_results}
\end{figure*}

\begin{figure*}
    \includegraphics[width=\linewidth]{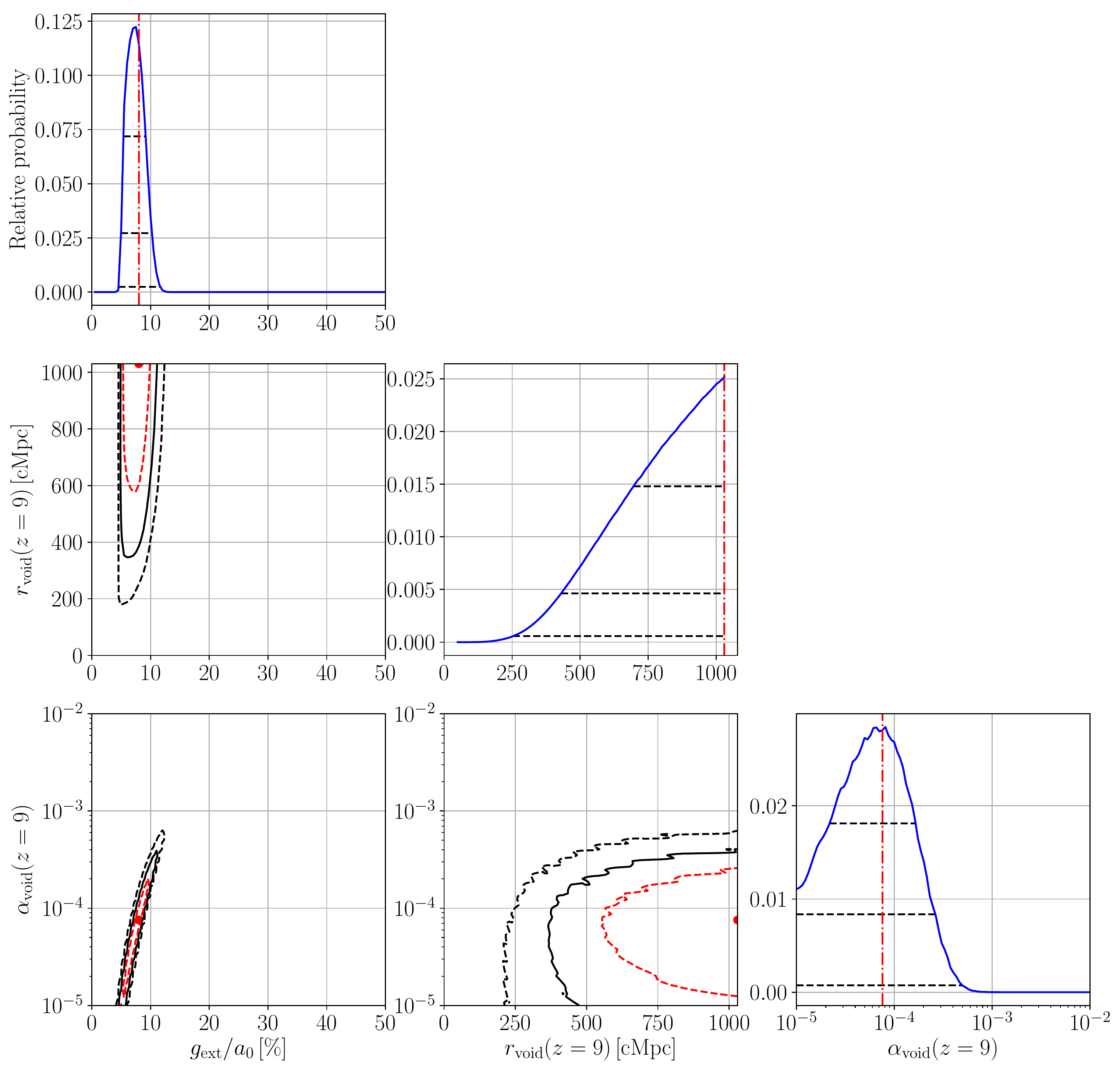}
    \caption{Similar to Figure~\ref{figure:marginalization_MBprofile}, but for a void modelled with an exponential profile (Equation~\ref{eq:governing_equation_Menc_EXP}). The red dashed, black solid, and black dashed lines mark the $1 \sigma$, $2 \sigma$, and $3 \sigma$ confidence levels, respectively. For 1D posteriors, these are shown using horizontal black lines. The red dot or vertical line marks the best-fitting model with an external field strength of $g_{\mathrm{ext}} = 0.080 \, a_{_0}$, an initial void size of $r_{\mathrm{void}} = 1030 \, \rm{cMpc}$ (the upper limit of the allowed range), and an initial void strength of $\alpha_{\mathrm{void}} = 7.56 \times 10^{-5}$ at $z = 9$.}
    \label{figure_MOND_EXP_results}
\end{figure*}

The marginalized posterior distribution of the model parameters based on $10^{6}$ MOND models for a Gaussian and an exponential initial void profile are shown in Figures~\ref{figure_MOND_GAUSSIAN_results} and \ref{figure_MOND_EXP_results}, respectively. All these models assume a time-independent EFE (i.e. $n_{\mathrm{EFE}} = 0$ in Equation~\ref{eq:time_dependent_external_field_history}). As with the Maxwell-Boltzmann profile, models with a very weak or a very strong EFE are ruled out, but the initial void parameters are only weakly constrained by local observations. In particular, models with a Gaussian and an exponential profile restrict $g_{\mathrm{ext}}$ to the range $\left(0.045 - 0.127 \right) a_{_0}$ and $\left(0.045 - 0.117 \right) a_{_0}$ at the $3 \sigma$ level, respectively.

The best-fitting model for a Gaussian void profile has an external field strength of $g_{\mathrm{ext}} = 0.070 \, a_{_0}$, an initial void size of $r_{\mathrm{void}} = 1030.0 \, \rm{cMpc}$ (the upper limit of the allowed parameter range), and an initial void strength of $\alpha_{\mathrm{void}} = 3.76\times10^{-5}$. This model is in $2.84\sigma$ ($0.45\%$) tension with local observations (Section~\ref{subsec:Observational constraints}).

For an exponential void profile, the best-fitting model has $g_{\mathrm{ext}} = 0.080 \, a_{_0}$, $r_{\mathrm{void}} = 1030.0 \, \rm{cMpc}$, and $\alpha_{\mathrm{void}} = 7.56 \times 10^{-5}$. The overall tension with observations is $2.93 \sigma$ ($0.34\%$).

The results for both models are listed and compared with observations in Table~\ref{table:comparison_bestfittingmodels_with_observations}. A time-dependent EFE and its implications for structure formation are studied in Section~\ref{subsubsec:Structure formation in MOND} for all three considered profiles.

\begin{table*}
    \caption{Similar to Table~\ref{table:comparison_bestfittingmodel_with_observations}, but now showing results for different void profiles. In all cases, we fix $n_{\mathrm{EFE}} = 0$.}
    \label{table:comparison_bestfittingmodels_with_observations}
    \begin{tabular}{lllllll} \hline 
     \multicolumn{7}{c}{Maxwell-Boltzmann density profile, $g_{\mathrm{ext}} = 0.055 \, a_{_0}$, $r_{\mathrm{void}} = 228.2 \, \rm{cMpc}$, $\alpha_{\mathrm{void}} = 3.76 \times 10^{-5}$, $v_{\mathrm{void}} = 1586 \, \rm{km\,s^{-1}}$, $r_{\mathrm{void}}^{\mathrm{rms}} = 528.7 \, \rm{Mpc}$, $n_{\mathrm{EFE}} = 0$} \\  \hline 
    Parameter & $H_{0}^{\mathrm{local}} \, [\rm{km\,s^{-1}\,Mpc^{-1}}]$ & $\overline{q}_{_0}^{\mathrm{local}}$ & $H_{0}^{\mathrm{lensing}} \, [\rm{km\,s^{-1}\,Mpc^{-1}}]$ & $v_{\mathrm{LG}} \, [\rm{km\,s^{-1}}]$ & $\delta_{\mathrm{in}}$ & $\delta_{\mathrm{out}}$ \\  
    Observations & $75.35 \pm 1.68$ & $1.08 \pm 0.29$ & $--$ & $627$ & $0.254 \pm 0.083$ & $-0.052 \pm 0.105$ \\  
    MOND model & $76.15$ & $1.07$ & See Figure~\ref{figure:best_fitting_model_Hubble_vs_redshift} & See Figure~\ref{figure_peculiar_velocity_results} & $0.172$ & $0.050$  \\
    $\chi^{2}$ & \multicolumn{2}{c}{$0.34$} & $14.66$ & $--$ & $0.99$ & $0.94$ \\
    Degrees of freedom & \multicolumn{2}{c}{$2$}  & $7$ & $--$ & $1$ & $1$ \\
    $\chi$ (1D Gaussian equivalent) & \multicolumn{2}{c}{$0.20$}  & $2.05$ & $2.34$ & $0.99$ & $0.97$ \\ \hline 
    \multicolumn{7}{c}{Gaussian density profile, $g_{\mathrm{ext}} = 0.070 \, a_{_0}$, $r_{\mathrm{void}} = 1030.0 \, \rm{cMpc}$, $\alpha_{\mathrm{void}} = 3.76 \times 10^{-5}$, $v_{\mathrm{void}} = 2018 \, \rm{km\,s^{-1}}$, $r_{\mathrm{void}}^{\mathrm{rms}} = 744.7 \, \rm{Mpc}$, $n_{\mathrm{EFE}} = 0$} \\ \hline 
    Parameter & $H_{0}^{\mathrm{local}} \, [\rm{km\,s^{-1}\,Mpc^{-1}}]$ & $\overline{q}_{_0}^{\mathrm{local}}$ & $H_{0}^{\mathrm{lensing}} \, [\rm{km\,s^{-1}\,Mpc^{-1}}]$& $v_{\mathrm{LG}} \, [\rm{km\,s^{-1}}]$ & $\delta_{\mathrm{in}}$ & $\delta_{\mathrm{out}}$ \\  
    Observations & $75.35 \pm 1.68$ & $1.08 \pm 0.29$ & $--$  & $627$ & $0.274 \pm 0.081$ & $-0.085 \pm 0.108$ \\
    MOND model & $77.24$ & $1.43$ & $--$ & $--$ & $0.155$ & $0.078$ \\
    $\chi^{2}$ & \multicolumn{2}{c}{$1.79$} & $12.74$ & $--$ & $2.19$ & $2.26$ \\
    Degrees of freedom & \multicolumn{2}{c}{$2$}  & $7$ & $--$ & $1$ & $1$ \\
    $\chi$ (1D Gaussian equivalent) & \multicolumn{2}{c}{$0.83$} & $1.76$ & $2.35$ & $1.48$ & $1.50$ \\ \hline 
    \multicolumn{7}{c}{Exponential density profile, $g_{\mathrm{ext}} = 0.080 \, a_{_0}$, $r_{\mathrm{void}} = 1030.0 \, \rm{cMpc}$, $\alpha_{\mathrm{void}} = 7.56 \times 10^{-5}$, $v_{\mathrm{void}} = 2307 \, \rm{km\,s^{-1}}$, $r_{\mathrm{void}}^{\mathrm{rms}} = 730.4 \, \rm{Mpc}$, $n_{\mathrm{EFE}} = 0$} \\ \hline 
    Parameter & $H_{0}^{\mathrm{local}} \, [\rm{km\,s^{-1}\,Mpc^{-1}}]$ & $\overline{q}_{_0}^{\mathrm{local}}$ & $H_{0}^{\mathrm{lensing}} \, [\rm{km\,s^{-1}\,Mpc^{-1}}]$ & $v_{\mathrm{LG}} \, [\rm{km\,s^{-1}}]$ & $\delta_{\mathrm{in}}$ & $\delta_{\mathrm{out}}$ \\ 
    Observations & $75.35 \pm 1.68$ & $1.08 \pm 0.29$ & $--$& $627$ & $0.276 \pm 0.080$ & $-0.078 \pm 0.108$ \\
    MOND model & $77.25$ & $1.46$ & $--$ & $--$ & $0.158$ & $0.073$ \\
    $\chi^{2}$ & \multicolumn{2}{c}{$1.98$}  & $13.19$ & $--$ & $2.17$ & $1.97$ \\
    Degrees of freedom & \multicolumn{2}{c}{$2$}  & $7$ & $--$ & $1$ & $1$ \\
    $\chi$ (1D Gaussian equivalent) & \multicolumn{2}{c}{$0.89$} & $1.83$ & $2.47$ & $1.47$ & $1.40$ \\ \hline
    \end{tabular}
\end{table*}

\end{appendix}

\bsp
\label{lastpage}
\end{document}